\documentclass[english]{article}
\usepackage[T1]{fontenc}
\usepackage[latin9]{inputenc}
\usepackage{color}
\usepackage{babel}
\usepackage{mathrsfs}
\usepackage{amsmath}
\usepackage{amssymb}
\usepackage{graphicx}
\usepackage{setspace}
\usepackage[unicode=true,pdfusetitle,
 bookmarks=true,bookmarksnumbered=false,bookmarksopen=false,
 breaklinks=false,pdfborder={0 0 1},backref=false,colorlinks=false]
 {hyperref}

\makeatletter

\providecommand{\tabularnewline}{\\}

\usepackage{cite}

\pdfoutput=1

\usepackage{mathabx}

\makeatother

\begin{document}
\noindent \textbf{\Large{}Pattern Storage, Bifurcations and Higher-Order
Correlation Structure of an Exactly Solvable Asymmetric Neural Network
Model }\\
{\Large \par}

\noindent \begin{flushleft}
Diego Fasoli$^{1,2,\ast}$, Anna Cattani$^{1}$, Stefano Panzeri$^{1}$ 
\par\end{flushleft}

\medskip{}

\noindent \begin{flushleft}
\textbf{{1} Laboratory of Neural Computation, Center for Neuroscience
and Cognitive Systems @UniTn, Istituto Italiano di Tecnologia, 38068
Rovereto, Italy}
\par\end{flushleft}

\noindent \begin{flushleft}
\textbf{{2} Center for Brain and Cognition, Computational Neuroscience
Group, Universitat Pompeu Fabra, 08002 Barcelona, Spain}
\par\end{flushleft}

\noindent \begin{flushleft}
\textbf{$\ast$ Corresponding Author. E-mail: diego.fasoli@upf.edu}
\par\end{flushleft}

\section*{Abstract}

Exactly solvable neural network models with asymmetric weights are
rare, and exact solutions are available only in some mean-field approaches.
In this article we find exact analytical solutions of an asymmetric
spin-glass-like model of arbitrary size and we perform a complete
study of its dynamical and statistical properties. The network has
discrete-time evolution equations, binary firing rates and can be
driven by noise with any distribution. We find analytical expressions
of the conditional and stationary joint probability distributions
of the membrane potentials and the firing rates. The conditional probability
distribution of the firing rates allows us to introduce a new learning
rule to store safely, under the presence of noise, point and cyclic
attractors, with important applications in the field of content-addressable
memories. Furthermore, we study the neuronal dynamics in terms of
the bifurcation structure of the network. We derive analytically examples
of the codimension one and codimension two bifurcation diagrams of
the network, which describe how the neuronal dynamics changes with
the external stimuli. In particular, we find that the network may
undergo transitions among multistable regimes, oscillatory behavior
elicited by asymmetric synaptic connections, and various forms of
spontaneous symmetry-breaking. On the other hand, the joint probability
distributions allow us to calculate analytically the higher-order
correlation structure of the network, which reveals neuronal regimes
where, statistically, the membrane potentials and the firing rates
are either synchronous or asynchronous. Our results are valid for
networks composed of an arbitrary number of neurons, but for completeness
we also derive the network equations in the mean-field limit and we
study analytically their local bifurcations. All the analytical results
are extensively validated by numerical simulations.

\section{Introduction \label{sec:Introduction}}

In biological neural networks, asymmetric synapses constitute $75-95\%$
of all synapses \cite{DeFelipe1999}. Yet, exactly solvable neural
network models with asymmetric synaptic connections are rare. Typically,
asymmetric models that admit exact solutions are spin-glass-like systems
of neural networks, such as the Little-Hopfield model \cite{Little1974,Hopfield1982}
and the Sherrington-Kirkpatrick model \cite{Sherrington1976,Morita1976}.
In neuroscience, these models were investigated in the '$70$s and
'$80$s, during a renewed interest in neural networks that followed
the pioneer work on the McCulloch-Pitts model \cite{McCulloch1943}
and the Perceptron \cite{Rosenblatt1958} ('$40$s and '$50$s respectively).
Due to their complexity, spin-glass-like models admit exact analytical
solutions only in some mean-field approaches. In \cite{Hertz1987},
Hertz et al. solved the Langevin dynamics of the $n$-component soft-spin
version of the asymmetric Sherrington-Kirkpatrick model in the limit
$n\rightarrow\infty$. In \cite{Crisanti1987}, Crisanti and Sompolinsky
found exact analytical expressions for the average autocorrelation
and susceptibility of a spherical asymmetric spin-glass model in the
mean-field limit of infinite network size. In \cite{Rieger1988},
Rieger et al. found exact solutions for the response and autocorrelation
functions of a fully asymmetric Sherrington-Kirkpatrick model, by
means of a path-integral approach. Moreover, in \cite{Derrida1987}
Derrida et al. found an analytical expression of the evolution of
a spin configuration having a finite overlap on one stored pattern
in a dilute version of the asymmetric Little-Hopfield model. Since
spin-glasses display features that are widespread in complex systems,
nowadays the interest in these models is still strong in fields of
research such as mathematics, physics, chemistry, and materials science
\cite{Bovier2012}.

The main difficulty in studying asymmetric neural networks is the
impossibility to apply the powerful methods of equilibrium statistical
mechanics, because no energy function exists for these systems. In
this work, we introduce an asymmetric spin-glass like neural network
model with random noise (Sec.~(\ref{sec:Materials-and-Methods}))
that can be solved exactly without adopting any mean-field approach
(Sec.~(\ref{sec:Results})). In particular, our solutions are valid
for networks composed of an arbitrary number of neurons, and do not
require additional constraints such as synaptic dilution.

Despite most of the work on spin-glass-like neural networks dates
back to the '$70$s and '$80$s, we set our study in the modern context
of bifurcation theory and functional connectivity analysis, which
nowadays are topics of central importance to systems neuroscience
\cite{Fasoli2016a,Fasoli2016b}. In SubSec.~(\ref{sub:Conditional-and-Joint-Probability-Distributions})
we derive exact analytical solutions for the conditional probability
distributions of the membrane potentials and the firing rates, as
well as for the joint probability distributions in the stationary
regime. In SubSec.~(\ref{sub:A-New-Learning-Rule-for-Storing-Point-and-Cyclic-Attractors})
we show that the conditional probability distribution of the firing
rates allows us to define a new learning rule to safely store stationary
and oscillating solutions even in presence of noise. In SubSec.~(\ref{sub:Bifurcations-in-the-Deterministic-Network})
we derive examples of exact codimension one and codimension two bifurcation
diagrams in the zero-noise limit of the network equations. Then, in
SubSec.~(\ref{sub:Higher-Order-Cross-Correlations}) we calculate
the higher-order correlation structure of the noisy network, for both
the membrane potentials and the firing rates. While all the results
are valid for networks composed of an arbitrary number of neurons,
for completeness in SubSec.~(\ref{sub:Mean-Field-Limit}) we also
derive the mean-field equations in the thermodynamic limit $N\rightarrow\infty$,
and we find exact analytical expressions of the local codimension
one bifurcations. To conclude, in Sec.~(\ref{sec:Discussion}) we
discuss the novelty and the biological implications of our results.

\section{Materials and Methods \label{sec:Materials-and-Methods}}

In this section we introduce the neural network model we study. For
the sake of clarity, in the main text of this article we suppose that
the network is driven by independent noise sources with Gaussian distribution,
while in the Supplementary Materials we consider noise with arbitrary
distribution.

Here we describe the neuronal activity by means of the following spin-glass-like
network model:

\begin{spacing}{0.8}
\begin{center}
{\small{}
\begin{equation}
V_{i}\left(t+1\right)=\frac{1}{M_{i}}\sum_{j=0}^{N-1}J_{ij}\mathscr{H}\left(V_{j}\left(t\right)-\theta_{j}\right)+I_{i}\left(t\right)+\sigma_{i}^{\mathcal{B}}\mathcal{B}_{i}\left(t\right),\quad i=0,...,N-1.\label{eq:discrete-time-voltage-based-equations}
\end{equation}
}
\par\end{center}{\small \par}
\end{spacing}

\noindent The network is synchronously updated, which is considered
a more realistic description of the biological dynamics \cite{Ermentrout1998,Peretto1984}.
Moreover, in Eq.\textcolor{blue}{~}(\ref{eq:discrete-time-voltage-based-equations})
$N\geq2$ represents the number of neurons in the network and is generally
finite. $V_{i}\left(t+1\right)$ is the membrane potential of the
$i$th neuron at the time instant $t+1$. The external current $I_{i}\left(t\right)$
is the deterministic component of the stimulus to the $i$th neuron.
$\sigma_{i}^{\mathcal{B}}\mathcal{B}_{i}$ is the stochastic component
of the stimulus, where $\mathcal{B}_{i}\left(t\right)$ for $i=0,...,N-1$
are independent normally distributed noise sources with unit variance.
$\sigma_{i}^{\mathcal{B}}$ represents the overall intensity (i.e.
the standard deviation) of the noisy term. Moreover, $\mathscr{H}\left(\cdot\right)$
is the Heaviside step function with threshold $\theta$:

\begin{spacing}{0.8}
\begin{center}
{\small{}
\begin{equation}
\mathscr{H}\left(V-\theta\right)=\begin{cases}
0 & \mathrm{if}\; V\leq\theta\\
\\
1 & \mathrm{otherwise},
\end{cases}\label{eq:Heaviside-step-function}
\end{equation}
}
\par\end{center}{\small \par}
\end{spacing}

\noindent which converts the membrane potential of the $j$th neuron
into its corresponding binary firing rate, $\nu_{j}\left(t\right)=\mathscr{H}\left(V_{j}\left(t\right)-\theta_{j}\right)\in\left\{ 0,1\right\} $.
Then, $J_{ij}$ is the synaptic weight from the $j$th (presynaptic)
neuron to the $i$th (postsynaptic) neuron. In this article, the matrix
$J$ is arbitrary, therefore it may be asymmetric and self-connections
(loops) may be present. $M_{i}$ is the total number of incoming connections
to the $i$th neuron, and represents a normalization factor that prevents
the divergence of the sum $\sum_{j=0}^{N-1}J_{ij}\mathscr{A}\left(V_{j}\left(t\right)\right)$
for large networks.

In the special case when $J$ is symmetric, invertible and with zeros
on the main diagonal ($J_{ii}=0$), while $I_{i}=\sigma_{i}^{\mathcal{B}}=0\;\forall i$,
under the following change of variables:

\begin{spacing}{0.8}
\begin{center}
{\small{}
\[
V_{i}\left(t\right)=\frac{1}{M_{i}}\sum_{j=0}^{N-1}J_{ij}A_{j}\left(t\right),
\]
}
\par\end{center}{\small \par}
\end{spacing}

\noindent Eq.~(\ref{eq:discrete-time-voltage-based-equations}) is
equivalent to the synchronous version of the network model introduced
by Hopfield in \cite{Hopfield1982}:

\begin{spacing}{0.8}
\begin{center}
{\small{}
\[
A_{i}\left(t+1\right)=\mathscr{H}\left(\frac{1}{M_{i}}\sum_{j=0}^{N-1}J_{ij}A_{j}\left(t\right)-\theta_{i}\right),\quad i=0,...,N-1.
\]
}
\par\end{center}{\small \par}
\end{spacing}

\noindent Important differences arise in the asymmetric model, as
we discuss in the next section.

\section{Results \label{sec:Results}}

In this section we study the dynamical and statistical properties
of the network equations (\ref{eq:discrete-time-voltage-based-equations}).
In SubSec.~(\ref{sub:Conditional-and-Joint-Probability-Distributions})
we report the exact solutions of the conditional probability distributions
of the membrane potentials and the firing rates, as well as the joint
probability distributions in the stationary regime. Then, in SubSec.~(\ref{sub:A-New-Learning-Rule-for-Storing-Point-and-Cyclic-Attractors})
we invert the formula of the conditional probability distribution
of the firing rates, which allows us to define a new learning rule
to safely store stationary and oscillatory patterns of neural activity
even in presence of noise. In SubSec.~(\ref{sub:Bifurcations-in-the-Deterministic-Network})
we study analytically the bifurcations of the network dynamics in
the zero-noise limit $\sigma^{\mathcal{B}}\rightarrow0$, in terms
of its codimension one and codimension two bifurcation diagrams. In
SubSec.~(\ref{sub:Higher-Order-Cross-Correlations}) we derive exact
analytical expressions for the higher-order cross-correlations of
the membrane potentials and the firing rates, for any noise intensity.
To conclude, in SubSec.~(\ref{sub:Mean-Field-Limit}) we report the
mean-field equations of the network, which are exact in the thermodynamic
limit $N\rightarrow\infty$, and we study analytically its local bifurcations.
All the analytical results are extensively validated by numerical
simulations.

\subsection{Conditional and Joint Probability Distributions \label{sub:Conditional-and-Joint-Probability-Distributions}}

We call $\boldsymbol{V}=\left[\begin{array}[t]{ccc}
V_{0}, & \ldots, & V_{N-1}\end{array}\right]^{T}$ the collection of the membrane potentials at time $t+1$, and $\boldsymbol{V}'$
that at time $t$. In the Supplementary Materials (see SubSec.~(S1.1.3))
we prove that if the membrane potentials evolve in time according
to Eq.~(\ref{eq:discrete-time-voltage-based-equations}), then their
conditional probability distribution is:

\begin{spacing}{0.8}
\begin{center}
{\small{}
\begin{equation}
p\left(\boldsymbol{V}|\boldsymbol{V}'\right)=\frac{1}{\left(2\pi\right)^{\frac{N}{2}}\prod_{i=0}^{N-1}\sigma_{i}^{\mathcal{B}}}\prod_{m=0}^{N-1}e^{-\frac{1}{2}\left(\frac{V_{m}-\frac{1}{M_{m}}\sum_{n=0}^{N-1}J_{mn}\mathscr{H}\left(V'_{n}-\theta_{n}\right)-I_{m}\left(t\right)}{\sigma_{m}^{\mathcal{B}}}\right)^{2}}.\label{eq:membrane-potentials-conditional-probability-distribution}
\end{equation}
}
\par\end{center}{\small \par}
\end{spacing}

\noindent Eq.~(\ref{eq:membrane-potentials-conditional-probability-distribution})
holds for any $t$ and also for time-varying stimuli $I_{i}\left(t\right)$.
Moreover, if the network statistics reach a stationary regime (typically
this occurs for $t\rightarrow\infty$ and for constant stimuli $I_{i}\left(t\right)=I_{i}\;\forall t$),
the joint probability distribution of the membrane potentials is:

\begin{spacing}{0.8}
\begin{center}
{\small{}
\begin{equation}
p\left(\boldsymbol{V}\right)=\frac{1}{\left(2\pi\right)^{\frac{N}{2}}\prod_{i=0}^{N-1}\sigma_{i}^{\mathcal{B}}}\sum_{j=0}^{2^{N}-1}F_{j}\prod_{m=0}^{N-1}e^{-\frac{1}{2}\left(\frac{V_{m}-\frac{1}{M_{m}}\sum_{n=0}^{N-1}J_{mn}\mathscr{B}_{j,n}^{\left(N\right)}-I_{m}}{\sigma_{m}^{\mathcal{B}}}\right)^{2}}.\label{eq:membrane-potentials-joint-probability-distribution}
\end{equation}
}
\par\end{center}{\small \par}
\end{spacing}

\noindent In Eq.~\ref{eq:membrane-potentials-joint-probability-distribution},
$\mathscr{B}_{j,n}^{\left(N\right)}\overset{\mathrm{def}}{=}\left[\pmb{\mathscr{B}}_{j}^{\left(N\right)}\right]_{n}$,
and $\pmb{\mathscr{B}}_{j}^{\left(N\right)}$ is the $N\times1$ vector
whose entries are the digits of the binary representation of $j$
(e.g. $\pmb{\mathscr{B}}_{5}^{\left(4\right)}=\left[\begin{array}{cccc}
0, & 1, & 0, & 1\end{array}\right]^{T}$, so that $\mathscr{B}_{5,0}^{\left(4\right)}=\mathscr{B}_{5,2}^{\left(4\right)}=0$
and $\mathscr{B}_{5,1}^{\left(4\right)}=\mathscr{B}_{5,3}^{\left(4\right)}=1$).
If we define $\widetilde{\boldsymbol{F}}\overset{\mathrm{def}}{=}\left[\begin{array}{ccc}
F_{0}, & \ldots, & F_{2^{N}-2}\end{array}\right]^{T}$, then:

\begin{spacing}{0.8}
\begin{center}
{\small{}
\[
\widetilde{\boldsymbol{F}}=\mathcal{A}^{-1}\widetilde{\boldsymbol{G}},
\]
}
\par\end{center}{\small \par}
\end{spacing}

\noindent while $F_{2^{N}-1}=1-\sum_{j=0}^{2^{N}-2}F_{j}$. Moreover,
$\mathcal{A}$ and $\widetilde{\boldsymbol{G}}$ are a $\left(2^{N}-1\right)\times\left(2^{N}-1\right)$
matrix and a $\left(2^{N}-1\right)\times1$ column vector respectively,
defined as follows:

\begin{spacing}{0.80000000000000004}
\begin{center}
{\small{}
\begin{align*}
 & \mathcal{A}_{i,j}=\delta_{i,j}+G_{i,2^{N}-1}-G_{i,j}\\
\\
 & \left[\widetilde{\boldsymbol{G}}\right]_{i}=G_{i,2^{N}-1}\\
\\
 & G_{i,j}=\frac{1}{2^{N}}\prod_{m=0}^{N-1}\left[1+\left(-1\right)^{\mathscr{B}_{i,m}^{\left(N\right)}}\mathrm{erf}\left(\frac{\theta_{m}-\frac{1}{M_{m}}\sum_{n=0}^{N-1}J_{mn}\mathscr{B}_{j,n}^{\left(N\right)}-I_{m}}{\sqrt{2}\sigma_{m}^{\mathcal{B}}}\right)\right],
\end{align*}
}
\par\end{center}{\small \par}
\end{spacing}

\noindent while $\delta_{i,j}$ is the Kronecker delta. $\mathcal{A}^{-1}$
can be calculated analytically from the cofactor matrix $\mathcal{C}$
of $\mathcal{A}$, through the relation $\mathcal{A}^{-1}=\frac{1}{\det\left(\mathcal{A}\right)}\mathcal{C}^{T}$,
with $\det\left(\mathcal{A}\right)=\sum_{j=0}^{N-1}\mathcal{C}_{ij}\mathcal{A}_{ij}$
(for any $i$) according to Laplace's formula. Then, from Eq.~(\ref{eq:membrane-potentials-joint-probability-distribution})
we obtain the following expression of the single-neuron marginal probability
distribution:

\begin{spacing}{0.8}
\begin{center}
{\small{}
\begin{equation}
p_{i}\left(V\right)=\frac{1}{\sqrt{2\pi}\sigma_{i}^{\mathcal{B}}}\sum_{j=0}^{2^{N}-1}F_{j}e^{-\frac{1}{2}\left(\frac{V-\frac{1}{M_{i}}\sum_{k=0}^{N-1}J_{ik}\mathscr{B}_{j,k}^{\left(N\right)}-I_{i}}{\sigma_{i}^{\mathcal{B}}}\right)^{2}},\quad i=0,\ldots N-1.\label{eq:membrane-potentials-marginal-probability-distribution}
\end{equation}
}
\par\end{center}{\small \par}
\end{spacing}

\begin{figure}
\begin{centering}
\includegraphics[scale=0.18]{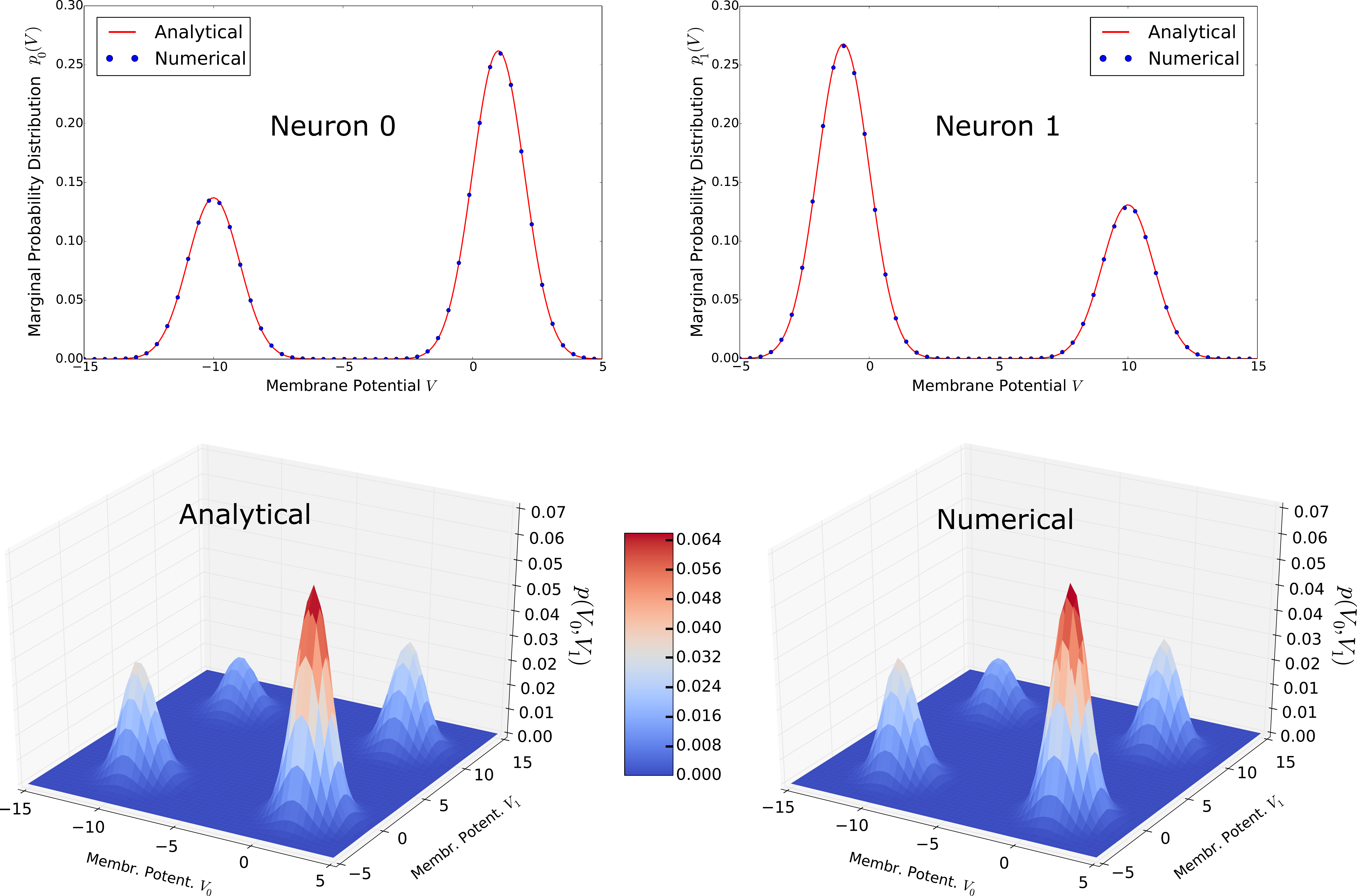}
\par\end{centering}

\protect\caption{\label{fig:two-neurons-0} \small\textbf{ Examples of probability
distributions for} $\boldsymbol{N=2}$\textbf{.} This figure is obtained
for the values of the parameters in Tab.~(\ref{tab:network-parameters-0}).
The red curves in the top panels show the single-neuron marginal probability
distributions of the two neurons, as given by Eq.~(\ref{eq:membrane-potentials-marginal-probability-distribution}).
The blue dots represent the numerical evaluation of the same distributions.
The bottom panels show the two-neurons joint probability distribution,
as given by Eq.~(\ref{eq:membrane-potentials-joint-probability-distribution})
(left), and numerically evaluated (right). The numerical distributions
of this figure have been obtained through a Monte Carlo method over
$10^{6}$ repetitions of the network dynamics, obtained by solving
iteratively Eq.~(\ref{eq:discrete-time-voltage-based-equations})
in the temporal interval $t=\left[0,100\right]$. Then, the probability
distributions have been evaluated from the collected data at $t=100$
through a kernel density estimator. We assume that at $t=100$ the
probability distributions have already reached a stationary regime,
which is confirmed by the good agreement between the analytical and
numerical results.}
\end{figure}

\begin{figure}
\begin{centering}
\includegraphics[scale=0.17]{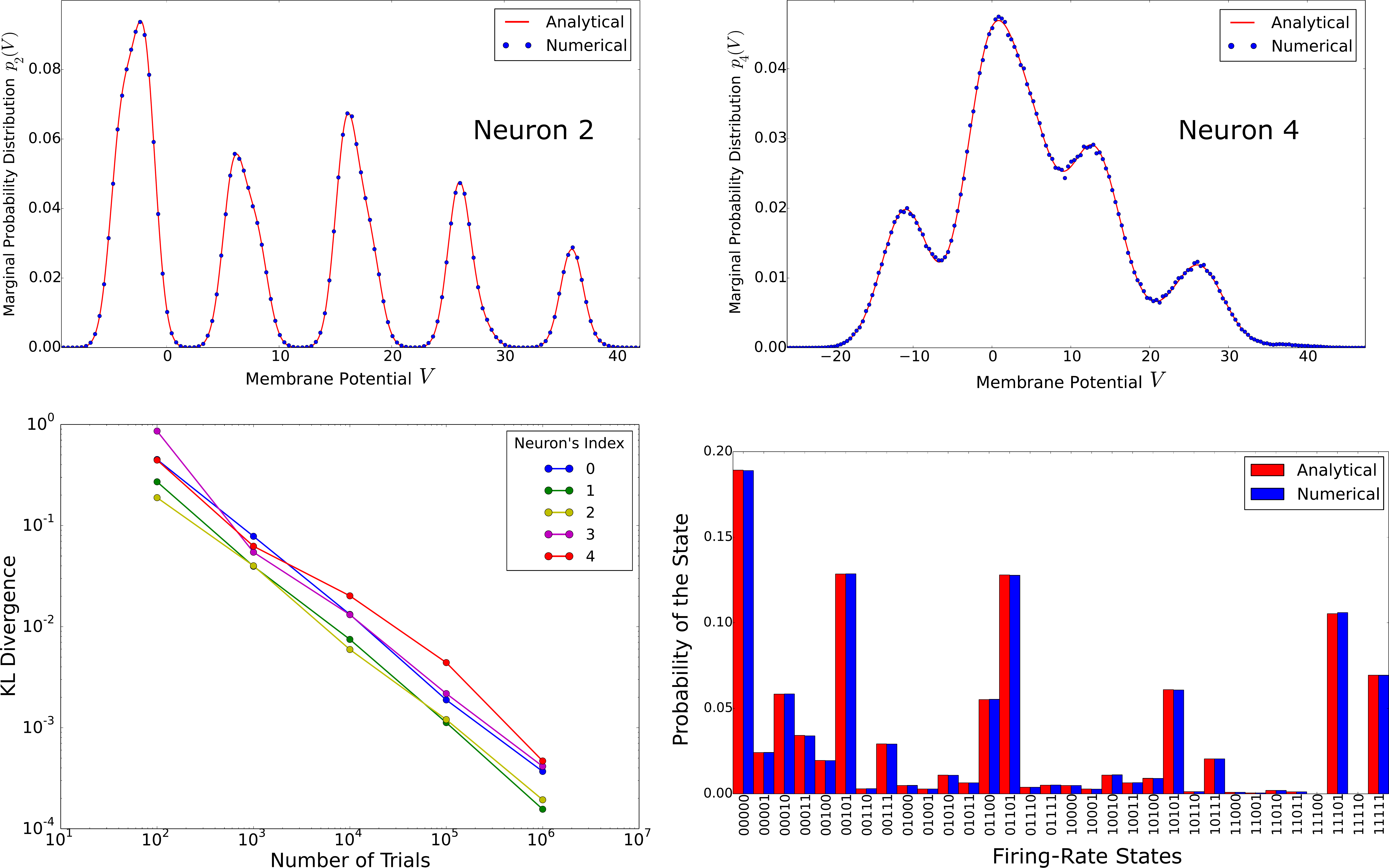}
\par\end{centering}

\protect\caption{\label{fig:five-neurons} \small\textbf{ Examples of probability
distributions for} $\boldsymbol{N=5}$\textbf{.} This figure is obtained
for the values of the parameters in Tab.~(\ref{tab:network-parameters-1}).
The top panels show two examples of the single-neuron marginal probability
distributions of the membrane potentials. The bottom-left panel shows
the Kullback\textendash Leibler divergence between the analytical
single-neuron distributions (as given by Eq.~(\ref{eq:membrane-potentials-marginal-probability-distribution})),
and the corresponding numerical distributions (calculated as in Fig.~(\ref{fig:two-neurons-0})),
for an increasing number of Monte Carlo repetitions (trials) of the
network dynamics. The divergence decreases with the number of repetitions,
therefore the numerical distribution is better approximated by its
analytical counterpart when the statistical error (due to the finite
number of repetitions) decreases. The bottom-right panel shows the
comparison between the analytical joint probability distribution of
the firing rates (red bars, calculated by Eq.~(\ref{eq:firing-rates-joint-probability-distribution})),
and the corresponding numerical distribution (blue bars), over the
$2^{N}=16$ states of the network activity.}
\end{figure}

\noindent Examples of the probability distributions of the membrane
potentials for $N=2$ and $N=5$ are shown in Fig.~(\ref{fig:two-neurons-0})
and in the top panels of Fig.~(\ref{fig:five-neurons}), respectively.
 
\begin{figure}
\begin{centering}
\includegraphics[scale=0.26]{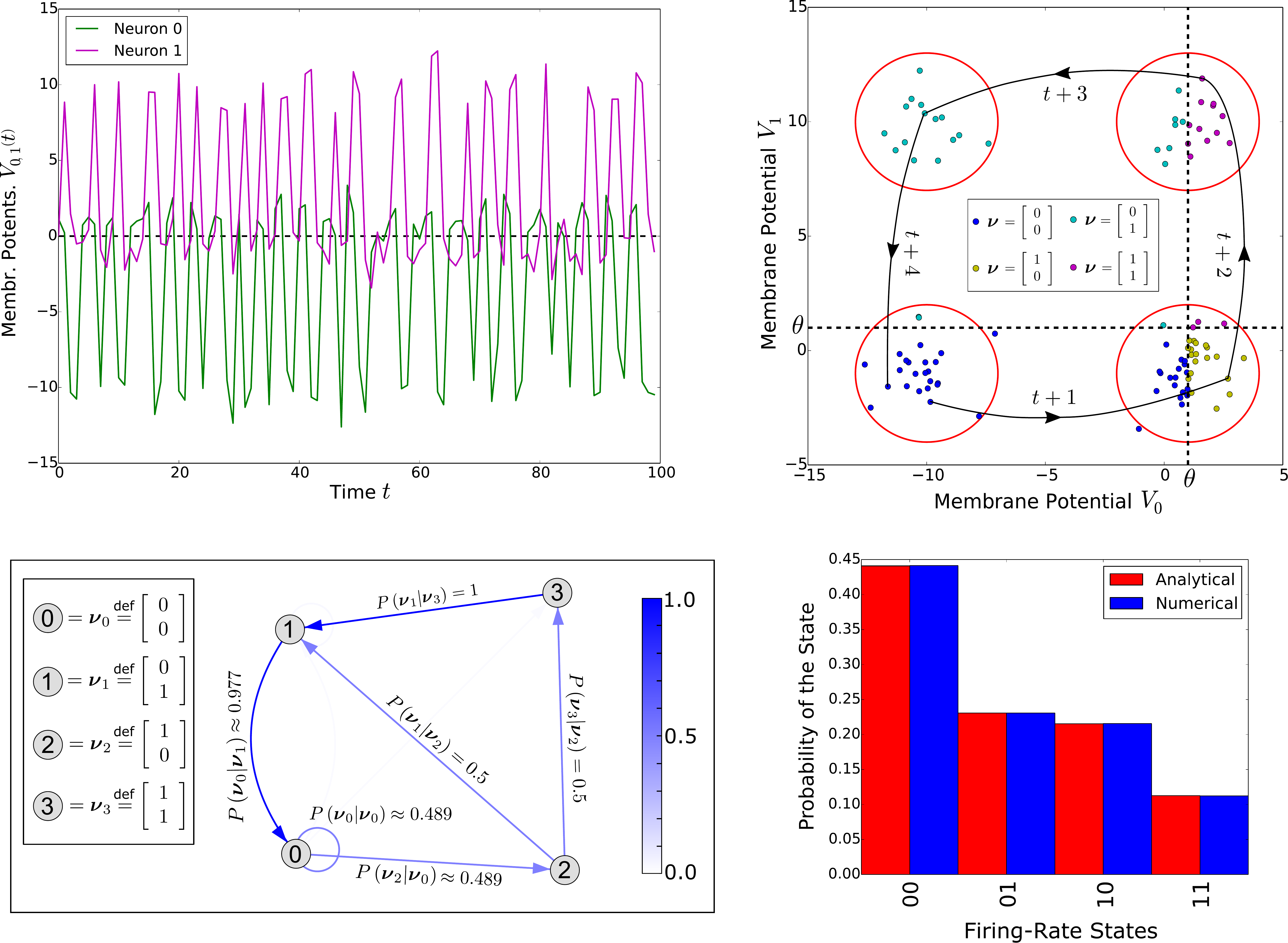}
\par\end{centering}

\protect\caption{\label{fig:two-neurons-1} \small \textbf{An example of oscillatory
dynamics for} $\boldsymbol{N=2}$\textbf{.} This figure is obtained
for the values of the parameters in Tab.~(\ref{tab:network-parameters-0}),
and shows an example of oscillatory dynamics that corresponds to the
stationary probability distributions of Fig.~(\ref{fig:two-neurons-0}).
The synchronous version of the Hopfield network (which has symmetric
connections) can sustain only oscillations with period $\mathcal{T}=2$,
known as \textit{two-cycles} \cite{GolesChacc1985}. Here we show
that the asymmetric network can undergo oscillations with period $\mathcal{T}=4$.
The oscillations are perturbed by the presence of the noisy terms
$\sigma_{i}^{\mathcal{B}}\mathcal{B}_{i}\left(t\right)$, while oscillations
with $\mathcal{T}>2$ in the noise-free version of the model are shown
in Fig.~(\ref{fig:example-of-codimension-one-bifurcation-diagram}).
The top-left panel shows an example of temporal dynamics of the membrane
potentials during the oscillation, while the top-right panel shows
the dynamics in the phase space of the model. The bottom-left panel
shows the transition probabilities between the four possible states
of the firing rates $\boldsymbol{\nu}$, according to Eq.~(\ref{eq:firing-rates-conditional-probability-distribution}).
Then, the bottom-right panel shows the comparison between the analytical
joint probability distribution of the firing rates (red bars, calculated
by Eq.~(\ref{eq:firing-rates-joint-probability-distribution})),
and the corresponding numerical distribution (blue bars).}
\end{figure}
 
\begin{table}
\begin{centering}
{\small{}}%
\begin{tabular}{|c|}
\hline 
\tabularnewline
{\small{}$J=\left[\begin{array}{cc}
0 & -11\\
11 & 0
\end{array}\right],\quad\boldsymbol{I}=\left[\begin{array}{c}
1\\
-1
\end{array}\right],\quad\boldsymbol{\sigma}^{\mathcal{B}}=\left[\begin{array}{c}
1\\
1
\end{array}\right]$}\tabularnewline
\tabularnewline
\hline 
\end{tabular}
\par\end{centering}{\small \par}

\protect\caption{\label{tab:network-parameters-0} \small Set of parameters used for
generating Figs.~(\ref{fig:two-neurons-0}) and (\ref{fig:two-neurons-1}).}
\end{table}
 
\begin{table}
\begin{centering}
{\small{}}%
\begin{tabular}{|c|}
\hline 
\tabularnewline
{\small{}$J=\left[\begin{array}{ccccc}
0 & 40 & -36 & 60 & -36\\
104 & 0 & -40 & 32 & -40\\
40 & 80 & 0 & 40 & -8\\
52 & 60 & -56 & 0 & -84\\
36 & 64 & -44 & 48 & 0
\end{array}\right],\quad\boldsymbol{I}=\left[\begin{array}{c}
-1\\
0\\
-2\\
2\\
0
\end{array}\right],\quad\boldsymbol{\sigma}^{\mathcal{B}}=\left[\begin{array}{c}
2\\
1\\
1\\
2\\
3
\end{array}\right]$}\tabularnewline
\tabularnewline
\hline 
\end{tabular}
\par\end{centering}{\small \par}

\protect\caption{\label{tab:network-parameters-1} \small Set of parameters used for
generating Fig.~(\ref{fig:five-neurons}).}
\end{table}

Neuroscientists often make use of measures of correlation between
firing rates, rather than between membrane potentials. For this reason,
it is interesting to evaluate also the probability distributions of
the firing rates, which in our model are binary quantities defined
as $\nu_{i}\left(t\right)\overset{\mathrm{def}}{=}\mathscr{H}\left(V_{i}\left(t\right)-\theta_{i}\right)$.
If we introduce the vector $\boldsymbol{\nu}=\left[\begin{array}[t]{ccc}
\nu_{0}, & \ldots, & \nu_{N-1}\end{array}\right]^{T}$, then from Eqs.~(\ref{eq:membrane-potentials-conditional-probability-distribution})
and (\ref{eq:membrane-potentials-joint-probability-distribution})
it is easy to prove (see SubSec.~(S1.2.1) of the Supplementary Materials)
that the conditional probability distribution, and the stationary
joint and single-neuron marginal distributions of the firing rates
are:

\begin{spacing}{0.8}
\begin{center}
{\small{}
\begin{align}
P\left(\boldsymbol{\nu}|\boldsymbol{\nu}'\right)= & \frac{1}{2^{N}}\prod_{m=0}^{N-1}\left[1+\left(-1\right)^{\nu_{m}}\mathrm{erf}\left(\frac{\theta_{m}-\frac{1}{M_{m}}\sum_{n=0}^{N-1}J_{mn}\nu_{n}'-I_{m}\left(t\right)}{\sqrt{2}\sigma_{m}^{\mathcal{B}}}\right)\right]\label{eq:firing-rates-conditional-probability-distribution}\\
\nonumber \\
P\left(\boldsymbol{\nu}\right)= & \frac{1}{2^{N}}\sum_{j=0}^{2^{N}-1}F_{j}\prod_{m=0}^{N-1}\left[1+\left(-1\right)^{\nu_{m}}\mathrm{erf}\left(\frac{\theta_{m}-\frac{1}{M_{m}}\sum_{n=0}^{N-1}J_{mn}\mathscr{B}_{j,n}^{\left(N\right)}-I_{m}}{\sqrt{2}\sigma_{m}^{\mathcal{B}}}\right)\right]\label{eq:firing-rates-joint-probability-distribution}\\
\nonumber \\
P_{i}\left(\nu\right)= & \frac{1}{2}\sum_{j=0}^{2^{N}-1}F_{j}\left[1+\left(-1\right)^{\nu}\mathrm{erf}\left(\frac{\theta_{i}-\frac{1}{M_{i}}\sum_{n=0}^{N-1}J_{in}\mathscr{B}_{j,n}^{\left(N\right)}-I_{i}}{\sqrt{2}\sigma_{i}^{\mathcal{B}}}\right)\right],\quad i=0,\ldots N-1\label{eq:firing-rates-marginal-probability-distribution}
\end{align}
}
\par\end{center}{\small \par}
\end{spacing}

\noindent respectively. We observe that while Eqs.~(\ref{eq:membrane-potentials-conditional-probability-distribution}),
(\ref{eq:membrane-potentials-joint-probability-distribution}) and
(\ref{eq:membrane-potentials-marginal-probability-distribution})
represent \textit{probability density functions} (pdfs), namely probability
distributions of continuous random variables (the membrane potentials),
Eqs.~(\ref{eq:firing-rates-conditional-probability-distribution}),
(\ref{eq:firing-rates-joint-probability-distribution}) and (\ref{eq:firing-rates-marginal-probability-distribution})
represent \textit{probability mass functions} (pmfs), namely probability
distributions of discrete random variables (the firing rates). In
particular, $P\left(\boldsymbol{\nu}|\boldsymbol{\nu}'\right)$ is
known as \textit{state-to-state transition probability matrix} in
the context of the theory of Markov processes\textcolor{blue}{{} }\cite{Feller1971}.
Examples of $P\left(\boldsymbol{\nu}\right)$ for $N=2$ and $N=5$
are shown in the bottom-right panels of Figs.~(\ref{fig:two-neurons-1})
and (\ref{fig:five-neurons}), respectively.

\subsection{A New Learning Rule for Storing Point and Cyclic Attractors \label{sub:A-New-Learning-Rule-for-Storing-Point-and-Cyclic-Attractors}}

At time $t$, the state of the neural network is described by the
vector of the firing rates, $\boldsymbol{\nu}\left(t\right)=\left[\begin{array}[t]{ccc}
\nu_{0}\left(t\right), & \ldots, & \nu_{N-1}\end{array}\left(t\right)\right]^{T}$, which represents the \textit{activity pattern} of the system at
time $t$. In the context of content-addressable memories, one aims
to determine a synaptic connectivity matrix $J$ that stores one or
more desired sequences of activity patterns. The way such matrix is
built defines a \textit{learning rule} for storing these patterns.

In particular, we suppose we want to store $D$ pattern sequences
$\boldsymbol{\nu}^{\left(i\right)}\left(t_{0}\right)\rightarrow\ldots\rightarrow\boldsymbol{\nu}^{\left(i\right)}\left(t_{L_{i}}\right)$
of length $L_{i}$, for $i=0,\ldots,D-1$. By inverting Eq.~(\ref{eq:firing-rates-conditional-probability-distribution}),
in Sec.~(S2) of the Supplementary Materials we prove that if the
network is fully-connected without loops, the matrix $J$ that stores
these pattern sequences satisfies the following sets of linear algebraic
equations:

\begin{spacing}{0.8}
\begin{center}
{\small{}
\begin{equation}
\Omega^{\left(j\right)}\boldsymbol{J}^{\left(j\right)}=\boldsymbol{u}^{\left(j\right)},\quad j=0,\ldots,N-1.\label{eq:learning-rule}
\end{equation}
}
\par\end{center}{\small \par}
\end{spacing}

\noindent In Eq.~(\ref{eq:learning-rule}), $\boldsymbol{J}^{\left(j\right)}$
is the $\left(N-1\right)\times1$ vector with entries $J_{jk}$ for
$k\neq j$ (the weights $J_{jj}$ are equal to zero, therefore they
are already known). Moreover, if we define $\mathcal{L}\overset{\mathrm{def}}{=}\sum_{i=0}^{D-1}L_{i}$
and $L_{-1}\overset{\mathrm{def}}{=}0$, then $\boldsymbol{u}^{\left(j\right)}$
is a $\mathcal{L}\times1$ vector with entries:

\begin{spacing}{0.8}
\begin{center}
{\small{}
\[
u_{L_{i-1}+n_{i}}^{\left(j\right)}=\left(N-1\right)\left[\theta_{j}-\left(-1\right)^{\nu_{j}^{\left(i\right)}\left(t_{n_{i}}+1\right)}K_{j}^{\left(i,n_{i}\right)}\sqrt{2}\sigma_{j}^{\mathcal{B}}-I_{j}\right],\quad n_{i}=0,\ldots,L_{i}-1,\quad i=0,\ldots,D-1,
\]
}
\par\end{center}{\small \par}
\end{spacing}

\noindent where $K_{j}^{\left(i,n_{i}\right)}$ is any sufficiently
large and positive constant. Moreover, $\Omega^{\left(j\right)}$
is the $\mathcal{L}\times\left(N-1\right)$ matrix obtained by removing
the $j$th column of the matrix:

\begin{spacing}{0.80000000000000004}
\noindent \begin{center}
{\small{}
\[
\Omega=\left[\begin{array}{ccc}
\nu_{0}^{\left(0\right)}\left(t_{0}\right) & \ldots & \nu_{N-1}^{\left(0\right)}\left(t_{0}\right)\\
\vdots & \ddots & \vdots\\
\nu_{0}^{\left(0\right)}\left(t_{L_{0}-1}\right) & \ldots & \nu_{N-1}^{\left(0\right)}\left(t_{L_{0}-1}\right)\\
\vdots & \ddots & \vdots\\
\nu_{0}^{\left(D-1\right)}\left(t_{0}\right) & \ldots & \nu_{N-1}^{\left(D-1\right)}\left(t_{0}\right)\\
\vdots & \ddots & \vdots\\
\nu_{0}^{\left(D-1\right)}\left(t_{L_{D-1}-1}\right) & \ldots & \nu_{N-1}^{\left(D-1\right)}\left(t_{L_{D-1}-1}\right)
\end{array}\right].
\]
}
\par\end{center}{\small \par}
\end{spacing}

\noindent Generally, each system in Eq.~(\ref{eq:learning-rule})
may be solved through the pseudoinverse of the matrix $\Omega^{\left(j\right)}$,
providing the set of synaptic weights that store the $D$ pattern
sequences $\boldsymbol{\nu}^{\left(i\right)}\left(t_{0}\right)\rightarrow\ldots\rightarrow\boldsymbol{\nu}^{\left(i\right)}\left(t_{L_{i}}\right)$
(even though solutions to Eq.~(\ref{eq:learning-rule}) do not always
exist, depending on $\Omega^{\left(j\right)}$ and $\boldsymbol{u}^{\left(j\right)}$).

\begin{figure}
\begin{centering}
\includegraphics[scale=0.21]{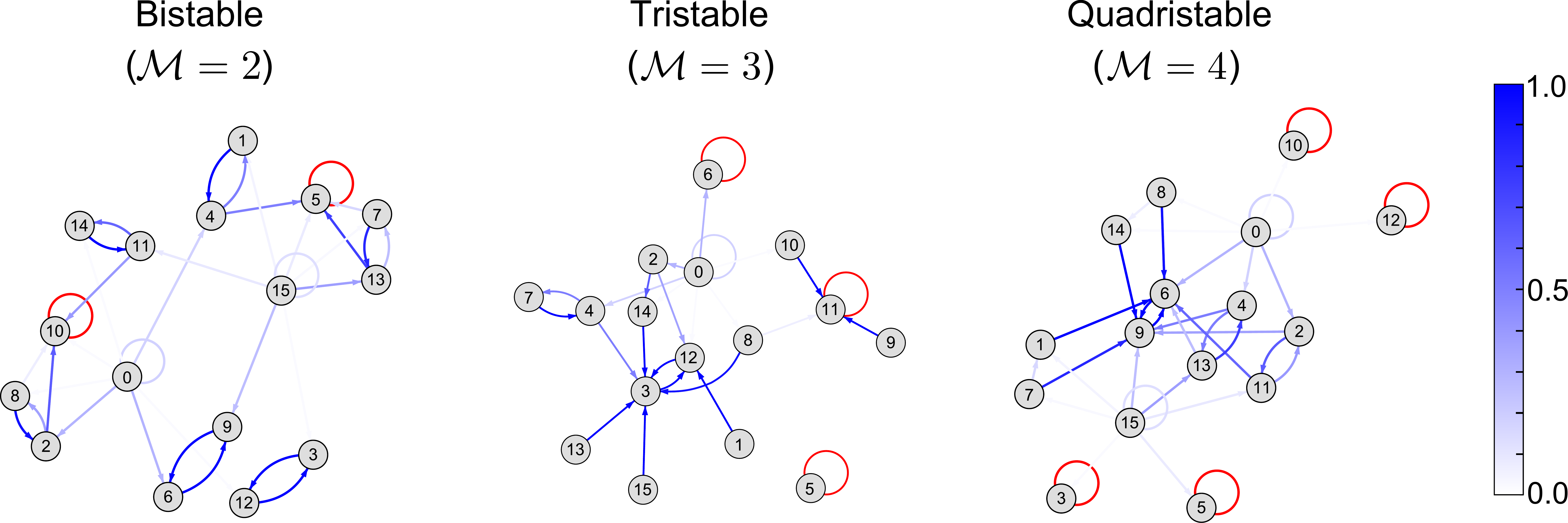}
\par\end{centering}

\protect\caption{\label{fig:multistability} \small \textbf{Neuronal multistability.}
Examples of stationary states stored in a network of size $N=4$,
by means of the learning rule (\ref{eq:learning-rule}). $\mathcal{M}$
represents the degree of multistability of the network, namely the
total number of stationary states. This figure plots the transitions
between the states of the network (from $0_{10}=0000_{2}$ to $15_{10}=1111_{2}$),
and the color of the arrows is determined by $P\left(\boldsymbol{\nu}|\boldsymbol{\nu}'\right)$
(see Eq.~(\ref{eq:firing-rates-conditional-probability-distribution}))
for the values of the parameters in Tab.~(\ref{tab:network-parameters-2}).
The stationary states are highlighted in red.}
\end{figure}

\begin{figure}
\begin{centering}
\includegraphics[scale=0.21]{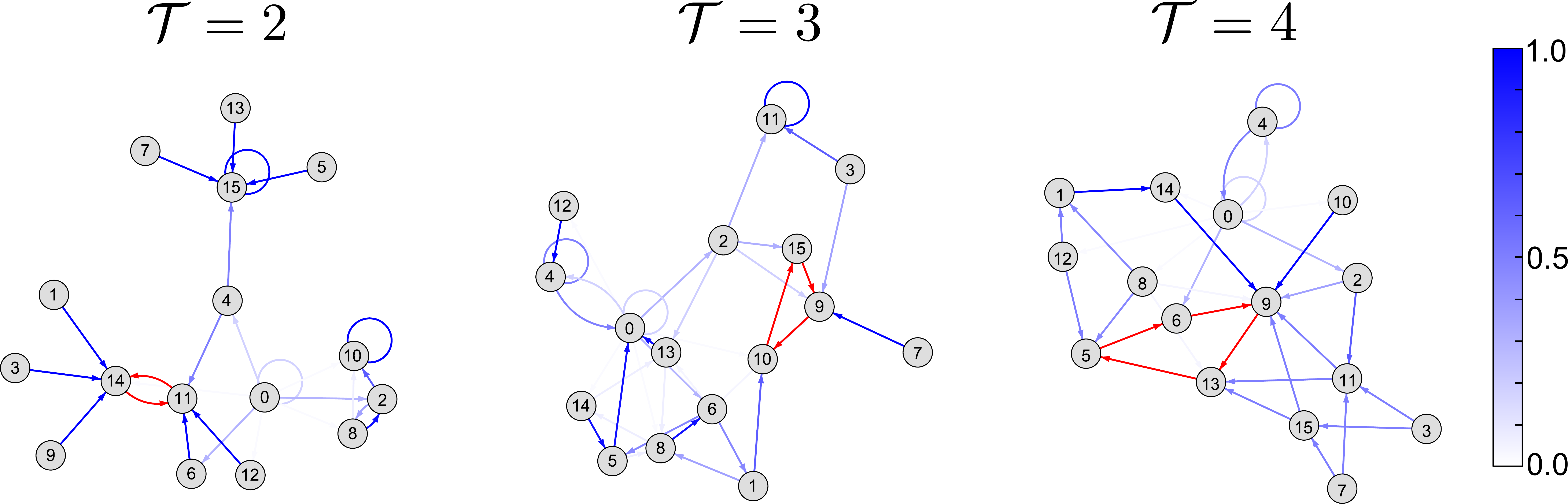}
\par\end{centering}

\protect\caption{\label{fig:oscillations} \small\textbf{ Neuronal oscillations.}
Examples of oscillatory states with period $\mathcal{T}$ stored in
a network of size $N=4$, by means of the learning rule (\ref{eq:learning-rule}).
This figure plots the conditional probability $P\left(\boldsymbol{\nu}|\boldsymbol{\nu}'\right)$
for the values of the parameters in Tab.~(\ref{tab:network-parameters-3}).
The oscillatory states are highlighted in red..}
\end{figure}

In particular, we observe that oscillations of period $\mathcal{T}$
correspond to the special case $L_{i}=\mathcal{T}$ with $\boldsymbol{\nu}^{\left(i\right)}\left(t\right)=\boldsymbol{\nu}^{\left(i\right)}\left(t+\mathcal{T}\right)$.
For $\mathcal{T}=1$ this condition allows us to store a stationary
state. Examples of $\mathcal{M}$-stable systems obtained through
this method are shown in Fig.~(\ref{fig:multistability}), for $\mathcal{M}=2,\,3,\,4$,
while examples of oscillatory dynamics with period $\mathcal{T}=2,\,3,\,4$
are shown in Fig.~(\ref{fig:oscillations}).

The learning rule Eq.~(\ref{eq:learning-rule}) can be easily extended
by including further constraints. For example, it is possible to relax
the full-connectivity assumption and to set a given portion of the
synaptic weights to zero. This allows us to define a learning rule
for sparse neural networks. The storage capacity depends on the number
of synaptic connections, but a detailed investigation is beyond the
purpose of this article. 
 
\begin{table}
\begin{centering}

\par\end{centering}

\begin{centering}
{\footnotesize{}}%
\begin{tabular}{|c|}
\hline 
\tabularnewline
{\footnotesize{}$J=\left[\begin{array}{cccc}
0 & -39 & 93 & -39\\
-84 & 0 & -84 & 171\\
123 & -66 & 0 & -66\\
-15 & 54 & -15 & 0
\end{array}\right],\quad J=\left[\begin{array}{cccc}
0 & -123 & 48 & 48\\
-510 & 0 & 171 & 171\\
378 & 123 & 0 & -255\\
138 & 54 & -84 & 0
\end{array}\right],\quad J=\left[\begin{array}{cccc}
0 & 93 & 93 & -171\\
171 & 0 & -339 & 171\\
123 & -255 & 0 & 123\\
-84 & 54 & 54 & 0
\end{array}\right]$}\tabularnewline
\tabularnewline
\hline 
\end{tabular}
\par\end{centering}{\footnotesize \par}

\smallskip{}

\begin{centering}
{\footnotesize{}}%
\begin{tabular}{|c|}
\hline 
\tabularnewline
{\footnotesize{}$\boldsymbol{I}=\left[\begin{array}{c}
-2\\
1\\
2\\
-3
\end{array}\right],\quad\boldsymbol{\sigma}^{\mathcal{B}}=\left[\begin{array}{c}
2\\
4\\
3\\
1
\end{array}\right]$}\tabularnewline
\tabularnewline
\hline 
\end{tabular}
\par\end{centering}{\footnotesize \par}

\protect\caption{\label{tab:network-parameters-2} \small Set of parameters used for
generating Fig.~(\ref{fig:multistability}). The synaptic connectivity
matrices $J$ have been obtained from Eq.~(\ref{eq:learning-rule})
for $K_{j}^{\left(i,n_{i}\right)}=10\;\forall i,\, j,\, n_{i}$.}
\end{table}
 
\begin{table}
\begin{centering}
{\footnotesize{}}%
\begin{tabular}{|c|}
\hline 
\tabularnewline
{\footnotesize{}$J=\left[\begin{array}{cccc}
0 & 30 & 63 & 30\\
-84 & 0 & -84 & 339\\
84 & 42 & 0 & 42\\
-15 & 84 & -15 & 0
\end{array}\right],\quad J=\left[\begin{array}{cccc}
0 & -93 & 93 & 93\\
171 & 0 & 3 & -339\\
123 & -255 & 0 & 3\\
-30 & 3 & 84 & 0
\end{array}\right],\quad J=\left[\begin{array}{cccc}
0 & -84 & 93 & -84\\
-171 & 0 & 171 & -171\\
42 & 42 & 0 & 42\\
-42 & -42 & 54 & 0
\end{array}\right]$}\tabularnewline
\tabularnewline
\hline 
\end{tabular}
\par\end{centering}{\footnotesize \par}

\protect\caption{\label{tab:network-parameters-3} \small Set of parameters used for
generating Fig.~(\ref{fig:oscillations}) ($\boldsymbol{I}$ and
$\boldsymbol{\sigma}^{\mathcal{B}}$ as in Tab.~(\ref{tab:network-parameters-2})).
The synaptic connectivity matrices $J$ have been obtained from Eq.~(\ref{eq:learning-rule})
for $K_{j}^{\left(i,n_{i}\right)}=10\;\forall i,\, j,\, n_{i}$.}
\end{table}

\subsection{Bifurcations in the Deterministic Network \label{sub:Bifurcations-in-the-Deterministic-Network}}

Bifurcation analysis is a mathematical technique for investigating
the qualitative change in the neuronal dynamics produced by varying
model parameters. Therefore it represents a fundamental tool for performing
a systematic analysis of the complexity of the neuronal activity patterns.
In particular, in this subsection we study how the dynamics of the
network depends on the external stimuli $I_{i}$. Moreover, we perform
our analysis in the zero-noise limit $\sigma^{\mathcal{B}}\rightarrow0$,
as is common practice in bifurcation theory (see e.g. \cite{Borisyuk1992,Grimbert2006,Haschke2005,Fasoli2016a}).
We observe that while local bifurcation analysis in graded models
is performed through the eigenvalues of the Jacobian matrix \cite{Kuznetsov1998},
in our case an alternative approach is required. Indeed, the Jacobian
matrix of our model is not defined at the discontinuity of the activation
function (\ref{eq:Heaviside-step-function}), thus preventing the
use of the powerful methods of bifurcation analysis developed for
graded models.

Since we study the bifurcations in terms of formation/destruction
of stationary and oscillatory solutions, the bifurcation diagrams
of Eq.~(\ref{eq:discrete-time-voltage-based-equations}) can be obtained
analytically from the conditional probability distribution of the
firing rates $P\left(\boldsymbol{\nu}|\boldsymbol{\nu}'\right)$ (see
Eq.~(\ref{eq:firing-rates-conditional-probability-distribution})),
which in the zero-noise limit $\sigma^{\mathcal{B}}\rightarrow0$
becomes:

\begin{spacing}{0.80000000000000004}
\begin{center}
{\small{}
\begin{equation}
P\left(\boldsymbol{\nu}|\boldsymbol{\nu}'\right)=\frac{1}{2^{N}}\prod_{j=0}^{N-1}\left[1+\left(-1\right)^{\nu_{j}}\mathrm{sgn}\left(\theta_{j}-\frac{1}{M_{j}}\sum_{k=0}^{N-1}J_{jk}\nu_{k}'-I_{j}\right)\right].\label{eq:conditional-probability-deterministic-limit}
\end{equation}
}
\par\end{center}{\small \par}
\end{spacing}

\noindent In Eq.~(\ref{eq:conditional-probability-deterministic-limit}),
$\mathrm{sgn}\left(\cdot\right)$ is the sign function, which is defined
as follows:

\begin{spacing}{0.80000000000000004}
\begin{center}
{\small{}
\[
\mathrm{sgn}\left(x\right)=\begin{cases}
-1 & \mathrm{if}\; x<0\\
\\
\hphantom{-}1 & \mathrm{otherwise}.
\end{cases}
\]
}
\par\end{center}{\small \par}
\end{spacing}

\noindent In particular, the stationary states $\boldsymbol{\nu}$
satisfy the condition $P\left(\boldsymbol{\nu}|\boldsymbol{\nu}\right)=1$.
Therefore, if the stationary states were known, it would be possible
to invert the condition $P\left(\boldsymbol{\nu}|\boldsymbol{\nu}\right)=1$
in the currents $I_{i}$, obtaining analytical expressions of the
range of the stimuli where the network admits the stationary solution
$\boldsymbol{\nu}$. In a similar way, each oscillatory solution of
the network has to satisfy the condition $P\left(\boldsymbol{\nu}|\boldsymbol{\nu}'\right)=1$
at the same time for each of its transitions $\boldsymbol{\nu}'\rightarrow\boldsymbol{\nu}$.
For example, given the oscillation $\boldsymbol{\nu}_{0}\rightarrow\boldsymbol{\nu}_{1}\rightarrow\boldsymbol{\nu}_{2}\rightarrow\boldsymbol{\nu}_{0}$,
it must be $P\left(\boldsymbol{\nu}_{0}|\boldsymbol{\nu}_{1}\right)=P\left(\boldsymbol{\nu}_{1}|\boldsymbol{\nu}_{2}\right)=P\left(\boldsymbol{\nu}_{2}|\boldsymbol{\nu}_{0}\right)=1$
for the same combination of stimuli. Again, by inverting analytically
these conditions, we get the range of the stimuli where the network
admits this oscillatory solution.

This approach requires prior knowledge of the stationary and oscillatory
solutions of the network. One possibility is to determine these solutions
numerically, for example by solving, for $\sigma_{i}^{\mathcal{B}}=0$,
Eq.~(\ref{eq:discrete-time-voltage-based-equations}) iteratively
for all the $2^{N}$ initial conditions of the firing rates and for
all the combinations of the currents $\left(I_{0},\ldots,I_{N-1}\right)$
on a sufficiently dense discretization of the stimulus space. Another
possibility is to perform a numerical calculation of the conditional
probability distribution (similarly to Figs.~(\ref{fig:multistability})
and (\ref{fig:oscillations})), from which the stationary and oscillatory
solutions can be detected through a search of the simple cycles of
length $L$ of the $2^{N}\times2^{N}$ binary matrix $P\left(\boldsymbol{\nu}|\boldsymbol{\nu}'\right)$.
In this approach, the stationary states correspond to loops of the
matrix $P\left(\boldsymbol{\nu}|\boldsymbol{\nu}'\right)$ (i.e. to
cycles of length $L=1$). In a similar way, oscillations correspond
to simple cycles of length $L=\mathcal{T}>1$, where $\mathcal{T}$
represents the period of the oscillation. More efficient techniques
will be considered in future work.

The bifurcation diagrams of the network strongly depend on its connectivity
matrix $J$. However, a detailed analysis of the relation between
the bifurcation structure and the network topology is beyond the purpose
of this article. For the sake of example, we apply our method to a
fully-connected network composed of $N_{E}=3$ excitatory and $N_{I}=3$
inhibitory neurons, even though this technique can be easily employed
for calculating the bifurcation diagrams of networks with any size
and topology. We suppose that each excitatory (respectively inhibitory)
neuron receives an external stimulus $I_{E}$ (respectively $I_{I}$),
and we derive the codimension two bifurcation diagram of the network
in the $I_{E}-I_{I}$ plane. The remaining parameters of the network
are reported in Tab.~(\ref{tab:network-parameters-4}).

\begin{figure}
\begin{centering}
\includegraphics[scale=0.27]{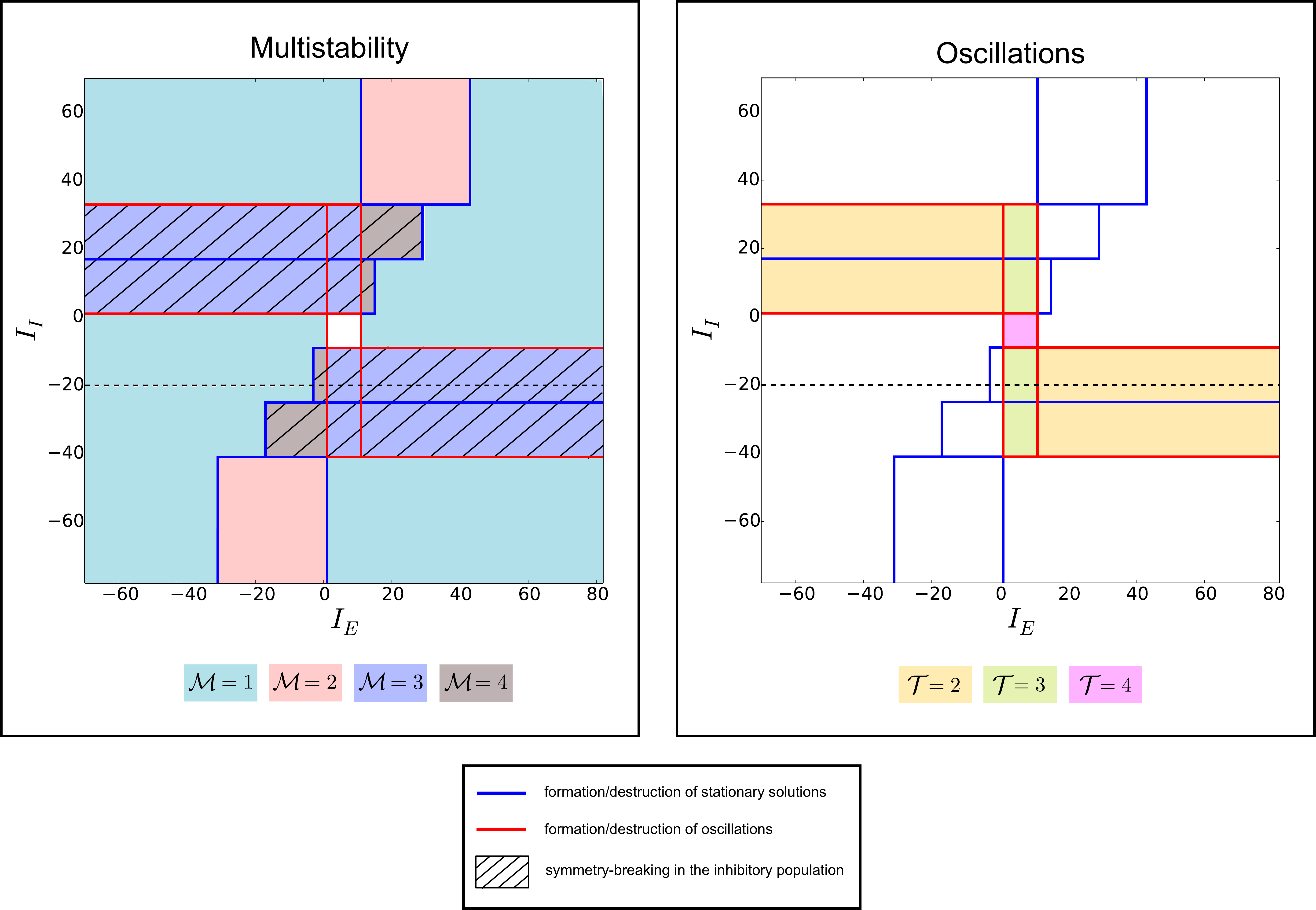}
\par\end{centering}

\protect\caption{\label{fig:example-of-codimension-two-bifurcation-diagram} \small\textbf{
An example of codimension two bifurcation diagram.} This diagram is
obtained for a a fully-connected network composed of $N_{\text{E}}=3$
excitatory and $N_{I}=3$ inhibitory neurons. Each excitatory (respectively
inhibitory) neuron receives an external stimulus $I_{E}$ (respectively
$I_{I}$), while the remaining parameters of the network are reported
in Tab.~(\ref{tab:network-parameters-4}). The two panels of the
figure refer to the same bifurcation diagram, but have been separated
for clarity in order to show how multistability (left panel) and oscillations
(right panel) depend on the external stimuli. In more detail, the
left panel shows how different combinations of the currents $I_{E,I}$,
for this specific network architecture, give rise to a number of stationary
solutions $\mathcal{M}$ that ranges from $1$ (monostability) to
$4$ (quadristability). Moreover, the right panel shows how, for different
values of the stimuli, the network undergoes oscillations with period
$\mathcal{T}=2$, $3$ or $4$. The blue lines in the diagram represent
the combinations of the current $I_{E,I}$ at which a bifurcation
occurs. Here, the stationary states lose their stability, turning
into different stationary states or into oscillatory solutions. In
a similar way, the red lines represent the bifurcations at which oscillations
turn into new oscillations or into stationary solutions. For this
reason, some blue and red lines in the diagram overlap. The readers
are referred to SubSecs.~(S3.1) and (S3.2) of the Supplementary Materials
for the derivation of their analytical formulas. The shaded areas
represent the regions of the $I_{E}-I_{I}$ plane where the symmetry
of the inhibitory neurons is broken, despite the symmetry of the underlying
neural equations. This occurs for example with the stationary state
$\boldsymbol{\nu}=\left[\protect\begin{array}[t]{cccccc}
0 & 0 & 0 & 1 & 0 & 0\protect\end{array}\right]^{T}$, since in this case only two inhibitory neurons over three do not
fire. On the contrary, in this example the symmetry in the excitatory
population is never broken, although this may occur with different
network sizes or with different topologies of the synaptic connections.
The horizontal dashed line in the panels corresponds to the value
of the current to the inhibitory population ($I_{I}=-20$) that we
have chosen for the calculation of the codimension one bifurcation
diagrams of Fig.~(\ref{fig:example-of-codimension-one-bifurcation-diagram}).}
\end{figure}

By representing the firing rates of the excitatory neurons through
the three top entries of the vector $\boldsymbol{\nu}$, we found
numerically that the network admits the stationary states $0-7$ and
$56-63$ (in decimal representation), for particular combinations
of $I_{E,I}$. Thus for example the state $\boldsymbol{\nu}=\left[\begin{array}[t]{cccccc}
0 & 0 & 0 & 1 & 0 & 1\end{array}\right]^{T}$ (i.e. the state $5$ in decimal representation), which is characterized
by two active inhibitory neurons (while the remaining neurons in the
network are not firing), is a stationary state for some values of
the stimuli. Moreover, we found numerically that the network undergoes
the oscillations $0\rightarrow7\rightarrow0$, $56\rightarrow63\rightarrow56$,
$0\rightarrow56\rightarrow63\rightarrow0$, $0\rightarrow63\rightarrow7\rightarrow0$
and $0\rightarrow56\rightarrow63\rightarrow7\rightarrow0$. Now, by
inverting the conditions provided by the conditional probability distribution
(\ref{eq:conditional-probability-deterministic-limit}), we get the
portions of the $I_{E}-I_{I}$ plane where each stationary state and
oscillation occurs. The details of the analytical calculations are
shown in SubSecs.~(S3.1) and (S3.2) of the Supplementary Materials,
for the stationary and oscillatory solutions respectively. The resulting
codimension two bifurcation diagram is shown in Fig.~(\ref{fig:example-of-codimension-two-bifurcation-diagram}).
 
\begin{table}
\begin{centering}
{\small{}}%
\begin{tabular}{|c|}
\hline 
\tabularnewline
{\small{}$\boldsymbol{\theta}=\left[\begin{array}{c}
1\\
\vdots\\
1
\end{array}\right],\quad J=\left[\begin{array}{cc}
\mathfrak{J}_{EE} & \mathfrak{J}_{EI}\\
\mathfrak{J}_{IE} & \mathfrak{J}_{II}
\end{array}\right],\quad\mathfrak{J}_{\alpha\alpha}=J_{\alpha\alpha}\left[\begin{array}{ccc}
0 & 1 & 1\\
1 & 0 & 1\\
1 & 1 & 0
\end{array}\right],\quad\mathfrak{J}_{\alpha\beta}=J_{\alpha\beta}\left[\begin{array}{ccc}
1 & 1 & 1\\
1 & 1 & 1\\
1 & 1 & 1
\end{array}\right]\;\mathrm{for}\;\alpha\neq\beta$}\tabularnewline
\tabularnewline
\hline 
\end{tabular}
\par\end{centering}{\small \par}

\smallskip{}

\begin{centering}
{\small{}}%
\begin{tabular}{|c|}
\hline 
\tabularnewline
{\small{}$J_{EE}=-J_{II}=80,\quad J_{IE}=-J_{EI}=70$}\tabularnewline
\tabularnewline
\hline 
\end{tabular}
\par\end{centering}{\small \par}

\protect\caption{\label{tab:network-parameters-4} \small Set of parameters used for
generating Figs.~(\ref{fig:example-of-codimension-two-bifurcation-diagram})
and (\ref{fig:example-of-codimension-one-bifurcation-diagram}).}
\end{table}

\begin{figure}
\begin{centering}
\includegraphics[scale=0.16]{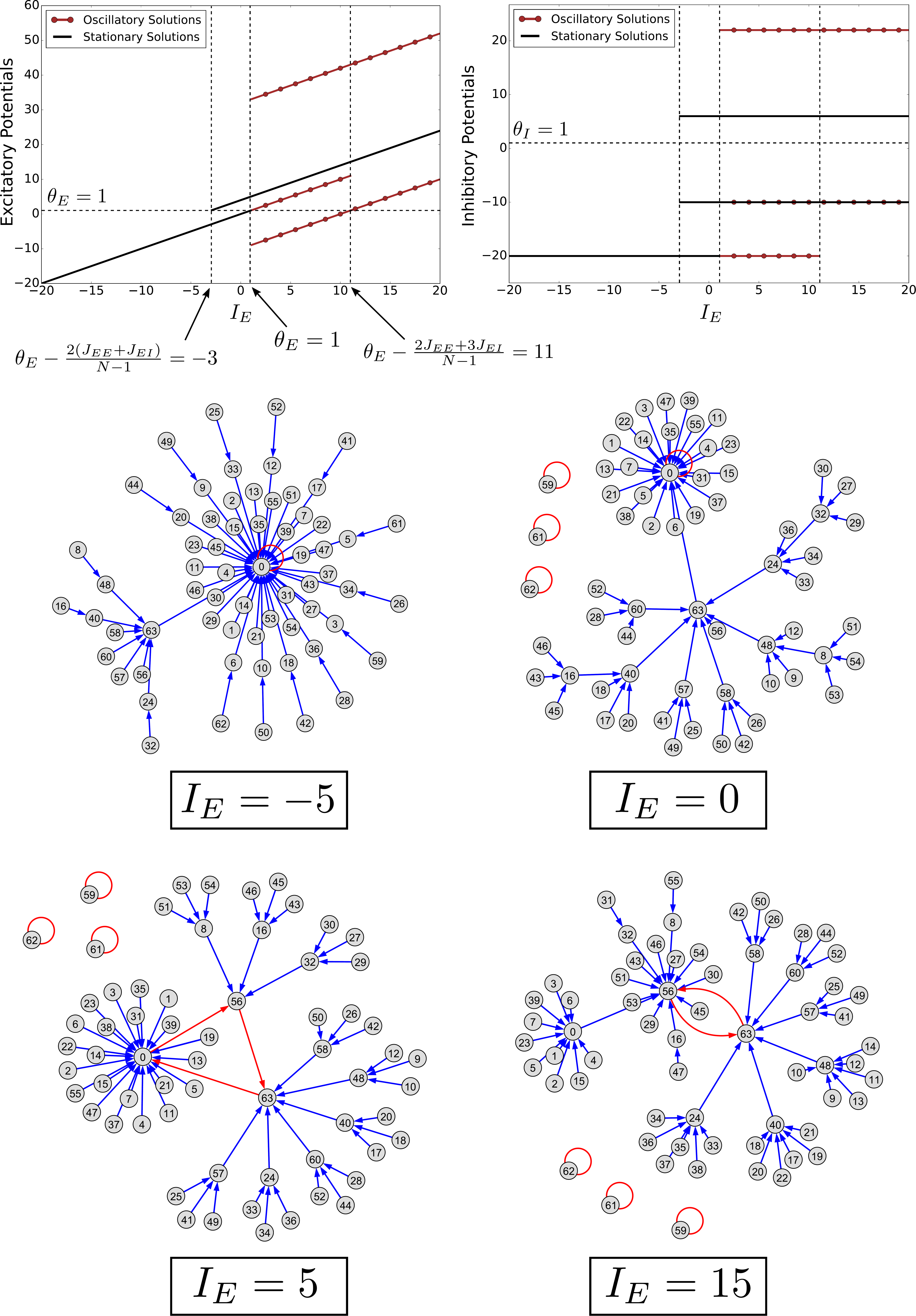}
\par\end{centering}

\protect\caption{\label{fig:example-of-codimension-one-bifurcation-diagram} \small\textbf{
Examples of codimension one bifurcation diagrams.} This figure is
obtained in the case of the fully-connected network discussed in SubSec.~(\ref{sub:Bifurcations-in-the-Deterministic-Network}),
for $I_{I}=-20$ (see the horizontal dashed line in Fig.~(\ref{fig:example-of-codimension-two-bifurcation-diagram})).
The top panels represent the codimension one bifurcation diagrams
of the excitatory (left) and inhibitory (right) neurons, obtained
from Eq.~(\ref{eq:codimension-one-bifurcation-diagram}) as a function
of the stimulus $I_{E}$. In particular, from the firing rates $\nu_{E}$
and $\nu_{I,j}$ of the stationary states, Eq.~(\ref{eq:codimension-one-bifurcation-diagram})
provides the fixed point solutions of the membrane potentials (black
lines). In a similar way, from the firing rates of the oscillatory
states we obtain the fixed point cycle solutions (brown lines). The
remaining panels of the figure show all the possible bifurcations
of the firing rates, for increasing $I_{E}$. The graphs have been
obtained from Eq.~(\ref{eq:conditional-probability-deterministic-limit}),
and highlight in red all the stationary and oscillatory solutions
of the network dynamics (compare with the areas crossed by the dashed
line in Fig.~(\ref{fig:example-of-codimension-two-bifurcation-diagram})
when moving from left to right).}
\end{figure}

The codimension one bifurcation diagrams can be derived analytically
from Eq.~(\ref{eq:discrete-time-voltage-based-equations}) for $\sigma_{i}^{\mathcal{B}}=0$,
by replacing the firing rates of the stationary solutions or those
of the oscillatory solutions, and then by calculating $V$ as a function
of the stimuli. More explicitly, in our example of a fully-connected
network we obtain:

\begin{spacing}{0.8}
\begin{center}
{\small{}
\begin{equation}
\begin{cases}
V_{E}\left(I_{E},I_{I}\right)=\frac{N_{E}-1}{N-1}J_{EE}\nu_{E}\left(I_{E},I_{I}\right)+\frac{J_{EI}}{N-1}\sum_{j=0}^{N_{I}-1}\nu_{I,j}\left(I_{E},I_{I}\right)+I_{E}\vphantom{{\displaystyle \sum_{\substack{j=0\\
j\neq i
}
}^{N_{I}-1}}}\\
V_{I,i}\left(I_{E},I_{I}\right)=\frac{N_{E}}{N-1}J_{IE}\nu_{E}\left(I_{E},I_{I}\right)+\frac{J_{II}}{N-1}{\displaystyle \sum_{\substack{j=0\\
j\neq i
}
}^{N_{I}-1}}\nu_{I,j}\left(I_{E},I_{I}\right)+I_{I},\quad i=0,1,2,
\end{cases}\label{eq:codimension-one-bifurcation-diagram}
\end{equation}
}
\par\end{center}{\small \par}
\end{spacing}

\noindent where $V_{E}\left(I_{E},I_{I}\right)$ and $V_{I,i}\left(I_{E},I_{I}\right)$
represent the (stimulus-dependent) membrane potentials in the excitatory
and inhibitory population, respectively. Moreover, $\nu_{E}\left(I_{E},I_{I}\right)$
and $\nu_{I,i}\left(I_{E},I_{I}\right)$ are the excitatory and inhibitory
firing rates of the stationary/oscillatory solutions, that we obtained
numerically as described above. The relation between $\nu_{E,I}$
and $I_{E,I}$ is calculated analytically in Sec.~(S3) of the Supplementary
Materials. In particular, we observe that in the excitatory population
the stationary firing rates (namely the three top entries of the binary
representation of the states $0-7$ and $56-63$) and the corresponding
membrane potentials are homogeneous ($\nu_{i}=\nu_{E}\left(I_{E},I_{I}\right)$
and $V_{i}=V_{E}\left(I_{E},I_{I}\right)$ for $i=0,1,2$), while
in the inhibitory population they are generally heterogeneous ($\nu_{N_{I}+i}=\nu_{I,i}$
and $V_{N_{I}+i}=V_{I,i}\left(I_{E},I_{I}\right)$ for $i=0,1,2$).
This is an example of \textit{symmetry-breaking} that occurs in the
inhibitory population (see the shaded areas in the left panel of Fig.~(\ref{fig:example-of-codimension-two-bifurcation-diagram})).
On the contrary, according to the numerical solutions there is no
symmetry-breaking during the oscillatory dynamics (see the right panel
of Fig.~(\ref{fig:example-of-codimension-two-bifurcation-diagram})),
therefore in this case the firing rates are homogeneous in both the
neural populations. Eq.~(\ref{eq:codimension-one-bifurcation-diagram})
is plotted in the top panels of Fig.~(\ref{fig:example-of-codimension-one-bifurcation-diagram}).

\subsection{Higher-Order Cross-Correlations \label{sub:Higher-Order-Cross-Correlations}}

The study of correlations among neurons is a topic of central importance
to systems neuroscience. Second-order and higher-order correlations
are key to understanding the information encoding capabilities of
neural populations \cite{Abbott1999,Pola2003,Pillow2008,Cohen2009,Moreno-Bote2014}
and to making inferences about how neurons exchange and integrate
information \cite{Singer1993,Tononi1994,David2004,Rogers2007,Friston2011}. 

In \cite{Fasoli2015} the authors introduced the following normalized
coefficient for quantifying the higher-order correlations among an
arbitrary number $n$ of neurons (groupwise correlation) in a network
of size $N$ (with $2\leq n\leq N$):

\begin{spacing}{0.80000000000000004}
\begin{center}
{\small{}
\begin{equation}
\mathrm{Corr}_{n}\left(x_{i_{0}}\left(t\right),\ldots,x_{i_{n-1}}\left(t\right)\right)=\frac{\overline{\prod_{m=0}^{n-1}\left(x_{i_{m}}\left(t\right)-\overline{x}_{i_{m}}\left(t\right)\right)}}{\sqrt[n]{\prod_{m=0}^{n-1}\overline{\left|x_{i_{m}}\left(t\right)-\overline{x}_{i_{m}}\left(t\right)\right|^{n}}}}.\label{eq:higher-order-correlations}
\end{equation}
}
\par\end{center}{\small \par}
\end{spacing}

\noindent The bar represents the statistical mean over trials computed
at time $t$. The variables $x$ in Eq.~(\ref{eq:higher-order-correlations})
can be either the membrane potentials or the firing rates. In the
first case, in SubSec.~(4.1) of the Supplementary Materials we prove
that in the stationary regime:

\begin{spacing}{0.80000000000000004}
\begin{center}
{\small{}
\begin{align}
\mathrm{Corr}_{n}\left(V_{i_{0}}\left(t\right),\ldots,V_{i_{n-1}}\left(t\right)\right)= & \frac{\sqrt{\pi}\sum_{j=0}^{2^{N}-1}F_{j}\prod_{m=0}^{n-1}\mathcal{R}_{j,i_{m}}^{\left(N\right)}}{2^{\frac{n}{2}}\Gamma\left(\frac{n+1}{2}\right)\sqrt[n]{\prod_{m=0}^{n-1}\left[\left(\sigma_{i_{m}}^{\mathcal{B}}\right)^{n}\sum_{j=0}^{2^{N}-1}F_{j}\Phi\left(-\frac{n}{2},\frac{1}{2};-\frac{1}{2}\left(\frac{\mathcal{R}_{j,i_{m}}^{\left(N\right)}}{\sigma_{i_{m}}^{\mathcal{B}}}\right)^{2}\right)\right]}}\nonumber \\
\label{eq:higher-order-correlations-of-the-memebrane-potentials}\\
\mathcal{R}_{j,i_{m}}^{\left(N\right)}= & \frac{1}{M_{i_{m}}}\sum_{l=0}^{N-1}\left[\left(\mathscr{B}_{j,l}^{\left(N\right)}-\sum_{k=0}^{2^{N}-1}F_{k}\mathscr{B}_{k,l}^{\left(N\right)}\right)J_{i_{m}l}\right],\nonumber 
\end{align}
}
\par\end{center}{\small \par}
\end{spacing}

\noindent where $\Gamma$ and $\Phi$ are the gamma function and Kummer's
confluent hypergeometric function of the first kind, respectively.
Moreover, in SubSec.~(S4.2) of the Supplementary Materials we prove
that the higher-order correlation structure of the firing rates is
given by the following formula:

\begin{spacing}{0.8}
\begin{center}
{\small{}
\begin{align}
\mathrm{Corr}_{n}\left(\nu_{i_{0}}\left(t\right),\ldots,\nu_{i_{n-1}}\left(t\right)\right)= & \frac{\sum_{j=0}^{2^{N}-1}F_{j}\prod_{m=0}^{n-1}\left(1-2\overline{\nu}_{i_{m}}-E_{j,i_{m}}\right)}{2^{n}\sqrt[n]{\prod_{m=0}^{n-1}Z_{n}\left(\overline{\nu}_{i_{m}}\right)}}\nonumber \\
\nonumber \\
Z_{n}\left(x\right)= & x^{n}\left(1-x\right)+x\left(1-x\right)^{n}\nonumber \\
\label{eq:higher-order-correlations-of-the-firing-rates}\\
\overline{\nu}_{i_{m}}= & \frac{1}{2}\left(1-\sum_{j=0}^{2^{N}-1}F_{j}E_{j,i_{m}}\right)\nonumber \\
\nonumber \\
E_{j,i_{m}}= & \mathrm{erf}\left(\frac{\theta_{i_{m}}-\frac{1}{M_{i_{m}}}\sum_{l=0}^{N-1}J_{i_{m}l}\mathscr{B}_{j,l}^{\left(N\right)}-I_{i_{m}}}{\sqrt{2}\sigma_{i_{m}}^{\mathcal{B}}}\right).\nonumber 
\end{align}
}
\par\end{center}{\small \par}
\end{spacing}

\begin{figure}
\begin{centering}
\includegraphics[scale=0.17]{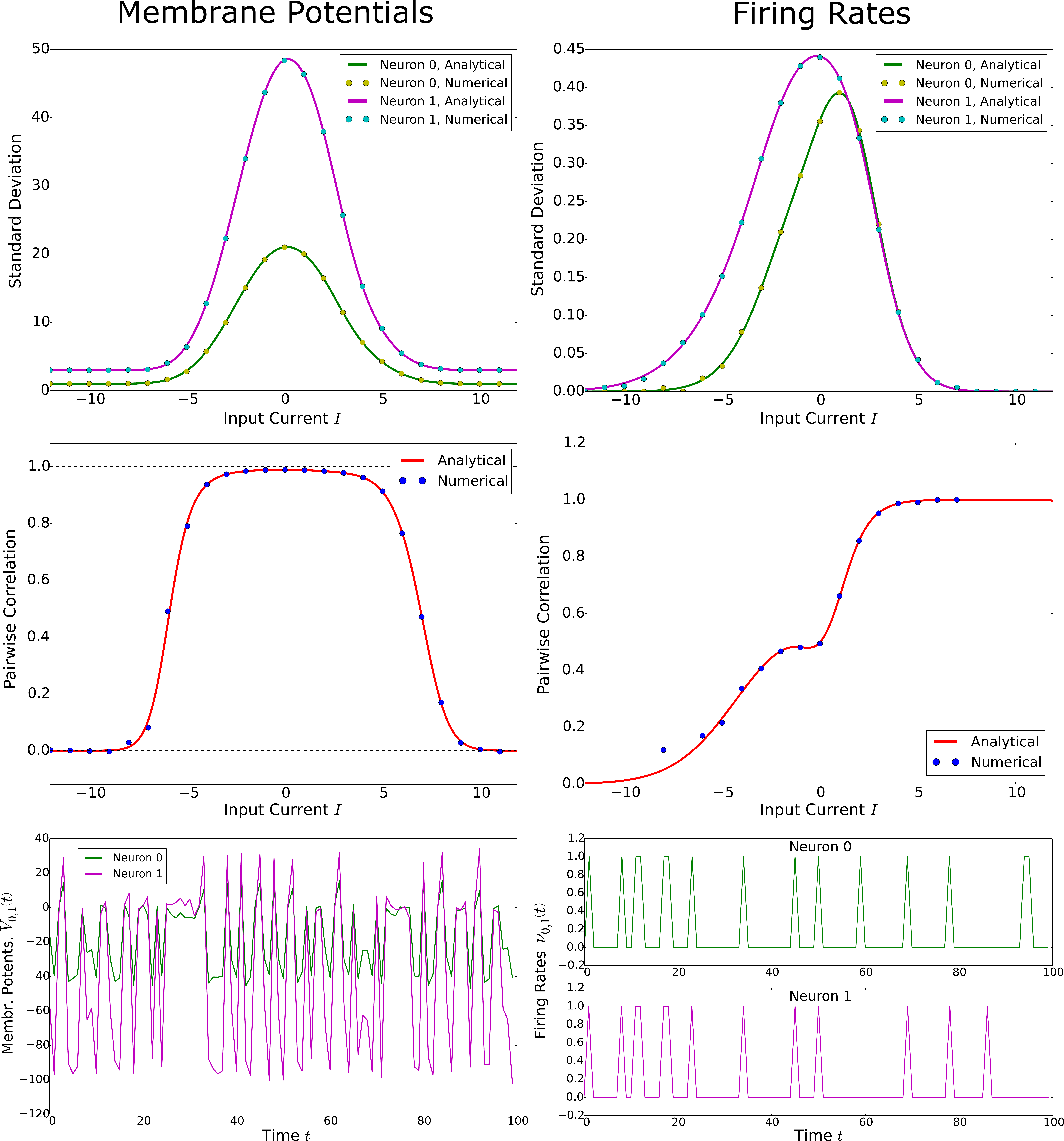}
\par\end{centering}

\protect\caption{\label{fig:cross-correlations} \small\textbf{ An example of cross-correlation
structure for $\boldsymbol{N=4}$.} This figure is obtained for the
values of the parameters in Tab.~(\ref{tab:network-parameters-5}).
The top-panels show the comparison between the analytical standard
deviations (given by Eq.~(\ref{eq:standard-deviations}), for $I\in\left[-12,12\right]$)
of the membrane potentials (left) and the firing rates (right), and
the corresponding numerical approximations. For each value of the
stimulus $I$, we calculated the network statistics through a Monte
Carlo method over $10^{5}$ repetitions. The middle panels show the
same comparison for the cross-correlations between neurons $0$ and
$1$ (the red curves are described by Eqs.~(\ref{eq:higher-order-correlations-of-the-memebrane-potentials}),
left, and (\ref{eq:higher-order-correlations-of-the-firing-rates}),
right). The bottom panels show examples of highly correlated activity
(synchronous states) between the membrane potentials (for $I=0$,
left) and between the firing rates (for $I=2$, right).}
\end{figure}

\noindent In Fig.~(\ref{fig:cross-correlations}) we show an example
of cross-correlations between pairs of neurons (i.e. $n=2$, in which
case Eq.~(\ref{eq:higher-order-correlations}) corresponds to the
Pearson's correlation coefficient). In the same figure we also show
the corresponding standard deviations of the membrane potentials and
the firing rates:

\begin{spacing}{0.80000000000000004}
\begin{center}
{\small{}
\begin{align}
\sigma_{i}^{V}= & \sqrt{\left(\sigma_{i}^{\mathcal{B}}\right)^{2}+\sum_{m=0}^{2^{N}-1}F_{m}\left(\mathcal{R}_{m,i}^{\left(N\right)}\right)^{2}}\nonumber \\
\label{eq:standard-deviations}\\
\sigma_{i}^{\nu}= & \sqrt{\overline{\nu}_{i}-\left(\overline{\nu}_{i}\right)^{2}}\nonumber 
\end{align}
}
\par\end{center}{\small \par}
\end{spacing}

\noindent (derived in Eqs.~(S29) and (S33) of the Supplementary Materials).
 
\begin{table}
\begin{centering}
{\small{}}%
\begin{tabular}{|c|}
\hline 
\tabularnewline
{\small{}$J=\left[\begin{array}{cccc}
0 & -15 & 45 & -120\\
15 & 0 & 90 & -285\\
90 & -105 & 0 & -3\\
165 & -60 & 75 & 0
\end{array}\right],\quad\boldsymbol{I}=I\left[\begin{array}{c}
1\\
1\\
1\\
1
\end{array}\right],\quad\boldsymbol{\sigma}=\left[\begin{array}{c}
1\\
3\\
2\\
2
\end{array}\right]$}\tabularnewline
\tabularnewline
\hline 
\end{tabular}
\par\end{centering}{\small \par}

\protect\caption{\label{tab:network-parameters-5} \small Set of parameters used for
generating Fig.~(\ref{fig:cross-correlations}). The values of the
stimulus $I$ are specified in the figure.}
\end{table}
 Examples of cross-correlations for $n>2$ are shown in Fig.~(S3)
of the Supplementary Materials. Our analysis shows that the external
stimulus $I$ dynamically switches the network between synchronous
(i.e. highly correlated) and asynchronous (i.e. uncorrelated) states.
The conditions under which these states may occur are discussed in
SubSecs.~(S4.1.1), (S4.1.2) and (S4.2.1) of the Supplementary Materials.
In general, we did not observe any relation between $\mathrm{Corr}_{n}\left(V_{i_{0}}\left(t\right),\ldots,V_{i_{n-1}}\left(t\right)\right)$
and $\mathrm{Corr}_{n}\left(\nu_{i_{0}}\left(t\right),\ldots,\nu_{i_{n-1}}\left(t\right)\right)$,
so that low (respectively high) correlations between the membrane
potentials do not necessarily correspond to low (respectively high)
correlations between the firing rates.

\subsection{Mean-Field Limit \label{sub:Mean-Field-Limit}}

In this section we study Eq.~(\ref{eq:discrete-time-voltage-based-equations})
in the thermodynamic limit $N\rightarrow\infty$ by means of Sznitman's
mean-field theory (see \cite{Touboul2012,Baladron2012} and references
therein). As discussed in \cite{Fasoli2015}, generally Sznitman's
theory can be applied only to networks with sufficiently dense synaptic
connections. For this reason, we suppose that the network is composed
of $\mathfrak{P}$ neural populations $\alpha$ (for $\alpha=0,\ldots,\mathfrak{P}-1$),
and that within each population the neurons have fully-connected topology
and homogeneous parameters. From this assumption it follows for example
that the stimuli are organized into $\mathfrak{P}$ vectors $\boldsymbol{I}_{\alpha}$,
one to each population, and such that:

\begin{spacing}{0.8}
\begin{center}
{\small{}
\begin{equation}
\boldsymbol{I}_{\alpha}\left(t\right)=I_{\alpha}\left(t\right)\boldsymbol{1}_{N_{\alpha}},\label{eq:external-stimuli}
\end{equation}
}
\par\end{center}{\small \par}
\end{spacing}

\noindent where $\boldsymbol{1}_{N_{\alpha}}$ is the $N_{\alpha}\times1$
all-ones vector and $N_{\alpha}$ is the size of the population $\alpha$.
In a similar way, the synaptic connectivity matrix can be written
as follows:

\begin{spacing}{0.8}
\begin{center}
{\small{}
\begin{equation}
\begin{array}{ccc}
J=\left[\begin{array}{cccc}
\mathfrak{J}_{00} & \mathfrak{J}_{01} & \cdots & \mathfrak{J}_{0,\mathfrak{P}-1}\\
\mathfrak{J}_{10} & \mathfrak{J}_{11} & \cdots & \mathfrak{J}_{1,\mathfrak{P}-1}\\
\vdots & \vdots & \ddots & \vdots\\
\mathfrak{J}_{\mathfrak{P}-1,0} & \mathfrak{J}_{\mathfrak{P}-1,1} & \cdots & \mathfrak{J}_{\mathfrak{P}-1,\mathfrak{P}-1}
\end{array}\right], &  & \mathfrak{J}_{\alpha\beta}=\begin{cases}
J_{\alpha\alpha}\left(\mathbb{I}_{N_{\alpha}}-\mathrm{Id}_{N_{\alpha}}\right), & \;\mathrm{for}\;\alpha=\beta\\
\\
J_{\alpha\beta}\mathbb{I}_{N_{\alpha},N_{\beta}}, & \;\mathrm{for}\;\alpha\neq\beta
\end{cases}\end{array}\label{eq:synaptic-connectivity-matrix}
\end{equation}
}
\par\end{center}{\small \par}
\end{spacing}

\noindent for $\alpha,\beta=0,\ldots,\mathfrak{P}-1$. The real numbers
$J_{\alpha\beta}$ are free parameters that describe the strength
of the synaptic connections from the population $\beta$ to the population
$\alpha$. Moreover, $\mathbb{I}_{N_{\alpha},N_{\beta}}$ is the $N_{\alpha}\times N_{\beta}$
all-ones matrix (here we use the simplified notation $\mathbb{I}_{N_{\alpha}}\overset{\mathrm{def}}{=}\mathbb{I}_{N_{\alpha},N_{\alpha}}$),
while $\mathrm{Id}_{N_{\alpha}}$ is the $N_{\alpha}\times N_{\alpha}$
identity matrix.

\begin{figure}
\begin{centering}
\includegraphics[scale=0.32]{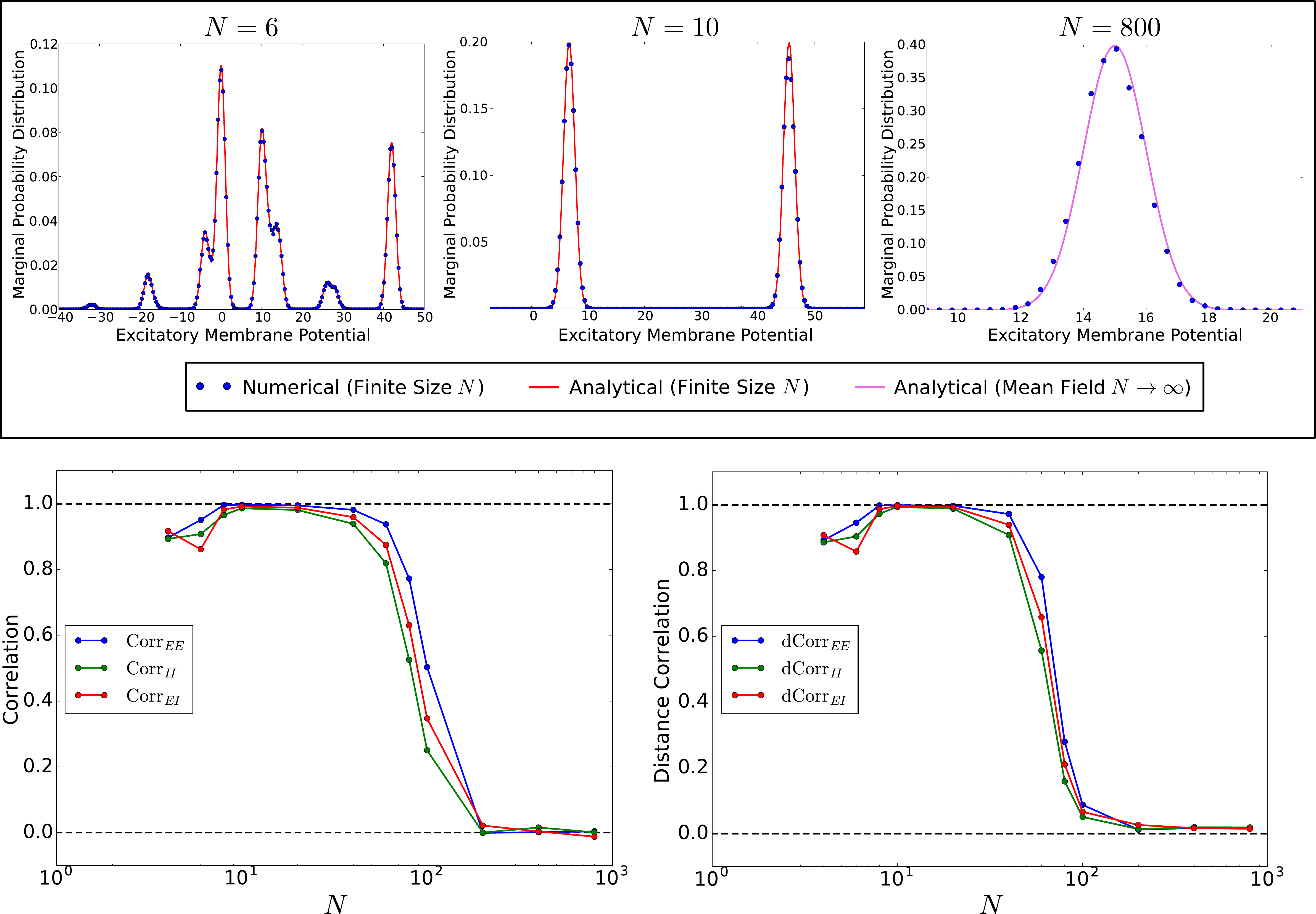}
\par\end{centering}

\protect\caption{\label{fig:thermodynamic-limit} \small\textbf{ Network statistics
in the thermodynamic limit.} This figure has been obtained for a fully-connected
network composed of one excitatory and one inhibitory population,
for $N_{E}=N_{I}=\frac{N}{2}$, \textbf{$I_{E}=-I_{I}=10$ }and the
values of the parameters reported in Tab.~(\ref{tab:network-parameters-6}).
The top panels show the fast convergence of the single-neuron marginal
probability distributions of the membrane potentials to the mean-field
normal distribution $\mathcal{N}\left(\overline{V}_{\alpha}\left(t\right),\left(\sigma_{\alpha}^{\mathcal{B}}\right)^{2}\right)$
for increasing network size $N$ (see text). In particular, the panels
show the evolution of the probability distribution of the excitatory
neurons, but this result holds also for the inhibitory population.
The bottom panels show the corresponding decrease of the pair-wise
correlation between the membrane potentials (left) and the decrease
of their distance correlation (right). $\mathrm{Corr}_{\alpha\beta}$
(respectively $\mathrm{dCorr}_{\alpha\beta}$) represents the correlation
(respectively the distance correlation) between the membrane potentials
of the populations $\alpha$ and $\beta$. In particular, the decrease
of the distance correlation with the network size numerically proves
that the neurons become increasingly independent in the thermodynamic
limit \cite{Szekely2007}. In the mathematical literature, this phenomenon
is known as \textit{propagation of chaos} \cite{Touboul2012,Baladron2012,Fasoli2015,Fasoli2016b},
and represents a key property of the network that allows us to derive
its mean-field equations (\ref{eq:mean-field-equations}). In this
figure, for each value of $N$ we calculated the numerical probability
distributions, the correlations and the distance correlations through
a Monte Carlo method over $10^{4}$ repetitions, so that the statistical
error is of the order of $10^{-2}$.}
\end{figure}

In the thermodynamic limit, the neurons become independent and normally
distributed, according to the law $V_{i}\left(i\right)\sim\mathcal{N}\left(\overline{V}_{\alpha}\left(t\right),\left(\sigma_{\alpha}^{\mathcal{B}}\right)^{2}\right)$
for every neuron $i$ in population $\alpha$ (see Fig.~(\ref{fig:thermodynamic-limit})
in the case $\mathfrak{P}=2$). In the mathematical literature, this
phenomenon is known as \textit{propagation of chaos} \cite{Touboul2012,Baladron2012,Fasoli2015,Fasoli2016b}.
As we show in SubSec.~(S5.1) of the Supplementary Materials, propagation
of chaos allows use to derive the following set of mean-field equations
for the mean membrane potentials $\overline{V}_{\alpha}$ :

\begin{spacing}{0.8}
\begin{center}
{\small{}
\begin{equation}
\overline{V}_{\alpha}\left(t+1\right)=\frac{1}{2}\sum_{\beta=0}^{\mathfrak{P}-1}R_{\beta}J_{\alpha\beta}\left[1-\mathrm{erf}\left(\frac{\theta_{\beta}-\overline{V}_{\beta}\left(t\right)}{\sqrt{2}\sigma_{\beta}^{\mathcal{B}}}\right)\right]+I_{\alpha}\left(t\right),\;\alpha=0,\ldots,\mathfrak{P}-1,\label{eq:mean-field-equations}
\end{equation}
}
\par\end{center}{\small \par}
\end{spacing}

\noindent where $R_{\alpha}=\underset{N\rightarrow\infty}{\lim}\frac{N_{\alpha}}{M_{\alpha}}$.
Therefore in the thermodynamic limit the stochastic network model
can be reduced to a set of $\mathfrak{P}$ deterministic equations
in the unknowns $\overline{V}_{\alpha}$. 
 
\begin{table}
\begin{centering}
{\small{}}%
\begin{tabular}{|ll|}
\hline 
 & \tabularnewline
{\small{}$J_{EE}=-J_{II}=80,$} & {\small{}$\theta_{E}=\theta_{I}=1$}\tabularnewline
{\small{}$J_{IE}=-J_{EI}=70,$} & {\small{}$\sigma_{E}^{\mathcal{B}}=1,\;\sigma_{I}^{\mathcal{B}}=2$}\tabularnewline
 & \tabularnewline
\hline 
\end{tabular}
\par\end{centering}{\small \par}

\protect\caption{\label{tab:network-parameters-6} \small Set of parameters used for
generating Fig.~(\ref{fig:thermodynamic-limit}) and the top panels
of Fig.~(\ref{fig:bifurcation-structure-in-the-mean-field-limit}).}
\end{table}

We observe that, unlike the original network equations (\ref{eq:discrete-time-voltage-based-equations}),
the activation functions in the mean-field equations (\ref{eq:mean-field-equations})
are differentiable everywhere for $\sigma^{\mathcal{B}}>0$. For this
reason, when noise is present in every population, the bifurcation
structure of Eq.~(\ref{eq:mean-field-equations}) can be studied
through the bifurcation theory of graded systems \cite{Kuznetsov1998}.
For the sake of example, we focus on the case of a network composed
of two neural populations, one excitatory and one inhibitory (even
though the bifurcation structure may be studied for every $\mathfrak{P}$).
From now on, it is convenient to change slightly the notation, and
to consider $\alpha=E,\, I$ rather than $\alpha=0,\,1$. Beyond global
bifurcations, which generally cannot be studied analytically, the
mean-field network may undergo limit-point, period-doubling and Neimark-Sacker
local bifurcations. By applying the technique developed in \cite{Haschke2005,Fasoli2016a},
in SubSec.~(S5.2) of the Supplementary Materials we prove that these
local codimension one bifurcations are analytically described by the
following set of parametric equations:

\begin{spacing}{0.80000000000000004}
\begin{center}
{\small{}
\begin{equation}
\begin{cases}
I_{E}\left(\mathfrak{v}\right)=\mathfrak{v}-\frac{1}{2}R_{E}J_{EE}\left[1+\mathrm{erf}\left(\frac{\mathfrak{v}-\theta_{E}}{\sqrt{2}\sigma_{E}^{\mathcal{B}}}\right)\right]-\frac{1}{2}R_{I}J_{EI}\left[1+\mathrm{erf}\left(\frac{\mu_{I}\left(\mathfrak{v}\right)-\theta_{I}}{\sqrt{2}\sigma_{I}^{\mathcal{B}}}\right)\right]\\
\\
I_{I}\left(\mathfrak{v}\right)=\mu_{I}\left(\mathfrak{v}\right)-\frac{1}{2}R_{E}J_{IE}\left[1+\mathrm{erf}\left(\frac{\mathfrak{v}-\theta_{E}}{\sqrt{2}\sigma_{E}^{\mathcal{B}}}\right)\right]-\frac{1}{2}R_{I}J_{II}\left[1+\mathrm{erf}\left(\frac{\mu_{I}\left(\mathfrak{v}\right)-\theta_{I}}{\sqrt{2}\sigma_{I}^{\mathcal{B}}}\right)\right]
\end{cases}\label{eq:parametric-equations}
\end{equation}
}
\par\end{center}{\small \par}
\end{spacing}

\noindent in the parameter $\mathfrak{v}$, where:

\begin{spacing}{0.8}
\begin{center}
{\small{}
\[
\mu_{I}\left(\mathfrak{v}\right)=\theta_{I}\pm\sqrt{-2\left(\sigma_{I}^{\mathcal{B}}\right)^{2}\ln\left(\sqrt{2\pi}\sigma_{I}^{\mathcal{B}}g_{I}\left(\mu_{I}\right)\right)}.
\]
}
\par\end{center}{\small \par}
\end{spacing}

\noindent For the limit-point and period-doubling bifurcations we
get:

\begin{spacing}{0.8}
\begin{center}
{\small{}
\[
g_{I}\left(\mu_{I}\right)=\frac{sR_{E}J_{EE}g_{E}\left(\mathfrak{v}\right)-s^{2}}{R_{E}R_{I}\left(J_{EE}J_{II}-J_{EI}J_{IE}\right)g_{E}\left(\mathfrak{v}\right)-sR_{I}J_{II}},
\]
}
\par\end{center}{\small \par}
\end{spacing}

\noindent with $s=1$ and $s=-1$ respectively, while for the Neimark-Sacker
bifurcation we obtain:

\begin{spacing}{0.8}
\begin{center}
{\small{}
\[
g_{I}\left(\mu_{I}\right)=\frac{1}{R_{E}R_{I}\left(J_{EE}J_{II}-J_{EI}J_{IE}\right)g_{E}\left(\mathfrak{v}\right)},
\]
}
\par\end{center}{\small \par}
\end{spacing}

\noindent where:

\begin{spacing}{0.8}
\begin{center}
{\small{}
\[
g_{E}\left(\mathfrak{v}\right)=\frac{1}{\sqrt{2\pi}\sigma_{E}^{\mathcal{B}}}e^{-\frac{\left(\mathfrak{v}-\theta_{E}\right)^{2}}{2\left(\sigma_{E}^{\mathcal{B}}\right)^{2}}}.
\]
}
\par\end{center}{\small \par}
\end{spacing}

\begin{figure}
\begin{centering}
\includegraphics[scale=0.43]{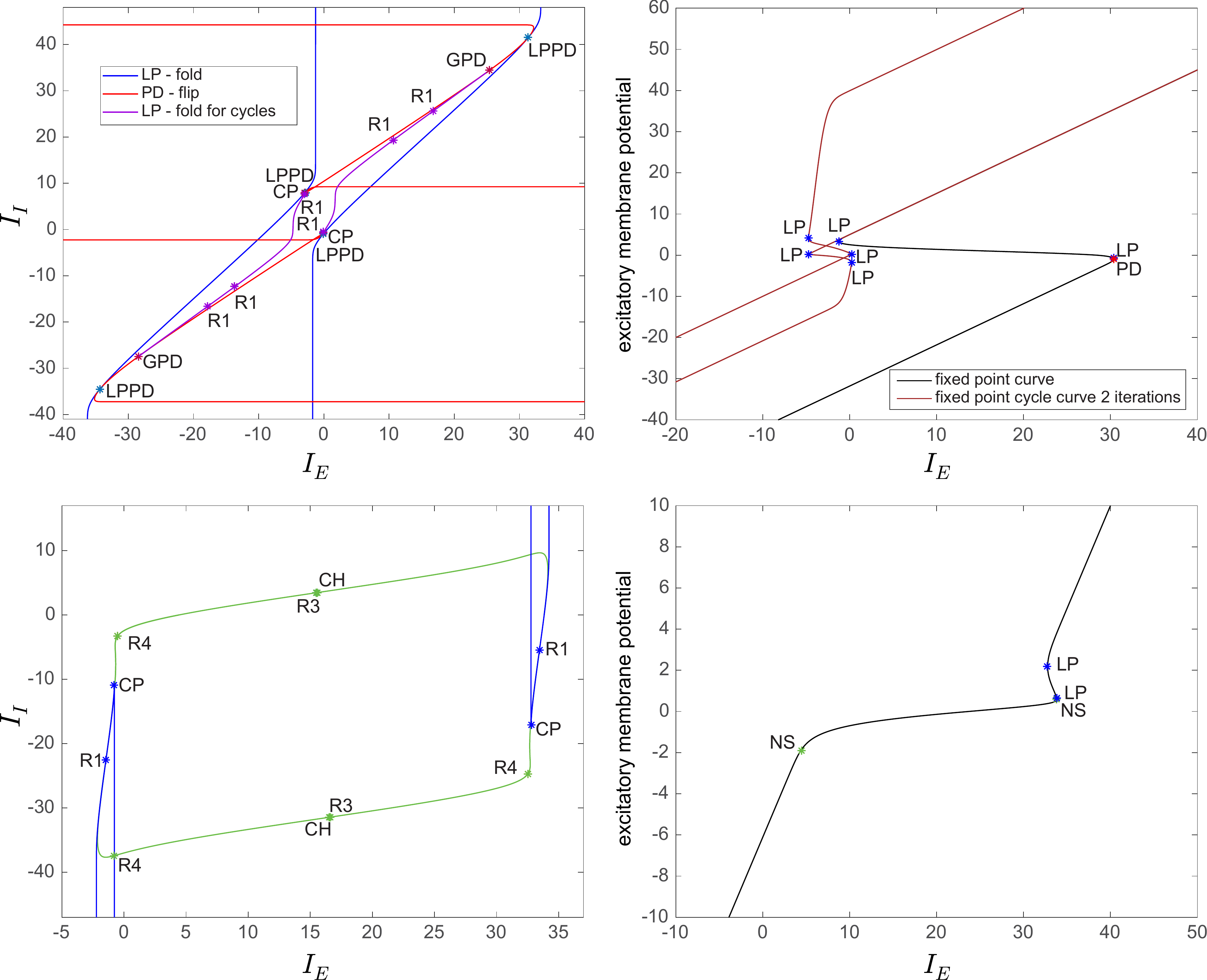}
\par\end{centering}

\protect\caption{\label{fig:bifurcation-structure-in-the-mean-field-limit} \small\textbf{
Bifurcation diagrams of the mean-field model.} This figure shows the
codimension two (left) and codimension one (right) bifurcation diagrams
of the model Eq.~(\ref{eq:mean-field-equations}), obtained numerically
by the MatCont Matlab toolbox \cite{Dhooge2003}. In the top panels,
which we obtained for $R_{E}=R_{I}=0.5$ and the parameters of Tab.~(\ref{tab:network-parameters-6}),
we show the formation of the limit-point ($\mathrm{LP}$) and period-doubling
($\mathrm{PD}$) bifurcations. These local codimension one bifurcations
correspond to the blue and red curves in the top-left panel of the
figure, and are described analytically by Eq.~(\ref{eq:parametric-equations}).
In this panel we also show the remaining bifurcations of the mean-field
model. In particular, the $\mathrm{LPPD}$ bifurcation is a local
codimension two bifurcation, which represents the simultaneous occurrence
of the limit-point and period-doubling bifurcations. Again, this bifurcation
can be described analytically by intersecting the $\mathrm{LP}$ and
$\mathrm{PD}$ curves obtained from Eq.~(\ref{eq:parametric-equations}).
Moreover, $\mathrm{LPC}$, $\mathrm{CP}$, $\mathrm{GPD}$ and $\mathrm{R1}$
are limit point of cycles, cusp, generalized period-doubling, and
$1:1$ strong resonance bifurcations, respectively. For more details,
the reader is referred to \cite{Kuznetsov1998}. In the top-right
panel we show the fixed point curves (black) and the fixed point cycle
curves (brown) obtained for $I_{I}=0$, which are mathematically described
by Eqs.~(S43) and (S52) in the Supplementary Materials, respectively.
In the bottom panels, which we obtained for $J_{EE}=-J_{II}=10$,
we show the formation of the Neimark-Sacker ($\mathrm{NS}$) bifurcation.
This local codimension one bifurcation corresponds to the green curve
in the bottom-left panel of the figure, and is described analytically
by Eq.~(\ref{eq:parametric-equations}). In this panel we also show
the formation of Chenciner ($\mathrm{CH}$) and of the $1:\mathrm{X}$
strong resonance bifurcations ($\mathrm{RX}$, for $\mathrm{X}=1,3,4$)
\cite{Kuznetsov1998}. In the bottom-right panel, obtained again for
$I_{I}=0$, we omitted the fixed point cycle curve from our analysis
because of its high complexity. This curve may be calculated numerically
as explained in SubSec.~(S5.2.3) of the Supplementary Materials,
if desired.}
\end{figure}

\noindent Eq.~(\ref{eq:parametric-equations}) describes the local
bifurcations in the codimension two bifurcation diagram of the mean-field
network. Since global bifurcations cannot be derived analytically,
we obtained the complete bifurcation structure of the mean-field network
by the MatCont Matlab toolbox \cite{Dhooge2003} (see Fig.~(\ref{fig:bifurcation-structure-in-the-mean-field-limit})).

\section{Discussion \label{sec:Discussion}}

We studied a synchronously updated firing-rate neural network model
with asymmetric synaptic weights and discrete-time evolution, that
allows exact analytical solutions for any network size $N$. The main
difficulty in studying asymmetric neural networks is the impossibility
to apply the powerful methods of equilibrium statistical mechanics,
because no energy function exists for these networks. For this reason,
exactly solvable neural network models with asymmetric weights are
still rare \cite{Hertz1987,Crisanti1987,Rieger1988,Derrida1987,Wang1990}.
Exact analytical solutions are available only in some mean-field approaches,
such as the limit of an infinite number of spin components, or the
thermodynamic limit of infinite network size. On the contrary, our
model allows exact solutions for any network size $N$. This is due
to the use of the stochastic recurrence relation (\ref{eq:discrete-time-voltage-based-equations}),
rather than Little's definition of temperature \cite{Little1974}
(in particular, compare our Eq.~(\ref{eq:firing-rates-marginal-probability-distribution})
with Little's Eq.~(4)). In this respect, our work follows the approach
described in \cite{Wang1990}, but with some important differences.
In \cite{Wang1990} the authors specifically considered first- and
second-order Hebbian synaptic connections in a diluted network with
Gaussian noise, and studied the dynamical evolution of the overlap
between the state of the network and the stored point attractors in
the thermodynamic limit. On the contrary, in our work we considered
first-order synaptic connections with an arbitrary synaptic matrix
$J$ and arbitrary noise statistics, and we derived exact solutions
for any network size $N$ without further assumptions, as described
below.

In particular, we derived exact solutions for the conditional probability
distributions of the membrane potentials and the firing rates, as
well as for the joint probability distributions in the stationary
regime. Due to the asymmetry of the synaptic weights, the network
we studied can undergo oscillations with period $\mathcal{T}\geq2$,
while synchronous Hopfield networks (which have symmetric connections)
can sustain only oscillations with period $\mathcal{T}=2$, known
as \textit{two-cycles} \cite{GolesChacc1985}. Moreover, compared
to small-size graded networks \cite{Fasoli2015,Fasoli2016b}, where
the impossibility to use statistical methods restricts the derivation
of (approximate) analytical formulas of the joint probability distributions
only to simple network topologies, here we derived an exact solution
which is valid for any connectivity matrix $J$.

The formula of the conditional probability distribution of the firing
rates allowed us to define a new learning rule to store point and
periodic attractors. Point attractors correspond to stable states,
while periodic attractors represent oscillatory solutions of the network
activity. The learning rule that we introduced can be seen as a variant
of the so called \textit{pseudoinverse learning rule} for discrete-time
systems \cite{Zhang2013}. While the pseudoinverse rule was introduced
in \cite{Personnaz1986} for deterministic Hopfield-type models, our
rule can also be used to safely store sequences of activity patterns
in noisy networks.

To complete our analysis on the formation of stable and oscillatory
solutions, we performed an analytical study of the bifurcations. The
method we proposed can be applied to networks with any topology, but
for the sake of example, we considered the case of a fully-connected
network. As is common practice, we performed the bifurcation analysis
in the zero-noise limit $\sigma^{\mathcal{B}}\rightarrow0$. We derived
analytical expressions for the codimension one and codimension two
bifurcation diagrams, showing how the external stimuli affect the
neuronal dynamics. It is important to observe that in graded networks
the local bifurcations are studied through the eigenvalues of the
Jacobian matrix of the network equations \cite{Fasoli2016a}, which
are not defined in our model due to the discontinuous activation function
(\ref{eq:Heaviside-step-function}). For this reason, we took advantage
of the conditional probability distribution of the firing rates, which
allowed us to determine for which combinations of the external stimuli
the network undergoes multistability, oscillations or symmetry-breaking.

Then, we derived exact expressions of the higher-order correlation
structure of noisy networks in the stationary regime, for both the
membrane potentials and the firing rates. In the case of the time-continuous
graded networks studied in \cite{Fasoli2015,Fasoli2016b}, the authors
found analytical (approximate) solutions of the correlation structure
through a perturbation analysis of the neural equations in the small-noise
limit $\sigma^{\mathcal{B}}\ll1$. A consequence of this approximation
is that the correlations between the membrane potentials and those
between the firing rates have the same mathematical expression in
the graded model. On the contrary, in this article we derived exact
expressions of the correlations for any noise intensity. Due to the
discontinuous activation function (\ref{eq:Heaviside-step-function}),
the two correlations structures are never identical, even in the small-noise
limit. Moreover, similarly to the case of graded networks \cite{Fasoli2016b},
we found that the external stimuli can dynamically switch the neural
activity from asynchronous (i.e. uncorrelated) to synchronous (i.e.
highly correlated) states, with two important differences. The first
is that low (respectively high) correlations between the membrane
potentials do not necessarily correspond to low (respectively high)
correlations between the firing rates. The second is that while in
graded networks synchronous states may occur through critical slowing
down \cite{Fasoli2016b}, the discrete network considered here relies
on different mechanisms for generating highly correlated activity,
that we have only partially covered. Indeed critical slowing down
is deeply related to the eigenvalues of the network, which are not
defined for a system with discontinuous activation function like ours.

For completeness and in order to link our results to previous work
on asymmetric models, we derived the mean-field equations of the network
in the thermodynamic limit $N\rightarrow\infty$. Due to the limitations
of Sznitman's mean-field theory, we derived these equations only for
sufficiently dense multi-population networks driven by independent
sources of noise. Then, by applying the methods developed in \cite{Haschke2005,Fasoli2016a},
we derived exact analytical expressions for the local codimension
one bifurcations in terms of the external stimuli. This method can
be applied to networks composed of an arbitrary number of populations,
but for the sake of example we considered the simple case of two populations.
This allowed us to describe analytically part of the codimension two
bifurcation diagram of the network, while we found the global bifurcations
numerically by the MatCont Matlab toolbox \cite{Dhooge2003}.

To conclude, we observe that solvable finite-size network models are
invaluable theoretical tools for studying the brain at its multiple
scales of spatial organization. Studying how the complexity of neuronal
activity changes for increasing network size is of fundamental importance
for unveiling the emergent properties of the brain. In this article,
we made an effort in this direction, trying to fill the gap in the
current neuroscientific literature. Asymmetric synaptic connections,
which are widely considered as a mathematically advanced task, increase
the biological plausibility of the model and allows a more complete
description of neural oscillations. While we think that these results
are of considerable theoretical interest by themselves, in future
work we will rigorously determine how the two main assumptions of
the model, namely the discrete-time evolution and the binary firing
rates, affect its capability to describe realistic neuronal activity.

\section*{Acknowledgments}

We thank Davide Corti for contributing to the initial stages of this
work. This research was supported by the Autonomous Province of Trento,
Call ``Grandi Progetti 2012,\textquotedbl{} project ``Characterizing
and improving brain mechanisms of attention\textemdash ATTEND\textquotedbl{},
and by the Future and Emerging Technologies (FET) programme within
the Seventh Framework Programme for Research of the European Commission,
under FET-Open grant FP7-600954 (VISUALISE).

\noindent The funders had no role in study design, data collection
and analysis, decision to publish, interpretation of results, or preparation
of the manuscript.

\bibliographystyle{plain}
\bibliography{Bibliography}

\end{document}


\noindent \textbf{\Large{}Pattern Storage, Bifurcations and Higher-Order
Correlation Structure of an Exactly Solvable Asymmetric Neural Network
Model }\\
\textbf{\Large{}}\\
\textbf{Supplementary Materials}\\

\noindent \begin{flushleft}
Diego Fasoli$^{1,2,\ast}$, Anna Cattani$^{1}$, Stefano Panzeri$^{1}$
\par\end{flushleft}

\medskip{}

\noindent \begin{flushleft}
\textbf{{1} Laboratory of Neural Computation, Center for Neuroscience
and Cognitive Systems @UniTn, Istituto Italiano di Tecnologia, 38068
Rovereto, Italy}
\par\end{flushleft}

\noindent \begin{flushleft}
\textbf{{2} Center for Brain and Cognition, Computational Neuroscience
Group, Universitat Pompeu Fabra, 08002 Barcelona, Spain}
\par\end{flushleft}

\noindent \begin{flushleft}
\textbf{$\ast$ Corresponding Author. E-mail: diego.fasoli@upf.edu}
\par\end{flushleft}

\section*{Structure of the Supplementary Materials \label{sec:Structure-of-the-Supplementary-Materials}}

The Supplementary Materials are organized as follows. In Sec.~\eqref{sec:Probability-Distributions}
we derive the analytical expressions of the probability distributions
of the membrane potentials (SubSec.~\eqref{sub:Probability-Density-Functions-of-the-Membrane-Potentials})
and the firing rates (SubSec.~\eqref{sub:Probability-Mass-Functions-of-the-Firing-Rates})
of a generalized version of the model introduced in Eq.~(1) of the
main text. In Sec.~\eqref{sec:Derivation-of-the-Learning-Rule} we
derive the learning rule introduced in SubSec.~(3.2). Then, in Sec.~\eqref{sec:An-Example-of-Codimension-Two-Bifurcation-Diagram}
we calculate analytically the codimension two bifurcation diagram
of the fully-connected network introduced in SubSec.~(3.3). Then,
in Sec.~\eqref{sec:Higher-Order-Cross-Correlation-Structure-of-the-Network}
we calculate the higher-order correlations among the membrane potentials
(SubSec.~\eqref{sub:Correlations-among-the-Membrane-Potentials})
and the firing rates (SubSec.~\eqref{sub:Correlations-among-the-Firing-Rates})
in the stationary regime. To conclude, in Sec.~\eqref{sec:Mean-Field-Theory}
we derive the mean-field limit of the multi-population model introduced
in SubSec.~(3.5) of the main text, and for the sake of example we
perform an analytical study of the local codimension one bifurcations
in the special case of a network composed of two populations.

\section{Probability Distributions \label{sec:Probability-Distributions}}

In this section we derive the analytical expressions of the probability
distributions of the following neural network model:

\begin{spacing}{0.8}
\begin{center}
{\small{}
\begin{equation}
V_{i}\left(t+1\right)=\frac{1}{M_{i}}\sum_{j=0}^{N-1}J_{ij}\mathscr{H}\left(V_{j}\left(t\right)-\theta_{j}\right)+I_{i}\left(t\right)+\sum_{j=0}^{N-1}\sigma_{ij}^{\mathcal{N}}\mathcal{N}_{i}\left(t\right),\quad i=0,...,N-1.\label{eq:generalized-discrete-time-voltage-based-equations}
\end{equation}
}
\par\end{center}{\small \par}
\end{spacing}

\noindent Compared to Eq.~(1) in the main text, we replaced the Gaussian
noise term $\sigma_{i}^{\mathcal{B}}\mathcal{B}_{i}\left(t\right)$
with a general term $\sum_{j=0}^{N-1}\sigma_{ij}^{\mathcal{N}}\mathcal{N}_{i}\left(t\right)$,
where the noise sources $\mathcal{N}_{i}\left(t\right)$ may have
any probability distribution.

In particular, in SubSec.~\eqref{sub:Probability-Density-Functions-of-the-Membrane-Potentials}
we derive the analytical expressions of the conditional and stationary
joint probability distributions of the membrane potentials $V_{i}$,
while in SubSec.~\eqref{sub:Probability-Mass-Functions-of-the-Firing-Rates}
we derive the corresponding expressions for the firing rates $\nu_{i}=\mathscr{H}\left(V_{i}-\theta_{i}\right)$.

\subsection{Probability Density Functions of the Membrane Potentials \label{sub:Probability-Density-Functions-of-the-Membrane-Potentials}}

If we define $\boldsymbol{V}\overset{\mathrm{def}}{=}\left[\begin{array}[t]{ccc}
V_{0}, & \ldots, & V_{N-1}\end{array}\right]^{T}$ as the collection of the membrane potentials at time $t+1$, and
$\boldsymbol{V}'$ that at time $t$, from Eq.~\eqref{eq:generalized-discrete-time-voltage-based-equations}
we get \footnote{Given a continuous random variable $X$ with probability distribution
$p_{X}\left(x\right)$ and two constants $a,b$, the random variable
$Y=aX+b$ has probability distribution $p_{Y}\left(y\right)=\frac{1}{\left|a\right|}p_{X}\left(\frac{y-b}{a}\right)$.
Eq.~\eqref{eq:membrane-potentials-conditional-probability-distribution}
is the multivariate generalization of this relation.}:

\begin{spacing}{0.8}
\begin{center}
{\small{}
\begin{equation}
p\left(\boldsymbol{V},t+1|\boldsymbol{V}',t\right)=\frac{1}{\left|\det\left(\Psi\right)\right|}p_{\mathcal{N}}\left(\Psi^{-1}\left[\boldsymbol{V}-\widehat{J}\pmb{\mathscr{H}}\left(\boldsymbol{V}'-\boldsymbol{\theta}\right)-\boldsymbol{I}\left(t\right)\right]\right),\label{eq:membrane-potentials-conditional-probability-distribution}
\end{equation}
}
\par\end{center}{\small \par}
\end{spacing}

\noindent where the matrix:

\begin{spacing}{0.8}
\begin{center}
{\small{}
\[
\Psi\overset{\mathrm{def}}{=}\left[\begin{array}{ccc}
\sigma_{00}^{\mathcal{N}} & \ldots & \sigma_{0,N-1}^{\mathcal{N}}\\
\vdots & \ddots & \vdots\\
\sigma_{N-1,0}^{\mathcal{N}} & \ldots & \sigma_{N-1,N-1}^{\mathcal{N}}
\end{array}\right]
\]
}
\par\end{center}{\small \par}
\end{spacing}

\noindent is invertible by hypothesis. In Eq.~\eqref{eq:membrane-potentials-conditional-probability-distribution}
we also defined:

\begin{spacing}{0.8}
\begin{center}
{\small{}
\[
\begin{array}{c}
\boldsymbol{\theta}\overset{\mathrm{def}}{=}\left[\begin{array}{c}
\theta_{0}\\
\vdots\\
\theta_{N-1}
\end{array}\right],\quad\boldsymbol{I}\left(t\right)\overset{\mathrm{def}}{=}\left[\begin{array}{c}
I_{0}\left(t\right)\\
\vdots\\
I_{N-1}\left(t\right)
\end{array}\right],\quad\pmb{\mathscr{H}}\left(\boldsymbol{V}'-\boldsymbol{\theta}\right)\overset{\mathrm{def}}{=}\left[\begin{array}{c}
\mathscr{H}\left(V{}_{0}'-\theta_{0}\right)\\
\vdots\\
\mathscr{H}\left(V{}_{N-1}'-\theta_{N-1}\right)
\end{array}\right],\\
\\
\\
\widehat{J}\overset{\mathrm{def}}{=}\left[\begin{array}{ccc}
\frac{J_{0,0}}{M_{0}} & \cdots & \frac{J_{0,N-1}}{M_{0}}\\
\vdots & \ddots & \vdots\\
\frac{J_{N-1,0}}{M_{N-1}} & \cdots & \frac{J_{N-1,N-1}}{M_{N-1}}
\end{array}\right],
\end{array}
\]
}
\par\end{center}{\small \par}
\end{spacing}

\noindent while $p_{\mathcal{N}}$ represents the joint probability
distribution of the stochastic process $\left[\begin{array}[t]{ccc}
\mathcal{N}_{0}\left(t\right), & \ldots, & \mathcal{N}_{N-1}\left(t\right)\end{array}\right]^{T}$. We observe that Eq.~\eqref{eq:membrane-potentials-conditional-probability-distribution}
holds for any time instant $t$.

Moreover, if the network statistics reach a stationary condition (typically
this occurs for $t\rightarrow\infty$ and for constant stimuli $I_{i}\left(t\right)=I_{i}\;\forall t$),
the joint probability distribution of the membrane potentials satisfies
the following condition:

\begin{spacing}{0.8}
\begin{center}
{\small{}
\begin{equation}
p\left(\boldsymbol{V},t+1\right)=p\left(\boldsymbol{V},t\right),\label{eq:stationarity-condition-0}
\end{equation}
}
\par\end{center}{\small \par}
\end{spacing}

\noindent which can be equivalently rewritten as follows:

\begin{spacing}{0.8}
\begin{center}
{\small{}
\begin{equation}
\int_{\mathbb{R}^{N}}p\left(\boldsymbol{V},t+1|\boldsymbol{V}',t\right)p\left(\boldsymbol{V}',t\right)d\boldsymbol{V}'=p\left(\boldsymbol{V},t\right).\label{eq:stationarity-condition-1}
\end{equation}
}
\par\end{center}{\small \par}
\end{spacing}

\noindent For simplicity, from now on we omit the symbol $t$ in the
formulas of the probability distributions. Then we observe that Eq.~\eqref{eq:stationarity-condition-1}
is a Fredholm integral equation of the second kind in $N$ dimensions,
and that its solution $p\left(\boldsymbol{V}\right)$ has also to
satisfy the normalization condition:

\begin{spacing}{0.8}
\begin{center}
{\small{}
\begin{equation}
\int_{\mathbb{R}^{N}}p\left(\boldsymbol{V}\right)d\boldsymbol{V}=1.\label{eq:normalization-condition}
\end{equation}
}
\par\end{center}{\small \par}
\end{spacing}

\noindent In what follows, we solve Eq.~\eqref{eq:stationarity-condition-1}
under Eqs.~\eqref{eq:membrane-potentials-conditional-probability-distribution}
and \eqref{eq:normalization-condition}. For the sake of clarity,
in SubSec.~\eqref{sub:Case-with-N=00003D2} we introduce our mathematical
approach in the case $N=2$, while in SubSec.~\eqref{sub:Case-with-Arbitrary-N}
we apply this technique to the case of a network of arbitrary size
$N$.

\subsubsection{Case with $N=2$ \label{sub:Case-with-N=00003D2}}

Eq.~\eqref{eq:stationarity-condition-1} can be solved for $N=2$
by decomposing the integration domain into four quadrants, namely
$\mathbb{R}^{2}=Q_{0}\cup Q_{1}\cup Q_{2}\cup Q_{3}$, where:

\begin{spacing}{0.8}
\begin{center}
{\small{}
\begin{align*}
Q_{0}= & \left\{ \left(x,y\right)\in\mathbb{R}^{2}:x\leq\theta_{0},y\leq\theta_{1}\right\} \\
\\
Q_{1}= & \left\{ \left(x,y\right)\in\mathbb{R}^{2}:x\leq\theta_{0},y>\theta_{1}\right\} \\
\\
Q_{2}= & \left\{ \left(x,y\right)\in\mathbb{R}^{2}:x>\theta_{0},y\leq\theta_{1}\right\} \\
\\
Q_{3}= & \left\{ \left(x,y\right)\in\mathbb{R}^{2}:x>\theta_{0},y>\theta_{1}\right\} .
\end{align*}
}
\par\end{center}{\small \par}
\end{spacing}

\noindent We observe that the index $i$ in $Q_{i}$ is the decimal
representation of $\mathscr{H}\left(x-\theta_{0}\right)\mathscr{H}\left(y-\theta_{1}\right)$
for any $\left(x,y\right)\in Q_{i}$. For example, for any $\left(x,y\right)\in Q_{3}$
we get $\mathscr{H}\left(x-\theta_{0}\right)\mathscr{H}\left(y-\theta_{1}\right)=11$,
whose decimal representation is $3$. On each quadrant $Q_{i}$, the
function $p\left(\boldsymbol{V}|\boldsymbol{V}'\right)$, as given
by Eq.~\eqref{eq:membrane-potentials-conditional-probability-distribution},
does not depend on $\boldsymbol{V}'$ anymore, because the term $\mathscr{H}\left(\boldsymbol{V}'-\boldsymbol{\theta}\right)$
is constant. Therefore Eq.~\eqref{eq:stationarity-condition-1} can
be equivalently rewritten as follows:

\begin{spacing}{0.8}
\begin{center}
{\small{}
\begin{equation}
\frac{1}{\left|\det\left(\Psi\right)\right|}\sum_{j=0}^{3}p_{\mathcal{N}}\left(\Psi^{-1}\left[\boldsymbol{V}-\widehat{J}\pmb{\mathscr{B}}_{j}^{\left(2\right)}-\boldsymbol{I}\right]\right)F_{j}=p\left(\boldsymbol{V}\right),\label{eq:equations-for-two-neurons-0}
\end{equation}
}
\par\end{center}{\small \par}
\end{spacing}

\noindent where:

\begin{spacing}{0.8}
\begin{center}
{\small{}
\[
\pmb{\mathscr{B}}_{0}^{\left(2\right)}=\left[\begin{array}{c}
0\\
0
\end{array}\right],\quad\pmb{\mathscr{B}}_{1}^{\left(2\right)}=\left[\begin{array}{c}
0\\
1
\end{array}\right],\quad\pmb{\mathscr{B}}_{2}^{\left(2\right)}=\left[\begin{array}{c}
1\\
0
\end{array}\right],\quad\pmb{\mathscr{B}}_{3}^{\left(2\right)}=\left[\begin{array}{c}
1\\
1
\end{array}\right]
\]
}
\par\end{center}{\small \par}
\end{spacing}

\noindent (namely ${\mathscr{B}}_{j}^{\left(2\right)}$ is the $2\times1$
vector whose entries are the digits of the binary representation of
$j$), and moreover:

\begin{spacing}{0.8}
\begin{center}
{\small{}
\[
F_{j}=\int_{Q_{j}}p\left(\boldsymbol{V}\right)d\boldsymbol{V}.
\]
}
\par\end{center}{\small \par}
\end{spacing}

\noindent We also observe that, due to Eq.~\eqref{eq:normalization-condition},
the terms $F_{j}$ satisfy the following relation:

\begin{spacing}{0.8}
\begin{center}
{\small{}
\[
\sum_{j=0}^{3}F_{j}=1.
\]
}
\par\end{center}{\small \par}
\end{spacing}

\noindent Therefore we can write, for example, $F_{3}$ as a function
of the remaining terms $F_{j}$. Now, if we integrate Eq.~\eqref{eq:equations-for-two-neurons-0}
over $Q_{i}$ for $i=0,1,2$ (thus we omit the case $i=3$), we get:

\begin{spacing}{0.8}
\begin{center}
{\small{}
\begin{equation}
\sum_{j=0}^{2}G_{i,j}F_{j}+G_{i,3}\left(1-\sum_{j=0}^{2}F_{j}\right)=F_{i},\;i=0,1,2,\label{eq:equations-for-two-neurons-1}
\end{equation}
}
\par\end{center}{\small \par}
\end{spacing}

\noindent where:

\begin{spacing}{0.8}
\begin{center}
{\small{}
\[
G_{i,j}=\frac{1}{\left|\det\left(\Psi\right)\right|}\int_{Q_{i}}p_{\mathcal{N}}\left(\Psi^{-1}\left[\boldsymbol{V}-\widehat{J}\pmb{\mathscr{B}}_{j}^{\left(2\right)}-\boldsymbol{I}\right]\right)d\boldsymbol{V}.
\]
}
\par\end{center}{\small \par}
\end{spacing}

\noindent Eq.~\eqref{eq:equations-for-two-neurons-1} is a system
of $3$ linear algebraic equations in the $3$ unknowns $F_{i}$ ($i=0,1,2$),
therefore it can be equivalently rewritten in the following matrix
form:

\begin{spacing}{0.8}
\begin{center}
{\small{}
\begin{equation}
\mathcal{A}\widetilde{\boldsymbol{F}}=\widetilde{\boldsymbol{G}},\label{eq:linear-algebraic-system}
\end{equation}
}
\par\end{center}{\small \par}
\end{spacing}

\noindent where:

\begin{spacing}{0.8}
\begin{center}
{\small{}
\begin{align*}
 & \mathcal{A}=\mathrm{Id}_{3}+\left[\begin{array}{ccc}
G_{0,3}-G_{0,0} & G_{0,3}-G_{0,1} & G_{0,3}-G_{0,2}\\
G_{1,3}-G_{1,0} & G_{1,3}-G_{1,1} & G_{1,3}-G_{1,2}\\
G_{2,3}-G_{2,0} & G_{2,3}-G_{2,1} & G_{2,3}-G_{2,2}
\end{array}\right]\\
\\
 & \widetilde{\boldsymbol{F}}=\left[\begin{array}{c}
F_{0}\\
F_{1}\\
F_{2}
\end{array}\right],\quad\widetilde{\boldsymbol{G}}=\left[\begin{array}{c}
G_{0,3}\\
G_{1,3}\\
G_{2,3}
\end{array}\right],
\end{align*}
}
\par\end{center}{\small \par}
\end{spacing}

\noindent while $\mathrm{Id}_{3}$ represents the $3\times3$ identity
matrix. If the matrix $\mathcal{A}$ is invertible, from Eq.~\eqref{eq:linear-algebraic-system}
we get $\widetilde{\boldsymbol{F}}=\mathcal{A}^{-1}\widetilde{\boldsymbol{G}}$.
Then, the entries of $\widetilde{\boldsymbol{F}}$ so obtained can
be replaced into Eq.~\eqref{eq:equations-for-two-neurons-0} (together
with $F_{3}=1-\sum_{j=0}^{2}F_{j}$) to get the function $p\left(\boldsymbol{V}\right)$.
This concludes the problem of calculating the joint probability density
function in the case $N=2$. In the next subsection we extend this
procedure to the case of neural networks composed of an arbitrary
number of neurons.

\subsubsection{Case with Arbitrary $N$ \label{sub:Case-with-Arbitrary-N}}

For an arbitrary $N$, we decompose $\mathbb{R}^{N}$ into the $2^{N}$
hypervolumes $\mathscr{V}_{i}$ that generalize the concept of quadrant.
For example, for $N=3$ we get $8$ octants:

\begin{spacing}{0.8}
\begin{center}
{\small{}
\begin{align*}
O_{0}= & \left\{ \left(x,y,z\right)\in\mathbb{R}^{3}:x\leq\theta_{0},y\leq\theta_{1},z\leq\theta_{2}\right\} \\
 & \vdots\\
O_{7}= & \left\{ \left(x,y,z\right)\in\mathbb{R}^{3}:x>\theta_{0},y>\theta_{1},z>\theta_{2}\right\} ,
\end{align*}
}
\par\end{center}{\small \par}
\end{spacing}

\noindent and so on. In this way, similarly to Eq.~\eqref{eq:equations-for-two-neurons-0},
we get:

\begin{spacing}{0.8}
\begin{center}
{\small{}
\begin{equation}
p\left(\boldsymbol{V}\right)=\frac{1}{\left|\det\left(\Psi\right)\right|}\sum_{j=0}^{2^{N}-1}p_{\mathcal{N}}\left(\Psi^{-1}\left[\boldsymbol{V}-\widehat{J}\pmb{\mathscr{B}}_{j}^{\left(N\right)}-\boldsymbol{I}\right]\right)F_{j},\label{eq:equations-for-N-neurons}
\end{equation}
}
\par\end{center}{\small \par}
\end{spacing}

\noindent where $\pmb{\mathscr{B}}_{j}^{\left(N\right)}$ is the $N\times1$
vector whose entries are the digits of the binary representation of
$j$ (e.g. $\pmb{\mathscr{B}}_{7}^{\left(5\right)}=\left[\begin{array}{ccccc}
0, & 0, & 1, & 1, & 1\end{array}\right]^{T}$). If we define $\widetilde{\boldsymbol{F}}=\left[\begin{array}{ccc}
F_{0}, & \ldots, & F_{2^{N}-2}\end{array}\right]^{T}$, then similarly to the case $N=2$ we get $\widetilde{\boldsymbol{F}}=\mathcal{A}^{-1}\widetilde{\boldsymbol{G}}$,
where:

\begin{spacing}{0.8}
\begin{center}
{\small{}
\begin{align}
 & \mathcal{A}_{i,j}=\delta_{i,j}+G_{i,2^{N}-1}-G_{i,j}\nonumber \\
\label{eq:matrix-A-case-with-N-neurons}\\
 & \left[\widetilde{\boldsymbol{G}}\right]_{i}=G_{i,2^{N}-1},\nonumber 
\end{align}
}
\par\end{center}{\small \par}
\end{spacing}

\noindent for $i,j=0,\ldots,2^{N}-2$, while $F_{2^{N}-1}=1-\sum_{j=0}^{2^{N}-2}F_{j}$
and $\delta_{i,j}$ is the Kronecker delta. The terms $G_{i,j}$ are
obtained by integrating the function $p_{\mathcal{N}}\left(\Psi^{-1}\left[\boldsymbol{V}-\widehat{J}\pmb{\mathscr{B}}_{j}^{\left(N\right)}-\boldsymbol{I}\right]\right)$
over the hypervolumes $\mathscr{V}_{i}$, while $\mathcal{A}^{-1}$
can be calculated analytically from the cofactor matrix $\mathcal{C}$
of $\mathcal{A}$, through the relation $\mathcal{A}^{-1}=\frac{1}{\det\left(\mathcal{A}\right)}\mathcal{C}^{T}$,
with $\det\left(\mathcal{A}\right)=\sum_{j=0}^{N-1}\mathcal{C}_{ij}\mathcal{A}_{ij}$
(for any $i$) according to Laplace's formula. Finally, from Eqs.~\eqref{eq:equations-for-N-neurons}
and \eqref{eq:matrix-A-case-with-N-neurons} we get the exact analytical
expression of the joint probability distribution $p\left(\boldsymbol{V}\right)$.

\begin{figure}
\begin{centering}
\includegraphics[scale=0.19]{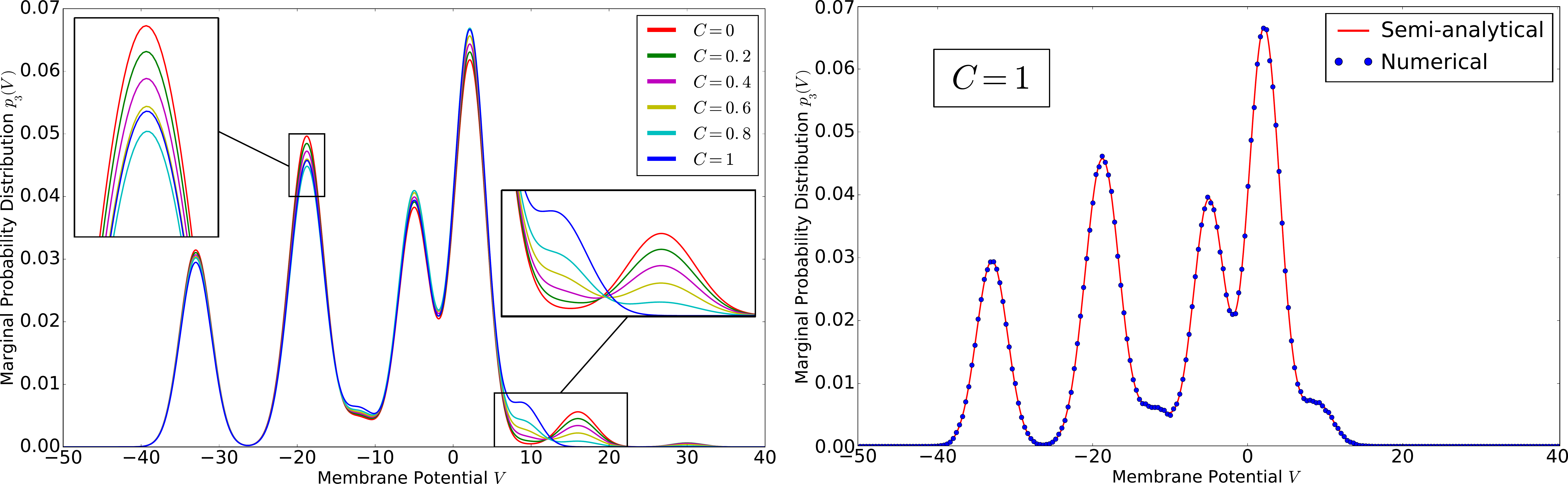}
\par\end{centering}

\protect\caption{\label{fig:Example-with-correlated-Brownian-motions} \small \textbf{Examples
of probability distributions in a network driven by correlated Gaussian
sources of noise.} The figure is obtained for the values of the parameters
of Tab.~(2) in the main text, while the noise sources are characterized
by homogeneous cross-correlation $C$ (namely $\mathrm{Corr}\left(\mathcal{B}_{i}\left(t\right),\mathcal{B}_{j}\left(t\right)\right)=C$,
$\mathrm{Corr}\left(\mathcal{B}_{i}\left(t\right),\mathcal{B}_{j}\left(s\right)\right)=0$
for $t\protect\neq s$ and $\forall i,\,j$). In particular, we focused
on the probability density of the $3$rd neuron. The left panel shows
the semi-analytical probability distribution, evaluated from Eq.~\eqref{eq:equations-for-N-neurons}
and by calculating the coefficients $G_{i,j}$ numerically for $C=0,\,0.2,\,0.4,\,0.6,\,0.8,\,1$.
The right panel shows the comparison between the semi-analytical probability
distribution in the special case $C=1$ (red line), and the same function
evaluated numerically through a Monte Carlo method over $10^{6}$
repetitions of the network dynamics (blue dots). We performed the
numerical simulations by solving iteratively Eq.~\eqref{eq:generalized-discrete-time-voltage-based-equations}
in the temporal interval $t=\left[0,100\right]$ and then by calculating
the probability distributions through a kernel density estimator at
$t=100$.}
\end{figure}

We observe that our approach works for arbitrary sources of noise
$\mathcal{N}_{i}\left(t\right)$ with any probability distribution.
The coefficients $G_{i,j}$ can be equivalently rewritten as follows:

\begin{spacing}{0.80000000000000004}
\begin{center}
{\small{}
\begin{equation}
G_{i,j}=\int_{\varphi_{j}\left(\mathscr{V}_{i}\right)}p_{\mathcal{N}}\left(\boldsymbol{x}\right)d\boldsymbol{x},\label{eq:cumulative-distribution-function}
\end{equation}
}
\par\end{center}{\small \par}
\end{spacing}

\noindent where:

\begin{spacing}{0.80000000000000004}
\begin{center}
{\small{}
\begin{align*}
\varphi_{j}: & \mathbb{R}^{N}\rightarrow\mathbb{R}^{N}\\
\\
 & \boldsymbol{y}\mapsto\Psi^{-1}\left[\boldsymbol{y}-\widehat{J}\pmb{\mathscr{B}}_{j}^{\left(N\right)}-\boldsymbol{I}\right],
\end{align*}
}
\par\end{center}{\small \par}
\end{spacing}

\noindent therefore in the special case when the matrix $\Psi$ is
diagonal the set $\varphi_{j}\left(\mathscr{V}_{i}\right)$ is hyperrectangular
and $G_{i,j}$ corresponds to the cumulative distribution function
of $p_{\mathcal{N}}$ on $\varphi_{j}\left(\mathscr{V}_{i}\right)$.
In principle the sources of noise may also be correlated, but in this
case closed-form expressions of the corresponding cumulative distribution
function are generally not known. For this reason, the calculation
of the probability distributions of the membrane potentials for correlated
sources of noise may require a numerical evaluation of the coefficients
$G_{i,j}$ (an example is shown in Fig.~\eqref{fig:Example-with-correlated-Brownian-motions}
for correlated sources of noise with Gaussian distribution) or analytical
approximations of the cumulative distribution. 

\begin{figure}
\begin{centering}
\includegraphics[scale=0.2]{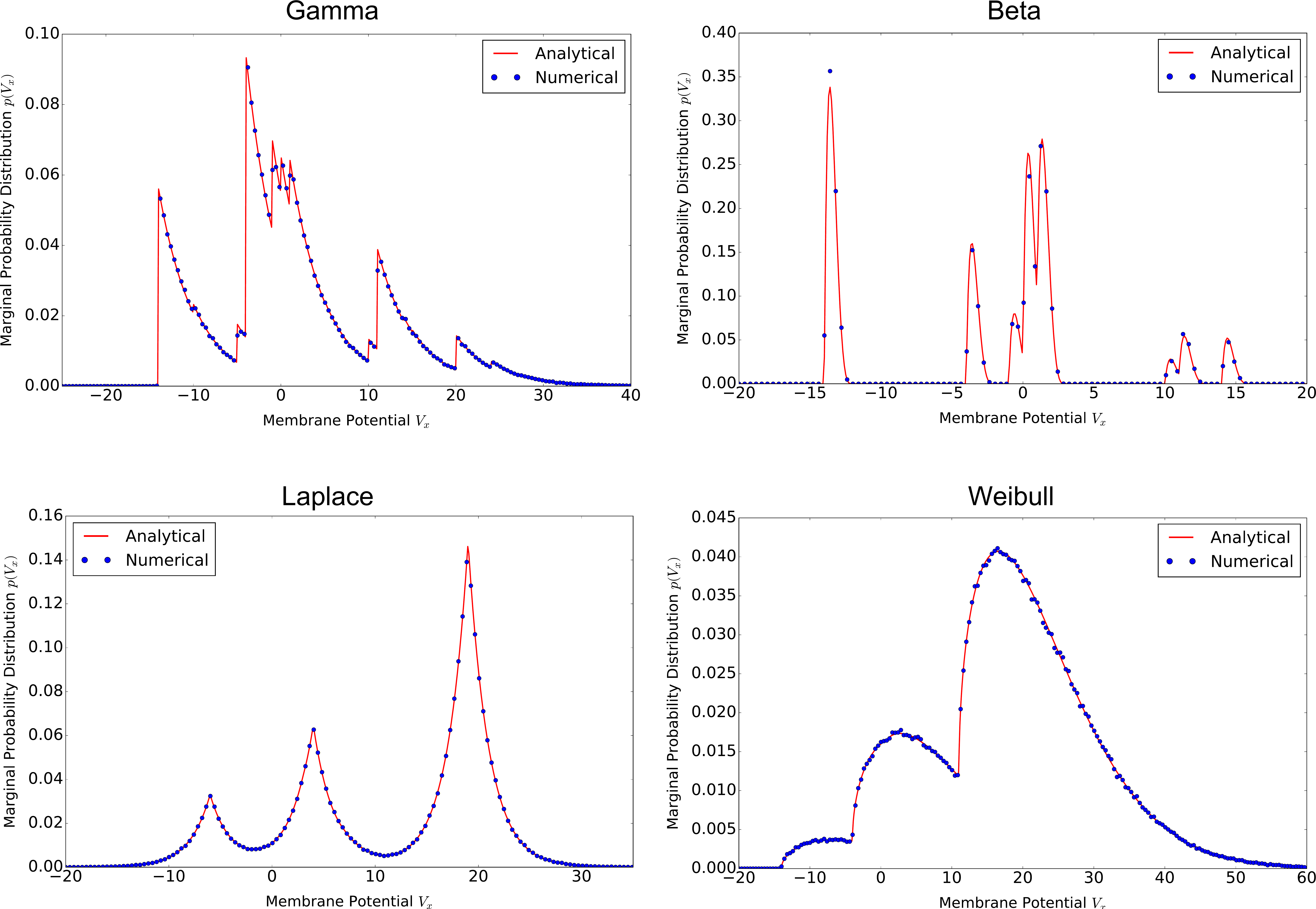}
\par\end{centering}

\protect\caption{\label{fig:Examples-with-independent-sources-of-noise} \small \textbf{Examples
of probability distributions in networks driven by independent non-Gaussian
sources of noise.} The figure is obtained for the network parameters
of Tab.~(2) in the main text and the noise distributions in Eq.~\eqref{eq:examples-of-non-normal-distributions}.
The parameters of the noise distributions are: Beta ($\alpha=2$,
$\beta=5$), Gamma ($\alpha=1$, $\beta=0.5$), Laplace ($\alpha=4$,
$\beta=1$) and Weibull ($\alpha=1.5$, $\beta=7$). In this figure
we focused on the probability distribution of the $0$th neuron, which
we calculated analytically from Eqs.~\eqref{eq:equations-for-N-neurons}
and \eqref{eq:factorized-cumulative-distribution-function} (red lines),
and numerically through a Monte Carlo method over $10^{6}$ repetitions
of the network dynamics (blue dots).}
\end{figure}

A special case is that of independent sources of noise ($p_{\mathcal{N}}\left(\mathbf{x}\right)=\prod_{i=0}^{N-1}p_{\mathcal{N}_{i}}\left(x_{i}\right)$),
because they allow closed-form expressions of the cumulative distribution
function. Indeed, in this case the cumulative distribution of the
vector process $\left[\begin{array}[t]{ccc}
\mathcal{N}_{0}\left(t\right), & \ldots, & \mathcal{N}_{N-1}\left(t\right)\end{array}\right]^{T}$ can be factorized into the product of the cumulative distribution
functions $\mathcal{F}_{i}$ of the single processes $\mathcal{N}_{i}\left(t\right)$.
Therefore Eq.~\eqref{eq:cumulative-distribution-function} can be
written as follows:

\begin{spacing}{0.80000000000000004}
\begin{center}
{\small{}
\begin{equation}
G_{i,j}=\prod_{p}\mathcal{F}_{p}\left(\frac{\theta_{p}-\frac{1}{M_{p}}\sum_{n=0}^{N-1}J_{pn}\mathscr{B}_{j,n}^{\left(N\right)}-I_{p}}{\sigma_{p}^{\mathcal{N}}}\right)\prod_{q}\left(1-\mathcal{F}_{q}\left(\frac{\theta_{q}-\frac{1}{M_{q}}\sum_{n=0}^{N-1}J_{qn}\mathscr{B}_{j,n}^{\left(N\right)}-I_{q}}{\sigma_{q}^{\mathcal{N}}}\right)\right),\label{eq:factorized-cumulative-distribution-function}
\end{equation}
}
\par\end{center}{\small \par}
\end{spacing}

\noindent where $\mathscr{B}_{j,n}^{\left(N\right)}\overset{\mathrm{def}}{=}\left[\pmb{\mathscr{B}}_{j}^{\left(N\right)}\right]_{n}$
is the $n$th entry of the vector $\pmb{\mathscr{B}}_{j}^{\left(N\right)}$.
Moreover, in Eq.~\eqref{eq:factorized-cumulative-distribution-function}
the index $p$ runs over the zeros of the binary representation of
the index $i$, while the index $q$ over the ones. For example, if
$N=5$ and $i=3$, the binary representation of $i$ is $00011$,
therefore $p=0,1,2$ and $q=3,4$. Examples of probability distributions
$p\left(\boldsymbol{V}\right)$ are shown in Fig.~\eqref{fig:Examples-with-independent-sources-of-noise},
given the following non-Gaussian distributions of the noise sources:

\begin{spacing}{0.80000000000000004}
\begin{center}
{\small{}
\begin{equation}
p_{\mathcal{N}_{i}}\left(x\right)=\begin{cases}
\frac{1}{\Gamma\left(\alpha\right)}\beta^{\alpha}x^{\alpha-1}e^{-\beta x},\;x\in\left[0,+\infty\right) & \left(\mathrm{Gamma}\right)\\
\\
\frac{1}{B\left(\alpha,\beta\right)}x^{\alpha-1}\left(1-x\right)^{\beta-1},\;x\in\left[0,1\right] & \left(\mathrm{Beta}\right)\\
\\
\frac{1}{2\beta}e^{-\frac{1}{\beta}\left|x-\alpha\right|},\;\forall x & \left(\mathrm{Laplace}\right)\\
\\
\frac{\alpha}{\beta}\left(\frac{x}{\beta}\right)^{\alpha-1}e^{-\left(\frac{x}{\beta}\right)^{\alpha}},\;x\in\left[0,+\infty\right) & \left(\mathrm{Weibull}\right),
\end{cases}\label{eq:examples-of-non-normal-distributions}
\end{equation}
}
\par\end{center}{\small \par}
\end{spacing}

\noindent whose cumulative distribution functions are:

\begin{spacing}{0.80000000000000004}
\begin{center}
{\small{}
\begin{equation}
\mathcal{F}_{i}\left(x\right)=\begin{cases}
\frac{1}{\Gamma\left(\alpha\right)}\gamma\left(\alpha,\beta x\right),\;x\in\left[0,+\infty\right) & \left(\mathrm{Gamma}\right)\\
\\
I\left(x;\alpha,\beta\right),\;x\in\left[0,1\right] & \left(\mathrm{Beta}\right)\\
\\
\begin{cases}
\frac{1}{2}e^{\frac{1}{\beta}\left(x-\alpha\right)},\;x<\alpha\\
1-\frac{1}{2}e^{-\frac{1}{\beta}\left(x-\alpha\right)},\;x\geq\alpha
\end{cases} & \left(\mathrm{Laplace}\right)\\
\\
1-e^{-\left(\frac{x}{\beta}\right)^{\alpha}},\;x\in\left[0,+\infty\right) & \left(\mathrm{Weibull}\right).
\end{cases}\label{eq:examples-of-cumulative-distribution-functions}
\end{equation}
}
\par\end{center}{\small \par}
\end{spacing}

\noindent In Eqs.~\eqref{eq:examples-of-non-normal-distributions}
and \ref{eq:examples-of-cumulative-distribution-functions}, $\Gamma$,
$\gamma$, $B$ and $I$ represent the gamma, incomplete gamma, beta
and incomplete beta functions, respectively \cite{Abramowitz1964}
To conclude, the case of independent Gaussian sources of noise is
considered in SubSec.~\eqref{sub:The-Case-of-Independent-Gaussian-Sources-of-Noise-for-the-Membrane-Potentials}.

\subsubsection{The Case of Independent Gaussian Sources of Noise \label{sub:The-Case-of-Independent-Gaussian-Sources-of-Noise-for-the-Membrane-Potentials}}

As in the main text, we consider the case of independent Gaussian
sources of noise $\mathcal{N}_{i}\left(t\right)=\mathcal{B}_{i}\left(t\right)$
with matrix:

\begin{spacing}{0.8}
\begin{center}
{\small{}
\[
\Psi=\mathrm{diag}\left(\sigma_{0}^{\mathcal{B}},\ldots,\sigma_{N-1}^{\mathcal{B}}\right).
\]
}
\par\end{center}{\small \par}
\end{spacing}

\noindent In this case $\sum_{j=0}^{N-1}\sigma_{ij}^{\mathcal{N}}\mathcal{N}_{i}\left(t\right)=\sigma_{i}^{\mathcal{B}}\mathcal{B}_{i}\left(t\right)$,
therefore the overall sources of noise to each neuron are independent.
Moreover, the joint probability distribution of the stochastic process
$\left[\begin{array}[t]{ccc}
\mathcal{B}_{0}\left(t\right), & \ldots, & \mathcal{B}_{N-1}\left(t\right)\end{array}\right]^{T}$ is:

\begin{spacing}{0.8}
\begin{center}
{\small{}
\[
p_{\mathcal{B}}\left(\boldsymbol{x}\right)=\frac{1}{\left(2\pi\right)^{\frac{N}{2}}}\prod_{m=0}^{N-1}e^{-\frac{x_{m}^{2}}{2}}.
\]
}
\par\end{center}{\small \par}
\end{spacing}

\noindent Therefore, according to Eq.~\eqref{eq:membrane-potentials-conditional-probability-distribution},
the conditional probability distribution of the membrane potentials
is:

\begin{spacing}{0.8}
\begin{center}
{\small{}
\begin{equation}
p\left(\boldsymbol{V}|\boldsymbol{V}'\right)=\frac{1}{\left(2\pi\right)^{\frac{N}{2}}\prod_{i=0}^{N-1}\sigma_{i}^{\mathcal{B}}}\prod_{m=0}^{N-1}e^{-\frac{1}{2}\left(\frac{V_{m}-\frac{1}{M_{m}}\sum_{n=0}^{N-1}J_{mn}\mathscr{H}\left(V'_{n}-\theta_{n}\right)-I_{m}\left(t\right)}{\sigma_{m}^{\mathcal{B}}}\right)^{2}},\label{eq:membrane-potentials-conditional-probability-distribution-independent-Gaussian-noise-case}
\end{equation}
}
\par\end{center}{\small \par}
\end{spacing}

\noindent while from Eq.~\eqref{eq:equations-for-N-neurons} we get
the following expression of the joint probability distribution:

\begin{spacing}{0.8}
\begin{center}
{\small{}
\begin{equation}
p\left(\boldsymbol{V}\right)=\frac{1}{\left(2\pi\right)^{\frac{N}{2}}\prod_{i=0}^{N-1}\sigma_{i}^{\mathcal{B}}}\sum_{j=0}^{2^{N}-1}F_{j}\prod_{m=0}^{N-1}e^{-\frac{1}{2}\left(\frac{V_{m}-\frac{1}{M_{m}}\sum_{n=0}^{N-1}J_{mn}\mathscr{B}_{j,n}^{\left(N\right)}-I_{m}}{\sigma_{m}^{\mathcal{B}}}\right)^{2}}.\label{eq:membrane-potentials-joint-probability-distribution-independent-Gaussian-noise-case}
\end{equation}
}
\par\end{center}{\small \par}
\end{spacing}

\noindent The coefficients $F_{j}$ for $j=0,\ldots,2^{N}-2$ are
obtained by inverting the matrix $\mathcal{A}$ (while $F_{2^{N}-1}=1-\sum_{j=0}^{2^{N}-2}F_{j}$).
In turn, $\mathcal{A}$ is calculated from Eq.~\eqref{eq:matrix-A-case-with-N-neurons},
where, according to Eq.~\eqref{eq:factorized-cumulative-distribution-function},
the coefficients $G_{i,j}$ have the following expression:

\begin{spacing}{0.8}
\begin{center}
{\small{}
\[
G_{i,j}=\frac{1}{2^{N}}\prod_{m=0}^{N-1}\left[1+\left(-1\right)^{\mathscr{B}_{i,m}^{\left(N\right)}}\mathrm{erf}\left(\frac{\theta_{m}-\frac{1}{M_{m}}\sum_{n=0}^{N-1}J_{mn}\mathscr{B}_{j,n}^{\left(N\right)}-I_{m}}{\sqrt{2}\sigma_{m}^{\mathcal{B}}}\right)\right].
\]
}
\par\end{center}{\small \par}
\end{spacing}

\noindent For graphical reasons, for $N>2$ it is convenient to plot
the single-neuron marginal distributions of the membrane potentials:

\begin{spacing}{0.80000000000000004}
\noindent \begin{center}
{\small{}
\begin{equation}
p_{i}\left(V\right)=\int_{\mathbb{R}^{N-1}}p\left(\boldsymbol{V}\right){\displaystyle \prod_{\substack{p=0\\
p\neq i
}
}^{N-1}}dV_{p}=\frac{1}{\sqrt{2\pi}\sigma_{i}^{\mathcal{B}}}\sum_{j=0}^{2^{N}-1}F_{j}e^{-\frac{1}{2}\left(\frac{V-\frac{1}{M_{i}}\sum_{k=0}^{N-1}J_{ik}\mathscr{B}_{j,k}^{\left(N\right)}-I_{i}}{\sigma_{i}^{\mathcal{B}}}\right)^{2}}.\label{eq:membrane-potentials-marginal-probability-distribution-independent-Gaussian-noise-case}
\end{equation}
}
\par\end{center}{\small \par}
\end{spacing}

\subsection{Probability Mass Functions of the Firing Rates \label{sub:Probability-Mass-Functions-of-the-Firing-Rates}}

We introduce the vector $\boldsymbol{\nu}\overset{\mathrm{def}}{=}\left[\begin{array}[t]{ccc}
\nu_{0}, & \ldots, & \nu_{N-1}\end{array}\right]^{T}$, where $\nu_{i}=\mathscr{H}\left(V_{i}-\theta_{i}\right)$ is the
firing rate of the $i$th neuron. Then, the conditional probability
mass function of the firing rates can be easily calculated by integrating
$p\left(\boldsymbol{V}|\boldsymbol{V}'\right)$ (see Eq.~\eqref{eq:membrane-potentials-conditional-probability-distribution})
with respect to the variable $\boldsymbol{V}$ over the $2^{N}$ hypervolumes
$\mathscr{V}_{i}$ of $\mathbb{R}^{N}$, that we introduced in SubSecs.~\eqref{sub:Case-with-N=00003D2}
and \eqref{sub:Case-with-Arbitrary-N}. For example, in the case $N=2$,
by integrating $p\left(\boldsymbol{V}|\boldsymbol{V}'\right)$ over
the quadrant $Q_{1}$ we obtain $P\left(\boldsymbol{\nu}=\left[\begin{array}{c}
0\\
1
\end{array}\right]|\boldsymbol{\nu}'\right)$, and so on. In this way, for an arbitrary $N$, we get:

\begin{spacing}{0.8}
\begin{center}
{\small{}
\begin{equation}
P\left(\boldsymbol{\nu}|\boldsymbol{\nu}'\right)=\frac{1}{\left|\det\left(\Psi\right)\right|}\int_{\mathscr{V}\left(\boldsymbol{\nu}\right)}p_{\mathcal{N}}\left(\Psi^{-1}\left[\boldsymbol{V}-\widehat{J}\boldsymbol{\nu}'-\boldsymbol{I}\right]\right)d\boldsymbol{V},\label{eq:firing-rates-conditional-probability-distribution}
\end{equation}
}
\par\end{center}{\small \par}
\end{spacing}

\noindent having used the relation $\boldsymbol{\nu}'=\pmb{\mathscr{H}}\left(\boldsymbol{V}'-\boldsymbol{\theta}\right)$.
The symbol $\mathscr{V}\left(\boldsymbol{\nu}\right)$ represents
the hypervolume corresponding to the firing rate $\boldsymbol{\nu}$,
e.g. $\mathscr{V}\left(\boldsymbol{\nu}=\left[\begin{array}{c}
0\\
0
\end{array}\right]\right)=Q_{0}$ and $\mathscr{V}\left(\boldsymbol{\nu}=\left[\begin{array}{c}
0\\
0\\
1
\end{array}\right]\right)=O_{1}$. In the same way, by integrating Eq.~\eqref{eq:equations-for-N-neurons}
we get the following expression of the joint probability mass function:

\begin{spacing}{0.8}
\begin{center}
{\small{}
\begin{equation}
P\left(\boldsymbol{\nu}\right)=\frac{1}{\left|\det\left(\Psi\right)\right|}\sum_{j=0}^{2^{N}-1}F_{j}\int_{\mathscr{V}\left(\boldsymbol{\nu}\right)}p_{\mathcal{N}}\left(\Psi^{-1}\left[\boldsymbol{V}-\widehat{J}\pmb{\mathscr{B}}_{j}^{\left(N\right)}-\boldsymbol{I}\right]\right)d\boldsymbol{V}.\label{eq:firing-rates-joint-probability-distribution}
\end{equation}
}
\par\end{center}{\small \par}
\end{spacing}

\subsubsection{The Case of Independent Gaussian Sources of Noise \label{sub:The-Case-of-Independent-Gaussian-Sources-of-Noise-for-the-Firing-Rates}}

\noindent In the special case of independent Gaussian sources of noise
discussed in SubSec.~\eqref{sub:The-Case-of-Independent-Gaussian-Sources-of-Noise-for-the-Membrane-Potentials},
by integrating Eqs.~\eqref{eq:membrane-potentials-conditional-probability-distribution-independent-Gaussian-noise-case}
and \eqref{eq:membrane-potentials-joint-probability-distribution-independent-Gaussian-noise-case}
over the hypervolumes $\mathscr{V}_{i}$ we get the following expressions
for the conditional and joint probability distributions of the firing
rates:

\begin{spacing}{0.80000000000000004}
\begin{center}
{\small{}
\begin{align}
P\left(\boldsymbol{\nu}|\boldsymbol{\nu}'\right)= & \frac{1}{2^{N}}\prod_{m=0}^{N-1}\left[1+\left(-1\right)^{\nu_{m}}\mathrm{erf}\left(\frac{\theta_{m}-\frac{1}{M_{m}}\sum_{n=0}^{N-1}J_{mn}\nu_{n}'-I_{m}\left(t\right)}{\sqrt{2}\sigma_{m}^{\mathcal{B}}}\right)\right]\label{eq:eq:firing-rates-conditional-probability-distribution-independent-Gaussian-noise-case}\\
\nonumber \\
P\left(\boldsymbol{\nu}\right)= & \frac{1}{2^{N}}\sum_{j=0}^{2^{N}-1}F_{j}\prod_{m=0}^{N-1}\left[1+\left(-1\right)^{\nu_{m}}\mathrm{erf}\left(\frac{\theta_{m}-\frac{1}{M_{m}}\sum_{n=0}^{N-1}J_{mn}\mathscr{B}_{j,n}^{\left(N\right)}-I_{m}}{\sqrt{2}\sigma_{m}^{\mathcal{B}}}\right)\right].\label{eq:firing-rates-joint-probability-distribution-independent-Gaussian-noise-case}
\end{align}
}
\par\end{center}{\small \par}
\end{spacing}

\noindent Similarly to the membrane potentials (see Eq.~\eqref{eq:membrane-potentials-marginal-probability-distribution-independent-Gaussian-noise-case}),
for completeness we report the single-neuron marginal distribution
of the firing rates:

\begin{spacing}{0.80000000000000004}
\begin{center}
{\small{}
\[
P_{i}\left(\nu\right)={\displaystyle \sum_{\substack{\nu_{p}\in\left\{ 0,1\right\} \\
\forall p\neq i
}
}}P\left(\boldsymbol{\nu}\right)=\frac{1}{2}\sum_{j=0}^{2^{N}-1}F_{j}\left[1+\left(-1\right)^{\nu}\mathrm{erf}\left(\frac{\theta_{i}-\frac{1}{M_{i}}\sum_{n=0}^{N-1}J_{in}\mathscr{B}_{j,n}^{\left(N\right)}-I_{i}}{\sqrt{2}\sigma_{i}^{\mathcal{B}}}\right)\right].
\]
}
\par\end{center}{\small \par}
\end{spacing}

\section{Derivation of the Learning Rule Eq.~(9) \label{sec:Derivation-of-the-Learning-Rule}}

We want to calculate the synaptic connectivity matrix $J$ that allows
us to store $D$ pattern sequences $\boldsymbol{\nu}^{\left(i\right)}\left(t_{0}\right)\rightarrow\ldots\rightarrow\boldsymbol{\nu}^{\left(i\right)}\left(t_{L_{i}}\right)$
of length $L_{i}$, for $i=0,\ldots,D-1$. We observe that for each
of the $L_{i}$ transitions $\boldsymbol{\nu}^{\left(i\right)}\left(t_{n_{i}}\right)\rightarrow\boldsymbol{\nu}^{\left(i\right)}\left(t_{n_{i}}+1\right)$
in this sequence (for $n_{i}=0,\ldots,L_{i}-1$), it must be $P\left(\boldsymbol{\nu}^{\left(i\right)}\left(t_{n_{i}}+1\right)|\boldsymbol{\nu}^{\left(i\right)}\left(t_{n_{i}}\right)\right)\approx1$
for the transition to occur. In the case of independent Gaussian sources
of noise discussed in the main text, according to Eq.~\eqref{eq:eq:firing-rates-conditional-probability-distribution-independent-Gaussian-noise-case}
this condition is equivalent to:

\begin{spacing}{0.8}
\begin{center}
{\small{}
\[
\mathrm{erf}\left(\frac{\theta_{j}-\frac{1}{M_{j}}\sum_{k=0}^{N-1}J_{jk}\nu_{k}^{\left(i\right)}\left(t_{n_{i}}\right)-I_{j}}{\sqrt{2}\sigma_{j}^{\mathcal{B}}}\right)\approx\left(-1\right)^{\nu_{j}^{\left(i\right)}\left(t_{n_{i}}+1\right)},
\]
}
\par\end{center}{\small \par}
\end{spacing}

\noindent or in other words:

\begin{spacing}{0.8}
\begin{center}
{\small{}
\begin{equation}
\frac{\theta_{j}-\frac{1}{M_{j}}\sum_{k=0}^{N-1}J_{jk}\nu_{k}^{\left(i\right)}\left(t_{n_{i}}\right)-I_{j}}{\sqrt{2}\sigma_{j}^{\mathcal{B}}}=\left(-1\right)^{\nu_{j}^{\left(i\right)}\left(t_{n_{i}}+1\right)}K_{j}^{\left(i,n_{i}\right)},\label{eq:system-of-equations-for-the-synaptic-weights-0}
\end{equation}
}
\par\end{center}{\small \par}
\end{spacing}

\noindent for any sufficiently large and positive constant $K_{j}^{\left(i,n_{i}\right)}$.
Here we specialize to the case of networks without self-connections
(i.e. $J_{jj}=0$ for $j=0,\ldots,N-1$), because they are more biologically
realistic. Moreover, for the sake of example, we suppose that the
network has a fully-connected topology, so that $M_{j}=N-1\;\forall j$.
Therefore, Eq.~\eqref{eq:system-of-equations-for-the-synaptic-weights-0}
can be interpreted as the following $N$ systems of linear algebraic
equations:

\begin{spacing}{0.8}
\begin{center}
{\small{}
\begin{equation}
\Omega^{\left(j\right)}\boldsymbol{J}^{\left(j\right)}=\boldsymbol{u}^{\left(j\right)},\quad j=0,\ldots,N-1,\label{eq:system-of-equations-for-the-synaptic-weights-1}
\end{equation}
}
\par\end{center}{\small \par}
\end{spacing}

\noindent where $\Omega^{\left(j\right)}$, $\boldsymbol{J}^{\left(j\right)}$
and $\boldsymbol{u}^{\left(j\right)}$ are defined as in SubSec.~(3.2)
of the main text.

\section{An Example of Codimension Two Bifurcation Diagram \label{sec:An-Example-of-Codimension-Two-Bifurcation-Diagram}}

Our technique for calculating the codimension two bifurcation diagram
can be easily applied to networks with any size and topology in the
zero-noise limit. However, for the sake of example, here we apply
the method on the fully-connected network introduced in SubSec.~(3.3)
of the main text.

\subsection{Stationary States \label{sub:Stationary-States}}

The state $\boldsymbol{\nu}$ is stationary if $P\left(\boldsymbol{\nu}|\boldsymbol{\nu}\right)=1$,
where the conditional probability distribution of the firing rates
is given by Eq.~(10) in the main text. This equation, when solved
for the stimuli $I_{E,I}$, describes analytically the formation of
the blue lines in Fig.~(6) of the main text, obtained in the specific
case of a fully-connected network with $N_{E}=3$ excitatory neurons
and $N_{I}=3$ inhibitory neurons.

For example, for the state $0$, after some algebra we get:

\begin{spacing}{0.80000000000000004}
\begin{center}
{\small{}
\[
\frac{1}{64}\left[1+\mathrm{sgn}\left(\theta_{E}-I_{E}\right)\right]^{3}\left[1+\mathrm{sgn}\left(\theta_{I}-I_{I}\right)\right]^{3}=1,
\]
}
\par\end{center}{\small \par}
\end{spacing}

\noindent whose solution is:

\begin{spacing}{0.80000000000000004}
\begin{center}
{\small{}
\[
\mathrm{State}\;0:\;\begin{cases}
I_{E}\leq\theta_{E}\\
\\
I_{I}\leq\theta_{I}.
\end{cases}
\]
}
\par\end{center}{\small \par}
\end{spacing}

\noindent In a similar way, the remaining stationary states are allowed
if the network's parameters satisfy the following conditions:

\begin{spacing}{0.80000000000000004}
\begin{center}
{\scriptsize{}
\begin{align}
\mathrm{States}\;1,2,4:\; & \begin{cases}
I_{E}\leq\theta_{E}-\frac{J_{EI}}{N-1}\\
\\
\theta_{I}<I_{I}\leq\theta_{I}-\frac{J_{II}}{N-1}
\end{cases} & \mathrm{States}\;3,5,6:\; & \begin{cases}
I_{E}\leq\theta_{E}-\frac{2J_{EI}}{N-1}\\
\\
\theta_{I}-\frac{J_{II}}{N-1}<I_{I}\leq\theta_{I}-\frac{2J_{II}}{N-1}
\end{cases}\nonumber \\
\nonumber \\
\mathrm{State}\;7:\; & \begin{cases}
I_{E}\leq\theta_{E}-\frac{3J_{EI}}{N-1}\\
\\
I_{I}>\theta_{I}-\frac{2J_{II}}{N-1}
\end{cases} & \mathrm{State}\;56:\; & \begin{cases}
I_{E}>\theta_{E}-\frac{2J_{EE}}{N-1}\\
\\
I_{I}\leq\theta_{I}-\frac{3J_{IE}}{N-1}
\end{cases}\nonumber \\
\nonumber \\
\mathrm{States}\;57,58,60:\; & \begin{cases}
I_{E}>\theta_{E}-\frac{2J_{EE}+J_{EI}}{N-1}\\
\\
\theta_{I}-\frac{3J_{IE}}{N-1}<I_{I}\leq\theta_{I}-\frac{3J_{IE}+J_{II}}{N-1}
\end{cases} & \mathrm{States}\;59,61,62:\; & \begin{cases}
I_{E}>\theta_{E}-\frac{2\left(J_{EE}+J_{EI}\right)}{N-1}\\
\\
\theta_{I}-\frac{3J_{IE}+J_{II}}{N-1}<I_{I}\leq\theta_{I}-\frac{3J_{IE}+2J_{II}}{N-1}
\end{cases}\nonumber \\
\nonumber \\
\mathrm{State}\;63:\; & \begin{cases}
I_{E}>\theta_{E}-\frac{2J_{EE}+3J_{EI}}{N-1}\\
\\
I_{I}>\theta_{I}-\frac{3J_{IE}+2J_{II}}{N-1}
\end{cases}\label{eq:conditions-for-the-formation-of-the-stationary-states}
\end{align}
}
\par\end{center}{\scriptsize \par}
\end{spacing}

\noindent for $N=6$. By substituting the values of the parameters
of Tab.~(5), these inequalities provide the areas delimited by the
blue lines in Fig.~(6) (see the main text).

We observe that all the stationary solutions are characterized by
homogeneous states in the excitatory population, while the inhibitory
population can be either in a homogeneous state (e.g. $\boldsymbol{\nu}=\left[\begin{array}[t]{cccccc}
0 & 0 & 0 & 1 & 1 & 1\end{array}\right]^{T}$ , which corresponds to the state $7$) or in a heterogeneous state
(e.g. $\boldsymbol{\nu}=\left[\begin{array}[t]{cccccc}
1 & 1 & 1 & 0 & 0 & 1\end{array}\right]^{T}$ , which corresponds to the state $57$). Heterogeneous stationary
states in the excitatory population are not allowed in this fully-connected
network (although they may occur for different topologies of the synaptic
connections). More generally, for a fully-connected network with $N_{E}$
excitatory neurons and $N_{I}$ of inhibitory neurons, with $N_{E,I}$
arbitrary, it is easy to prove that the stationary states are of the
form:

\begin{spacing}{0.8}
\begin{center}
{\small{}
\[
\boldsymbol{\nu}=\left[\begin{array}{cc}
\overset{N_{E}-\mathrm{times}}{\overbrace{\begin{array}{ccc}
\nu_{E} & \ldots & \nu_{E}\end{array}}} & \begin{array}{ccc}
\nu_{I,0} & \ldots & \nu_{I,N_{I}-1}\end{array}\end{array}\right]^{T},
\]
}
\par\end{center}{\small \par}
\end{spacing}

\noindent where $\nu_{E}\in\left\{ 0,1\right\} $ is the firing rate
of each neuron in the excitatory population, while $\nu_{I,i}\in\left\{ 0,1\right\} $
for $i=0,\ldots,N_{I}-1$ are the (generally heterogeneous) firing
rates of the $N_{I}$ neurons in the inhibitory population. The formation
of heterogeneous states in the inhibitory population was observed
also in the graded and time-continuous version of the network model
\cite{Fasoli2016}, with two important differences. The first is that
while in the graded model the heterogeneous states occur through the
destabilization of the primary (i.e. homogeneous) branch of the stationary
solutions at the branching-point bifurcations of the network \cite{Fasoli2016},
in the discrete model the homogeneous solution generally is still
allowed after the formation of the heterogeneous branches. This can
be seen from Fig.~(7) in the main text (middle panels), where we
showed that for $I_{I}=-20$ and $I_{E}\leq-3$ the network has only
one stationary state (i.e. the state $0$), while for $I_{E}>-3$
new stationary solutions occur (i.e. $59$, $61$, $62$), even though
the state $0$ is still a valid stationary solution. The second difference
is that while in the graded model the heterogeneous states occur only
for a sufficiently strong self-inhibitory weight $J_{II}$ \cite{Fasoli2016},
in the discrete model they may occur for any $J_{II}<0$. However,
the parameter $J_{II}$ determines the portion of the $I_{E}-I_{I}$
plane where the network undergoes the formation of the heterogeneous
states. The heterogeneous states always occur in a range of $I_{I}$
with length $\frac{\left|J_{II}\right|}{N-1}$, therefore for $J_{II}\rightarrow0^{-}$
they are less and less likely to occur. This can be seen for example
from Eq.~\eqref{eq:conditions-for-the-formation-of-the-stationary-states}
for the states $1-6$ and $57-62$. In a similar way, this result
also proves that in the thermodynamic limit $N\rightarrow\infty$
the heterogeneous states are not allowed anymore, as for the graded
version of the network model \cite{Fasoli2016}.

\subsection{Oscillations \label{sub:Oscillations}}

The network undergoes oscillatory dynamics for a given pair of stimuli
$I_{E,I}$ if $P\left(\boldsymbol{\nu}|\boldsymbol{\nu}'\right)=1$
at the same time for each of the transitions $\boldsymbol{\nu}'\rightarrow\boldsymbol{\nu}$
that define the given oscillation. By solving these conditions for
the stimuli $I_{E,I}$, we get:

\begin{spacing}{0.80000000000000004}
\begin{center}
{\small{}
\begin{align}
\mathrm{Oscill.}\;0\rightarrow7\rightarrow0:\; & \begin{cases}
I_{E}\leq\theta_{E}\\
\\
\theta_{I}<I_{I}\leq\theta_{I}-\frac{2J_{II}}{N-1}
\end{cases}\nonumber \\
\nonumber \\
\mathrm{Oscill.}\;56\rightarrow63\rightarrow56:\; & \begin{cases}
I_{E}>\theta_{E}-\frac{2J_{EE}+3J_{EI}}{N-1}\\
\\
\theta_{I}-\frac{3J_{IE}}{N-1}<I_{I}\leq\theta_{I}-\frac{3J_{IE}+2J_{II}}{N-1}
\end{cases}\nonumber \\
\nonumber \\
\mathrm{Oscill.}\;0\rightarrow56\rightarrow63\rightarrow0:\; & \begin{cases}
\theta_{E}<I_{E}\leq\theta_{E}-\frac{2J_{EE}+3J_{EI}}{N-1}\\
\\
\theta_{I}-\frac{3J_{IE}}{N-1}<I_{I}\leq\min\left(\theta_{I},\theta_{I}-\frac{3J_{IE}+2J_{II}}{N-1}\right)
\end{cases}\label{eq:conditions-for-the-formation-of-oscillations}\\
\nonumber \\
\mathrm{Oscill.}\;0\rightarrow63\rightarrow7\rightarrow0:\; & \begin{cases}
\theta_{E}<I_{E}\leq\theta_{E}-\frac{2J_{EE}+3J_{EI}}{N-1}\\
\\
\max\left(\theta_{I},\theta_{I}-\frac{3J_{IE}+2J_{II}}{N-1}\right)<I_{I}\leq\theta_{I}-\frac{2J_{II}}{N-1}
\end{cases}\nonumber \\
\nonumber \\
\mathrm{Oscill.}\;0\rightarrow56\rightarrow63\rightarrow7\rightarrow0:\; & \begin{cases}
\theta_{E}<I_{E}\leq\theta_{E}-\frac{2J_{EE}+3J_{EI}}{N-1}\\
\\
\theta_{I}-\frac{3J_{IE}+2J_{II}}{N-1}<I_{I}\leq\theta_{I}
\end{cases}\nonumber 
\end{align}
}
\par\end{center}{\small \par}
\end{spacing}

\noindent for $N=6$. By substituting the values of the parameters
of Tab.~(5), the inequalities \eqref{eq:conditions-for-the-formation-of-oscillations}
provide the areas delimited by the red lines in Fig.~(6) of the main
text.

In general, we found that the maximum period of the oscillations depends
on the parameters $N_{E,I}$. For example, for $N_{E}=3$ and $N_{I}=2$
the network undergoes only oscillations with period $2$, while for
$N_{E}=2$ and $N_{I}=4$ it undergoes oscillations with period $2$
or $3$, depending on the value of the stimuli $I_{E,I}$. We never
observed oscillations with period larger than $4$ for this network
topology, even though in general they can occur for more complicated
topologies and heterogeneous synaptic weights. We observe that, depending
on its parameters, a two-populations fully-connected network of arbitrary
size shows a large variety of oscillatory firing rates with period
$2$, such as those characterized by one of the following aspects:
\begin{itemize}
\item homogeneous and stationary excitatory neurons, homogeneous and oscillating
inhibitory neurons;
\item heterogeneous and oscillating excitatory neurons, stationary inhibitory
neurons (either homogeneous or heterogeneous);
\item heterogeneous and oscillating excitatory neurons, homogeneous and
oscillating inhibitory neurons;
\item synchronous oscillations between the states $0$ (all neurons not
firing) and $2^{N}-1$ (all neurons firing), etc.
\end{itemize}
In particular, we observe that the network undergoes the formation
of mixed states, where oscillating and stationary populations coexist.
This represents another important difference between the network model
\eqref{eq:generalized-discrete-time-voltage-based-equations} and
its corresponding graded version \cite{Fasoli2016}. Indeed, in the
discrete model sub- and super-threshold oscillations of the membrane
potentials do not result in oscillations of the corresponding firing
rates, while this phenomenon does not occur in the graded model as
a consequence of the continuous activation function.

\section{Higher-Order Cross-Correlation Structure of the Network \label{sec:Higher-Order-Cross-Correlation-Structure-of-the-Network}}

In this section we calculate the higher-order correlations of the
network, i.e. the statistical dependencies among an arbitrary number
of neurons (groupwise correlation). We perform this calculation for
both the membrane potentials (SubSec.~\eqref{sub:Correlations-among-the-Membrane-Potentials})
and the firing rates (SubSec.~\eqref{sub:Correlations-among-the-Firing-Rates}).

In \cite{Fasoli2015} the authors introduced the following normalized
coefficient for quantifying the higher-order correlations among an
arbitrary number $n$ of neurons in a network of size $N$ (with $2\leq n\leq N$):

\begin{spacing}{0.80000000000000004}
\begin{center}
{\small{}
\begin{equation}
\mathrm{Corr}_{n}\left(x_{i_{0}}\left(t\right),\ldots,x_{i_{n-1}}\left(t\right)\right)=\frac{\overline{\prod_{m=0}^{n-1}\left(x_{i_{m}}\left(t\right)-\overline{x}_{i_{m}}\left(t\right)\right)}}{\sqrt[n]{\prod_{m=0}^{n-1}\overline{\left|x_{i_{m}}\left(t\right)-\overline{x}_{i_{m}}\left(t\right)\right|^{n}}}},\label{eq:higher-order-correlations}
\end{equation}
}
\par\end{center}{\small \par}
\end{spacing}

\noindent where the bar represents the statistical mean over trials
computed at time $t$. In particular, we observe that Eq.~\eqref{eq:higher-order-correlations}
is a generalization of the Pearson's correlation coefficient, which
is obtained in the special case $n=2$.

We also observe that $\mathrm{Corr}_{n}\left(x_{i_{0}}\left(t\right),\ldots,x_{i_{n-1}}\left(t\right)\right)$
consists mainly of statistical combinations of lower-order correlations,
similarly to the \textit{$n$-particle correlation function} studied
in physics \cite{Botet2002}. In order to study genuine $n$-particle
correlations, we need to calculate the \textit{$n$-particle cumulants}
\cite{Botet2002}. Their derivation is very cumbersome and goes beyond
the purpose of the article. Notwithstanding, the cumulants can be
obtained from the results of SubSecs.~\eqref{sub:Correlations-among-the-Membrane-Potentials}
and \eqref{sub:Correlations-among-the-Firing-Rates}, if desired.
For simplicity, here we focus on the coefficient $\mathrm{Corr}_{n}\left(x_{i_{0}}\left(t\right),\ldots,x_{i_{n-1}}\left(t\right)\right)$,
since it provides compact formulas for every $n$. Moreover, we will
derive the correlation structure of a network driven by independent
Gaussian sources of noise (even though the correlations can be calculated
for arbitrary noise distributions from the results of Sec.~\eqref{sec:Probability-Distributions},
if desired), and among neurons with distinct neural indexes. Therefore
the formulas we will derive in SubSecs.~\eqref{sub:Correlations-among-the-Membrane-Potentials}
and \eqref{sub:Correlations-among-the-Firing-Rates} represent higher-order
correlations such as $\mathrm{Corr}_{3}\left(x_{0}\left(t\right),x_{1}\left(t\right),x_{2}\left(t\right)\right)$,
but not correlations such as $\mathrm{Corr}_{3}\left(x_{0}\left(t\right),x_{0}\left(t\right),x_{1}\left(t\right)\right)$.
Again, our results could be extended for calculating the higher-order
correlations among any combination of neural indexes, if desired.

\subsection{Correlations among the Membrane Potentials \label{sub:Correlations-among-the-Membrane-Potentials}}

In this subsection we derive the higher-order correlations among the
membrane potentials according to Eq.~\eqref{eq:higher-order-correlations}.
From Eq.~\eqref{eq:membrane-potentials-joint-probability-distribution-independent-Gaussian-noise-case}
and the integral:

\begin{spacing}{0.8}
\begin{center}
{\small{}
\[
\int_{-\infty}^{+\infty}\left(x-c\right)\frac{1}{\sqrt{2\pi}b}e^{-\frac{\left(x-a\right)^{2}}{2b^{2}}}dx=a-c,
\]
}
\par\end{center}{\small \par}
\end{spacing}

\noindent we get the following expression for the numerator of Eq.~\eqref{eq:higher-order-correlations}:

\begin{spacing}{0.8}
\begin{center}
{\small{}
\begin{align*}
\overline{\prod_{m=0}^{n-1}\left(V_{i_{m}}-\overline{V}_{i_{m}}\right)} & =\int_{\mathbb{R}^{N}}\left[\prod_{m=0}^{n-1}\left(V_{i_{m}}-\overline{V}_{i_{m}}\right)\right]p\left(\boldsymbol{V}\right)d\boldsymbol{V}\\
\\
 & =\sum_{j=0}^{2^{N}-1}F_{j}\prod_{m=0}^{n-1}\left(\frac{1}{M_{i_{m}}}\sum_{l=0}^{N-1}\mathscr{B}_{j,l}^{\left(N\right)}J_{i_{m}l}+I_{i_{m}}-\overline{V}_{i_{m}}\right).
\end{align*}
}
\par\end{center}{\small \par}
\end{spacing}

\noindent In a similar way we get:

\begin{spacing}{0.8}
\begin{center}
{\small{}
\[
\overline{V}_{i_{m}}=\frac{1}{M_{i_{m}}}\sum_{j=0}^{2^{N}-1}F_{j}\sum_{l=0}^{N-1}\mathscr{B}_{j,l}^{\left(N\right)}J_{i_{m}l}+I_{i_{m}},
\]
}
\par\end{center}{\small \par}
\end{spacing}

\noindent having also used the relation $\sum_{j=0}^{2^{N}-1}F_{j}=1$.
Moreover, according to the integral:

\begin{spacing}{0.8}
\begin{center}
{\small{}
\[
\int_{-\infty}^{+\infty}\left|x-c\right|^{n}\frac{1}{\sqrt{2\pi}b}e^{-\frac{\left(x-a\right)^{2}}{2b^{2}}}dx=\frac{2^{\frac{n}{2}}b^{n}}{\sqrt{\pi}}\Gamma\left(\frac{n+1}{2}\right)\Phi\left(-\frac{n}{2},\frac{1}{2};-\frac{\left(a-c\right)^{2}}{2b^{2}}\right),
\]
}
\par\end{center}{\small \par}
\end{spacing}

\noindent where $\Gamma$ and $\Phi$ are the gamma function and the
Kummer's confluent hypergeometric function of the first kind respectively
\cite{Abramowitz1964}, we get:

\begin{spacing}{0.8}
\begin{center}
{\scriptsize{}
\[
\overline{\left|V_{i_{m}}-\overline{V}_{i_{m}}\right|^{n}}=\frac{2^{\frac{n}{2}}\left(\sigma_{i_{m}}^{\mathcal{B}}\right)^{n}}{\sqrt{\pi}}\Gamma\left(\frac{n+1}{2}\right)\sum_{j=0}^{2^{N}-1}F_{j}\Phi\left(-\frac{n}{2},\frac{1}{2};-\frac{1}{2}\left(\frac{\mathcal{R}_{j,i_{m}}^{\left(N\right)}}{\sigma_{i_{m}}^{\mathcal{B}}}\right)^{2}\right),
\]
}
\par\end{center}{\scriptsize \par}
\end{spacing}

\noindent where:

\begin{spacing}{0.80000000000000004}
\begin{center}
{\small{}
\begin{equation}
\mathcal{R}_{j,i_{m}}^{\left(N\right)}=\frac{1}{M_{i_{m}}}\sum_{l=0}^{N-1}\left[\left(\mathscr{B}_{j,l}^{\left(N\right)}-\sum_{k=0}^{2^{N}-1}F_{k}\mathscr{B}_{k,l}^{\left(N\right)}\right)J_{i_{m}l}\right].\label{eq:argument-of-the-confluent-hypergeometric-function}
\end{equation}
}
\par\end{center}{\small \par}
\end{spacing}

\noindent From this result we can calculate the denominator of Eq.~\eqref{eq:higher-order-correlations},
therefore finally the higher-order correlation coefficient of the
membrane potentials is:

\begin{spacing}{0.80000000000000004}
\begin{center}
{\small{}
\begin{equation}
\mathrm{Corr}_{n}\left(V_{i_{0}}\left(t\right),\ldots,V_{i_{n-1}}\left(t\right)\right)=\frac{\sqrt{\pi}\sum_{j=0}^{2^{N}-1}F_{j}\prod_{m=0}^{n-1}\mathcal{R}_{j,i_{m}}^{\left(N\right)}}{2^{\frac{n}{2}}\Gamma\left(\frac{n+1}{2}\right)\sqrt[n]{\prod_{m=0}^{n-1}\left[\left(\sigma_{i_{m}}^{\mathcal{B}}\right)^{n}\sum_{j=0}^{2^{N}-1}F_{j}\Phi\left(-\frac{n}{2},\frac{1}{2};-\frac{1}{2}\left(\frac{\mathcal{R}_{j,i_{m}}^{\left(N\right)}}{\sigma_{i_{m}}^{\mathcal{B}}}\right)^{2}\right)\right]}},\label{eq:higher-order-correlations-of-the-membrane-potentials}
\end{equation}
}
\par\end{center}{\small \par}
\end{spacing}

\noindent where $\mathcal{R}_{j,i_{m}}^{\left(N\right)}$ is given
by Eq.~\eqref{eq:argument-of-the-confluent-hypergeometric-function}.

In the special case of $n$ even we get \cite{Abramowitz1964}:

\begin{spacing}{0.80000000000000004}
\begin{center}
{\small{}
\[
\Phi\left(-\frac{n}{2},\frac{1}{2};x\right)=\left(\frac{n}{2}\right)!\frac{\sqrt{\pi}}{\Gamma\left(\frac{n+1}{2}\right)}L_{\frac{n}{2}}^{\left(-\frac{1}{2}\right)}\left(x\right),
\]
}
\par\end{center}{\small \par}
\end{spacing}

\noindent where $L_{\frac{n}{2}}^{\left(-\frac{1}{2}\right)}$ is
a generalized Laguerre polynomial. In particular:

\begin{spacing}{0.80000000000000004}
\begin{center}
{\small{}
\[
L_{1}^{\left(-\frac{1}{2}\right)}\left(x\right)=\frac{1}{2}-x,
\]
}
\par\end{center}{\small \par}
\end{spacing}

\noindent therefore the Pearson's correlation coefficient of the membrane
potentials (which corresponds to the case $n=2$) can be written as
follows:

\begin{spacing}{0.80000000000000004}
\begin{center}
{\small{}
\begin{align}
\mathrm{Corr}_{2}\left(V_{i}\left(t\right),V_{j}\left(t\right)\right)= & \frac{\mathrm{Cov}\left(V_{i}\left(t\right),V_{j}\left(t\right)\right)}{\sqrt{\mathrm{Var}\left(V_{i}\left(t\right)\right)\mathrm{Var}\left(V_{j}\left(t\right)\right)}}\nonumber \\
\nonumber \\
\mathrm{Cov}\left(V_{i}\left(t\right),V_{j}\left(t\right)\right)= & \sum_{m=0}^{2^{N}-1}F_{m}\mathcal{R}_{m,i}^{\left(N\right)}\mathcal{R}_{m,j}^{\left(N\right)}\label{eq:Pearson-correlation-coefficient-of-the-membrane-potentials}\\
\nonumber \\
\mathrm{Var}\left(V_{i}\left(t\right)\right)= & \left(\sigma_{i}^{\mathcal{B}}\right)^{2}+\sum_{m=0}^{2^{N}-1}F_{m}\left(\mathcal{R}_{m,i}^{\left(N\right)}\right)^{2}.\nonumber 
\end{align}
}
\par\end{center}{\small \par}
\end{spacing}

\begin{figure}
\begin{centering}
\includegraphics[scale=0.25]{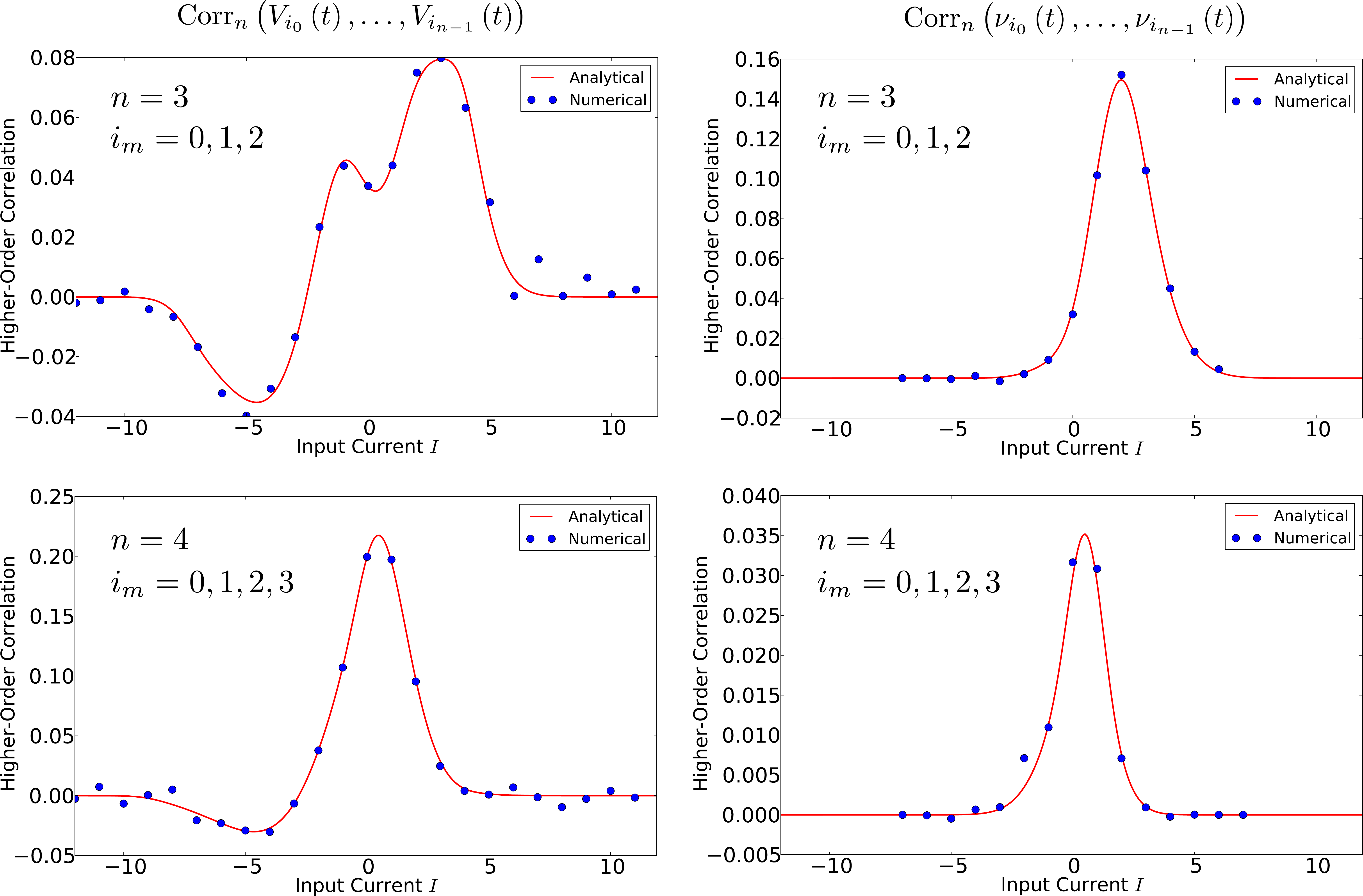}
\par\end{centering}

\protect\caption{\label{fig:Higher-order-correlations} \small \textbf{Examples of
higher-order correlations.} The figure is obtained for the network's
parameters of Tab.~(6) in the main text. The top panels show the
higher-order correlation among a triplet of neurons ($n=3$), while
the bottom panels show the correlation among a quadruplet ($n=4$).
In the left panels we report the correlations among the membrane potentials,
while in the right panels the correlations among the firing rates
(compare with Fig.~(8) in the main text, which shows the case $n=2$).
The analytical curves are obtained from Eqs.~\eqref{eq:higher-order-correlations-of-the-membrane-potentials}
and \eqref{eq:higher-order-correlations-of-the-firing-rates}, for
the membrane potentials and the firing rates respectively. The numerical
solutions are obtained through a Monte Carlo method over $10^{5}$
repetitions of the network dynamics for each value of the stimulus.}
\end{figure}

\noindent We plotted the variance and the Pearson's correlation coefficient
of the membrane potentials as a function of the stimulus in the top-left
and middle-left panels of Fig.~(8) in the main text, in the case
of the network's parameters of Tab.~(6). Moreover, we plotted examples
of higher-order correlations in the left panels of Fig.~\eqref{fig:Higher-order-correlations},
for $n=3$ and $n=4$. 

\subsubsection{Limit $\frac{\mathcal{R}_{j,i_{m}}^{\left(N\right)}}{\sigma_{i_{m}}^{\mathcal{B}}}\rightarrow\infty$
\label{sub:Small-Noise-Limit}}

Due to the high complexity of Eq.~\eqref{eq:higher-order-correlations-of-the-membrane-potentials},
which represents an exact analytical result, it may be useful to derive
a simplified expression of the higher-order correlations. Since \cite{Abramowitz1964}:

\begin{spacing}{0.80000000000000004}
\begin{center}
{\small{}
\[
\Phi\left(a,b;x\right)\approx\frac{\Gamma\left(b\right)}{\Gamma\left(b-a\right)}\left(-x\right)^{-a}
\]
}
\par\end{center}{\small \par}
\end{spacing}

\noindent for $x\rightarrow-\infty$, then if $\frac{\mathcal{R}_{j,i_{m}}^{\left(N\right)}}{\sigma_{i_{m}}^{\mathcal{B}}}\rightarrow\infty\;\forall i_{m}$
(e.g. in the small-noise limit), Eq.~\eqref{eq:higher-order-correlations-of-the-membrane-potentials}
tends to:

\begin{spacing}{0.80000000000000004}
\begin{center}
{\small{}
\begin{equation}
\mathrm{Corr}_{n}\left(V_{i_{0}}\left(t\right),\ldots,V_{i_{n-1}}\left(t\right)\right)=\frac{\sum_{j=0}^{2^{N}-1}F_{j}\prod_{m=0}^{n-1}\mathcal{R}_{j,i_{m}}^{\left(N\right)}}{\sqrt[n]{\prod_{m=0}^{n-1}\left[\sum_{j=0}^{2^{N}-1}F_{j}\left(\mathcal{R}_{j,i_{m}}^{\left(N\right)}\right)^{n}\right]}}.\label{eq:higher-order-correlations-of-the-membrane-potentials-for-weak-noise}
\end{equation}
}
\par\end{center}{\small \par}
\end{spacing}

\noindent The reader can easily check that for $n=2$ Eq.~\eqref{eq:higher-order-correlations-of-the-membrane-potentials-for-weak-noise}
can be obtained from Eq.~\eqref{eq:Pearson-correlation-coefficient-of-the-membrane-potentials}
in the limit $\frac{\mathcal{R}_{j,i_{m}}^{\left(N\right)}}{\sigma_{i_{m}}^{\mathcal{B}}}\rightarrow\infty$.

Now we call $\chi_{i_{0},\ldots i_{n-1}}^{\left(n\right)}$ the right-hand
side of Eq.~\eqref{eq:higher-order-correlations-of-the-membrane-potentials-for-weak-noise}.
For some combinations of the network's parameters $J$ and $\boldsymbol{I}$
and regardless of $\sigma_{i_{m}}^{\mathcal{B}}$, the term $\chi_{i_{0},\ldots i_{n-1}}^{\left(n\right)}$
may tend to one independently of the order of magnitude of the ratio
$\frac{\mathcal{R}_{j,i_{m}}^{\left(N\right)}}{\sigma_{i_{m}}^{\mathcal{B}}}$
(the latter being regulated by $\sigma_{i_{m}}^{\mathcal{B}}$). Therefore
the membrane potentials become highly correlated if the conditions
$\frac{\mathcal{R}_{j,i_{m}}^{\left(N\right)}}{\sigma_{i_{m}}^{\mathcal{B}}}\rightarrow\infty$
and $\chi_{i_{0},\ldots i_{n-1}}^{\left(n\right)}\rightarrow1$ occur
at the same time. This phenomenon can be observed for example in Fig.~(8)
of the main text, whose middle-left panel shows that the Pearson's
coefficient between the neurons $0$ and $1$ is close to one when
$I\in\left[-4,4\right]$. Due to the complexity of Eq.~\eqref{eq:higher-order-correlations-of-the-membrane-potentials},
the cross-correlations among the membrane potentials may tend to one
also in other ways, but their investigation is beyond the purpose
of this article.

\subsubsection{Strong-Current Limit \label{sub:Strong-Current-Limit-for-the-Membrane-Potentials}}

If the network is in a state $\boldsymbol{\nu}$, and the total current
$\frac{1}{M_{i}}\sum_{j=0}^{N-1}J_{ij}\nu_{j}+I_{i}$ to the $i$th
neuron is much larger than $\sigma_{i}^{\mathcal{B}}$, for $i=0,\ldots,N-1$,
then the membrane potentials fluctuate far from the thresholds $\theta_{i}$.
For this reason, under this hypothesis the firing rate $\boldsymbol{\nu}$
is more likely to settle in a stationary state, rather than switching
among several binary states. If in the strong-current limit $\boldsymbol{\nu}$
settles in a binary state whose decimal representation is $z$, then
the coefficient $F_{j}$ tends to the Kronecker delta $\delta_{0,z}$.
This implies:

\begin{spacing}{0.8}
\begin{center}
{\small{}
\[
\mathcal{R}_{j,i_{m}}^{\left(N\right)}\rightarrow\frac{1}{M_{i_{m}}}\sum_{l=0}^{N-1}\left[\left(\mathscr{B}_{j,l}^{\left(N\right)}-\mathscr{B}_{z,l}^{\left(N\right)}\right)J_{i_{m}l}\right],
\]
}
\par\end{center}{\small \par}
\end{spacing}

\noindent therefore in particular $\mathcal{R}_{z,i_{m}}^{\left(N\right)}\rightarrow0$.
In turn, the numerator of Eq.~\eqref{eq:higher-order-correlations-of-the-membrane-potentials}
becomes:

\begin{spacing}{0.8}
\begin{center}
{\small{}
\[
\sum_{j=0}^{2^{N}-1}F_{j}\prod_{m=0}^{n-1}\mathcal{R}_{j,i_{m}}^{\left(N\right)}\rightarrow\prod_{m=0}^{n-1}\mathcal{R}_{z,i_{m}}^{\left(N\right)}\rightarrow0,
\]
}
\par\end{center}{\small \par}
\end{spacing}

\noindent while for the argument of the $n$th root at the denominator
we get:

\begin{spacing}{0.8}
\begin{center}
{\small{}
\[
\prod_{m=0}^{n-1}\left[\left(\sigma_{i_{m}}^{\mathcal{B}}\right)^{n}\sum_{j=0}^{2^{N}-1}F_{j}\Phi\left(-\frac{n}{2},\frac{1}{2};-\frac{1}{2}\left(\frac{\mathcal{R}_{j,i_{m}}^{\left(N\right)}}{\sigma_{i_{m}}^{\mathcal{B}}}\right)^{2}\right)\right]\rightarrow\prod_{m=0}^{n-1}\left(\sigma_{i_{m}}^{\mathcal{B}}\right)^{n},
\]
}
\par\end{center}{\small \par}
\end{spacing}

\noindent since $\Phi\left(a,b;0\right)=1$. Therefore we conclude
that, in the strong-current limit, $\mathrm{Corr}_{n}\left(V_{i_{0}}\left(t\right),\ldots,V_{i_{n-1}}\left(t\right)\right)\rightarrow0\;\forall n$
whenever $\sigma_{i_{m}}^{\mathcal{B}}>0\;\forall m$. This result
proves the formation of asynchronous states for the membrane potentials.
An example of this phenomenon is shown in the middle-left panel of
Fig.~(8) in the main text. More generally, from Eq.~\eqref{eq:membrane-potentials-joint-probability-distribution-independent-Gaussian-noise-case}
we observe that for $F_{j}\rightarrow\delta_{0,z}$ the joint probability
distribution of the membrane potentials factorizes into the product
of the single-neuron marginal distributions:

\begin{spacing}{0.8}
\begin{center}
{\small{}
\[
p\left(\boldsymbol{V}\right)\rightarrow\frac{1}{\left(2\pi\right)^{\frac{N}{2}}\prod_{i=0}^{N-1}\sigma_{i}^{\mathcal{B}}}\prod_{m=0}^{N-1}e^{-\frac{1}{2}\left(\frac{V_{m}-\frac{1}{M_{m}}\sum_{n=0}^{N-1}J_{mn}\mathscr{B}_{z,n}^{\left(N\right)}-I_{m}}{\sigma_{m}^{\mathcal{B}}}\right)^{2}}=\prod_{m=0}^{N-1}p_{m}\left(V_{m}\right),
\]
}
\par\end{center}{\small \par}
\end{spacing}

\noindent where $p_{m}\left(V_{m}\right)$ is given by Eq.~\eqref{eq:membrane-potentials-marginal-probability-distribution-independent-Gaussian-noise-case}.
For this reason, in the strong-current limit the neurons become not
only uncorrelated, but also independent.

\subsection{Correlations among the Firing Rates \label{sub:Correlations-among-the-Firing-Rates}}

In this subsection we derive the higher-order correlations among the
firing rates, according to Eq.~\eqref{eq:higher-order-correlations}.
From Eq.~\eqref{eq:firing-rates-joint-probability-distribution-independent-Gaussian-noise-case},
we get the following expression for the numerator of Eq.~\eqref{eq:higher-order-correlations}:

\begin{spacing}{0.8}
\begin{center}
{\small{}
\[
\overline{\prod_{m=0}^{n-1}\left(\nu_{i_{m}}-\overline{\nu}_{i_{m}}\right)}=\sum_{\boldsymbol{\nu}\in\left\{ 0,1\right\} ^{N}}\left[\prod_{m=0}^{n-1}\left(\nu_{i_{m}}-\overline{\nu}_{i_{m}}\right)\right]P\left(\boldsymbol{\nu}\right)=\frac{1}{2^{n}}\sum_{j=0}^{2^{N}-1}F_{j}\prod_{m=0}^{n-1}\left(1-2\overline{\nu}_{i_{m}}-E_{j,i_{m}}\right),
\]
}
\par\end{center}{\small \par}
\end{spacing}

\noindent where:

\begin{spacing}{0.8}
\begin{center}
{\small{}
\begin{align}
\overline{\nu}_{i_{m}}= & \frac{1}{2}\left(1-\sum_{j=0}^{2^{N}-1}F_{j}E_{j,i_{m}}\right)\nonumber \\
\label{eq:mean-firing-rate}\\
E_{j,i_{m}}= & \mathrm{erf}\left(\frac{\theta_{i_{m}}-\frac{1}{M_{i_{m}}}\sum_{l=0}^{N-1}J_{i_{m}l}\mathscr{B}_{j,l}^{\left(N\right)}-I_{i_{m}}}{\sqrt{2}\sigma_{i_{m}}^{\mathcal{B}}}\right).\nonumber 
\end{align}
}
\par\end{center}{\small \par}
\end{spacing}

\noindent Moreover, the denominator of Eq.~\eqref{eq:higher-order-correlations}
can be calculated from the following formula:

\begin{spacing}{0.8}
\begin{center}
{\small{}
\[
\overline{\left|\nu_{i_{m}}-\overline{\nu}_{i_{m}}\right|^{n}}=\frac{1}{2}\left\{ \left(\overline{\nu}_{i_{m}}\right)^{n}+\left(1-\overline{\nu}_{i_{m}}\right)^{n}+\left[\left(\overline{\nu}_{i_{m}}\right)^{n}-\left(1-\overline{\nu}_{i_{m}}\right)^{n}\right]\sum_{j=0}^{2^{N}-1}F_{j}E_{j,i_{m}}\right\} .
\]
}
\par\end{center}{\small \par}
\end{spacing}

\noindent Finally, the higher-order correlation coefficient of the
firing rates is:

\begin{spacing}{0.8}
\begin{center}
{\small{}
\begin{align}
\mathrm{Corr}_{n}\left(\nu_{i_{0}}\left(t\right),\ldots,\nu_{i_{n-1}}\left(t\right)\right)= & \frac{\sum_{j=0}^{2^{N}-1}F_{j}\prod_{m=0}^{n-1}\left(1-2\overline{\nu}_{i_{m}}-E_{j,i_{m}}\right)}{2^{n}\sqrt[n]{\prod_{m=0}^{n-1}Z_{n}\left(\overline{\nu}_{i_{m}}\right)}}\nonumber \\
\label{eq:higher-order-correlations-of-the-firing-rates}\\
Z_{n}\left(x\right)= & x^{n}\left(1-x\right)+x\left(1-x\right)^{n},\nonumber 
\end{align}
}
\par\end{center}{\small \par}
\end{spacing}

\noindent where $\overline{\nu}_{i_{m}}$ and $E_{j,i_{m}}$ are given
by Eq.~\eqref{eq:mean-firing-rate}.

In the special case $n=2$, from Eq.~\eqref{eq:higher-order-correlations-of-the-firing-rates}
we get the following expression of the Pearson's correlation coefficient
of the firing rates:

\begin{spacing}{0.80000000000000004}
\begin{center}
{\small{}
\begin{align}
\mathrm{Corr}_{2}\left(\nu_{i}\left(t\right),\nu_{j}\left(t\right)\right)= & \frac{\mathrm{Cov}\left(\nu_{i}\left(t\right),\nu_{j}\left(t\right)\right)}{\sqrt{\mathrm{Var}\left(\nu_{i}\left(t\right)\right)\mathrm{Var}\left(\nu_{j}\left(t\right)\right)}}\nonumber \\
\nonumber \\
\mathrm{Cov}\left(\nu_{i}\left(t\right),\nu_{j}\left(t\right)\right)= & \frac{1}{4}\sum_{m=0}^{2^{N}-1}F_{m}\left(1-2\overline{\nu}_{i}-E_{m,i}\right)\left(1-2\overline{\nu}_{j}-E_{m,j}\right)\label{eq:Pearson-correlation-coefficient-of-the-firing-rates}\\
\nonumber \\
\mathrm{Var}\left(\nu_{i}\left(t\right)\right)= & \overline{\nu}_{i}-\left(\overline{\nu}_{i}\right)^{2},\nonumber 
\end{align}
}
\par\end{center}{\small \par}
\end{spacing}

\noindent where as usual $\overline{\nu}_{i}$ is given by Eq.~\eqref{eq:mean-firing-rate}.
We plotted the variance and the Pearson's correlation coefficient
of the firing rates as a function of the stimulus in the top-right
and middle-right panels of Fig.~(8) in the main text, in the case
of the network's parameters of Tab~(6). Moreover, we plotted examples
of higher-order correlations $\mathrm{Corr}_{n}\left(\nu_{i_{0}}\left(t\right),\ldots,\nu_{i_{n-1}}\left(t\right)\right)$
in the right panels of Fig.~\eqref{fig:Higher-order-correlations},
for $n=3$ and $n=4$.

Unlike the case of the membrane potentials (see SubSec.~\eqref{sub:Small-Noise-Limit}),
the small-noise limit of Eq.~\eqref{eq:higher-order-correlations-of-the-firing-rates}
does not provide any useful simplification of the statistical properties
of the firing rates, due to the discrete nature of the vector $\boldsymbol{\nu}$.
For this reason, in what follows we focus only on the strong-current
limit of the network.

\subsubsection{Strong-Current Limit \label{sub:Strong-Current-Limit-for-the-Firing-Rates}}

Unlike the case of the membrane potentials (see SubSec.~\eqref{sub:Strong-Current-Limit-for-the-Membrane-Potentials}),
in the strong-current limit the cross-correlations among the firing
rates do not necessarily tend to zero. For example, from Eq.~\eqref{eq:Pearson-correlation-coefficient-of-the-firing-rates},
after some algebra it is possible to prove that whenever $\overline{\nu}_{i}\approx\overline{\nu}_{j}$
and $\psi_{i}\overset{\mathrm{def}}{=}\frac{1-\sum_{m=0}^{2^{N}-1}F_{m}E_{m,i}^{2}}{1-\left(\sum_{m=0}^{2^{N}-1}F_{m}E_{m,i}\right)^{2}}\rightarrow0$
in the strong-current limit, we get:

\begin{spacing}{0.8}
\begin{center}
{\small{}
\begin{equation}
\mathrm{Corr}_{2}\left(\nu_{i}\left(t\right),\nu_{j}\left(t\right)\right)\approx\frac{\frac{1}{4}\sum_{m=0}^{2^{N}-1}F_{m}\left(1-2\overline{\nu}_{i}-E_{m,i}\right)^{2}}{\overline{\nu}_{i}-\left(\overline{\nu}_{i}\right)^{2}}=1-\psi_{i}\rightarrow1.\label{eq:strong-correlation-limit-for-the-firing-rates}
\end{equation}
}
\par\end{center}{\small \par}
\end{spacing}

\noindent The formation of strong correlation through this mechanism
can be observed for example in the middle-right panel of Fig.~(8)
of the main text, for $I>3$. In particular, the overlap of the curves
of the standard deviations of the neurons $0$ and $1$ shown in the
top-right panel of the figure for $I>3$ is a consequence of the fact
that $\overline{\nu}_{0}\approx\overline{\nu}_{1}$. This is one of
the assumptions required to obtain the limit \eqref{eq:strong-correlation-limit-for-the-firing-rates},
while the reader may check through the analytical expressions of $F_{m}$
and $E_{m,i}$ that the assumption $\psi_{i}\rightarrow0$ is also
satisfied. However, we stress that this is just an example of the
conditions that lead to strong correlations among the firing rates.
Similarly to the membrane potentials, the cross-correlations among
the firing rates may also grow under more complex conditions, whose
full characterization is beyond the purpose of this article.

\section{Mean-Field Theory \label{sec:Mean-Field-Theory}}

We derive the mean-field equations of the multi-population network
topology introduced in SubSec.~(3.5) of the main text. Then, for
the sake of example, we study analytically their local bifurcations
in the case of two neural populations.

\subsection{Equations \label{sub:Equations}}

For simplicity, we focus again on the case of a network driven by
noise with Gaussian distribution, as in the main text. Since the noise
sources $\mathcal{B}_{i}\left(t\right)$ are independent for any $N$,
the membrane potentials become independent in the thermodynamic limit
$N\rightarrow\infty$ \footnote{The proof is technically complex and beyond the purpose of this article.
For more details, the interested reader is referred to the work of
McKean, Tanaka, Sznitman and others (see \cite{Touboul2012,Baladron2012}
and references therein). For simplicity, we show the increase of independence
between the membrane potentials for increasing network size only numerically
(see the bottom-right panel of Fig.~(9) in the main text). The study
of infinite-size neural networks driven by correlated sources of noise,
which typically requires more advanced techniques such as \textit{large
deviations theory} \cite{Faugeras2015}, is not considered here.}. This implies that the neurons within each population become independent
and identically distributed, therefore according to the law of large
numbers we get:

\begin{spacing}{0.8}
\begin{center}
{\small{}
\begin{equation}
\kappa_{\alpha}\left(t\right)\overset{\mathrm{def}}{=}\underset{N\rightarrow\infty}{\lim}\frac{1}{M_{i}}\sum_{j=0}^{N-1}J_{ij}\mathscr{H}\left(V_{j}\left(t\right)-\theta_{j}\right)=\sum_{\beta=0}^{\mathfrak{P}-1}R_{\beta}J_{\alpha\beta}\int_{\theta_{\beta}}^{+\infty}p_{\beta}\left(V,t\right)dV,\label{eq:factor-kappa}
\end{equation}
}
\par\end{center}{\small \par}
\end{spacing}

\noindent given the $i$th neuron is in the population $\alpha$,
while $p_{\beta}\left(V,t\right)$ is the marginal probability distribution
of each neuron in the population $\beta$, and $R_{\alpha}=\underset{N\rightarrow\infty}{\lim}\frac{N_{\alpha}}{M_{\alpha}}$.
From Eqs.~\eqref{eq:generalized-discrete-time-voltage-based-equations}
and \eqref{eq:factor-kappa} it follows that:

\begin{spacing}{0.8}
\begin{center}
{\small{}
\[
V_{i}\left(t+1\right)=\kappa_{\alpha}\left(t\right)+I_{\alpha}\left(t\right)+\sigma_{\alpha}^{\mathcal{B}}\mathcal{B}_{i}\left(t\right),
\]
}
\par\end{center}{\small \par}
\end{spacing}

\noindent for all the neurons $i$ in population $\alpha$, therefore
$V_{i}\left(t+1\right)$ is normally distributed with mean:

\begin{spacing}{0.8}
\begin{center}
{\small{}
\begin{equation}
\overline{V}_{\alpha}\left(t+1\right)=\kappa_{\alpha}\left(t\right)+I_{\alpha}\left(t\right)\label{eq:mean-membrane-potential}
\end{equation}
}
\par\end{center}{\small \par}
\end{spacing}

\noindent and standard deviation $\sigma_{\alpha}^{\mathcal{B}}$.
It follows that:

\begin{spacing}{0.8}
\begin{center}
{\small{}
\begin{equation}
p_{\beta}\left(V,t\right)=\frac{1}{\sqrt{2\pi}\sigma_{\beta}^{\mathcal{B}}}e^{-\frac{\left(V-\kappa_{\beta}\left(t-1\right)-I_{\beta}\left(t-1\right)\right)^{2}}{2\left(\sigma_{\beta}^{\mathcal{B}}\right)^{2}}},\label{eq:marginal-probability-distribution-of-the-membrane-potentials-mean-field}
\end{equation}
}
\par\end{center}{\small \par}
\end{spacing}

\noindent therefore from Eqs.~\eqref{eq:factor-kappa} and \eqref{eq:marginal-probability-distribution-of-the-membrane-potentials-mean-field}
we obtain the following self-consistency mean-field equations:

\begin{spacing}{0.8}
\begin{center}
{\small{}
\begin{equation}
\kappa_{\alpha}\left(t\right)=\frac{1}{2}\sum_{\beta=0}^{\mathfrak{P}-1}R_{\beta}J_{\alpha\beta}\left[1-\mathrm{erf}\left(\frac{\theta_{\beta}-\kappa_{\beta}\left(t-1\right)-I_{\beta}\left(t-1\right)}{\sqrt{2}\sigma_{\beta}^{\mathcal{B}}}\right)\right].\label{eq:mean-field-equations-0}
\end{equation}
}
\par\end{center}{\small \par}
\end{spacing}

\noindent To conclude, by means of Eq.~\eqref{eq:mean-membrane-potential},
the mean-field equations \eqref{eq:mean-field-equations-0} can be
equivalently rewritten as follows:

\begin{spacing}{0.8}
\begin{center}
{\small{}
\begin{equation}
\overline{V}_{\alpha}\left(t+1\right)=\mathfrak{F}_{\alpha}\left(\overline{\boldsymbol{V}}\left(t\right)\right)\overset{\mathrm{def}}{=}\frac{1}{2}\sum_{\beta=0}^{\mathfrak{P}-1}R_{\beta}J_{\alpha\beta}\left[1-\mathrm{erf}\left(\frac{\theta_{\beta}-\overline{V}_{\beta}\left(t\right)}{\sqrt{2}\sigma_{\beta}^{\mathcal{B}}}\right)\right]+I_{\alpha}\left(t\right),\;\alpha=0,\ldots,\mathfrak{P}-1,\label{eq:mean-field-equations-1}
\end{equation}
}
\par\end{center}{\small \par}
\end{spacing}

\noindent where $\overline{\boldsymbol{V}}\overset{\mathrm{def}}{=}\left[\begin{array}[t]{ccc}
\overline{V}_{0}, & \ldots, & \overline{V}_{\mathfrak{P}-1}\end{array}\right]^{T}$. Eq.~\eqref{eq:mean-field-equations-1} defines a system of recurrence
relations for the temporal evolution of the mean membrane potentials
$\overline{V}_{\alpha}$, as a function of the network's parameters.
In particular, in SubSec.~\eqref{sub:Bifurcations} we will study
how the dynamical properties of this system depend on the stimuli
$I_{\alpha}$ in a network composed of two neural populations.

To conclude this subsection, we observe that the joint probability
distribution of the membrane potentials of a set of neurons with indexes
$i_{0},\ldots,i_{n-1}$(where $n$ is finite) can be written as follows,
in the thermodynamic limit:

\begin{spacing}{0.8}
\begin{center}
{\small{}
\begin{equation}
p\left(V_{i_{0}},\ldots,V_{i_{n-1}},t\right)=\prod_{m=0}^{n-1}p_{i_{m}}\left(V_{i_{m}},t\right)\quad\forall t,\label{eq:joint-probability-distribution-of-the-membrane-potentials-mean-field}
\end{equation}
}
\par\end{center}{\small \par}
\end{spacing}

\noindent since the neurons become independent. In Eq.~\eqref{eq:joint-probability-distribution-of-the-membrane-potentials-mean-field},
$p_{i_{m}}\left(\cdot,t\right)=p_{\alpha}\left(\cdot,t\right)$ if
the $i_{m}$th neuron belongs to the population $\alpha$. Moreover,
by integrating Eq.~\eqref{eq:joint-probability-distribution-of-the-membrane-potentials-mean-field}
over the hypervolume $\mathscr{V}\left(\left[\begin{array}[t]{ccc}
\nu_{i_{0}}, & \ldots, & \nu_{i_{n-1}}\end{array}\right]^{T}\right)$, we get the joint probability distribution of the firing rates:

\begin{spacing}{0.8}
\begin{center}
{\small{}
\begin{equation}
P\left(\nu_{i_{0}},\ldots,\nu_{i_{n-1}},t\right)=\prod_{m=0}^{n-1}P_{i_{m}}\left(\nu_{i_{m}},t\right)\quad\forall t.\label{eq:joint-probability-distribution-of-the-firing-rates-mean-field}
\end{equation}
}
\par\end{center}{\small \par}
\end{spacing}

\noindent In Eq.~\eqref{eq:joint-probability-distribution-of-the-firing-rates-mean-field},
$P_{i_{m}}\left(\cdot,t\right)=P_{\alpha}\left(\cdot,t\right)$ if
the $i_{m}$th neuron belongs to the population $\alpha$, where according
to Eq.~\eqref{eq:marginal-probability-distribution-of-the-membrane-potentials-mean-field}:

\begin{spacing}{0.8}
\begin{center}
{\small{}
\[
P_{\alpha}\left(\nu,t\right)=\frac{1}{2}\left[1+\left(-1\right)^{\nu}\mathrm{erf}\left(\frac{\theta_{\alpha}-\overline{V}_{\alpha}\left(t\right)}{\sqrt{2}\sigma_{\alpha}^{\mathcal{B}}}\right)\right].
\]
}
\par\end{center}{\small \par}
\end{spacing}

\subsection{Bifurcations \label{sub:Bifurcations}}

For the sake of example, in this subsection we focus on the case of
a network composed of two neural populations, one excitatory and one
inhibitory. It is convenient to change slightly the notation, and
to consider $\alpha=E,\,I$ rather than $\alpha=0,\,1$. By applying
the methods developed in \cite{Haschke2005,Fasoli2016}, we provide
an analytical study of the local codimension one bifurcations shown
in Fig.~(10) of the main text. Local bifurcations can be calculated
from the eigenvalues of the Jacobian matrix, evaluated at the stationary
solutions of the system. From Eq.~\eqref{eq:mean-field-equations-1}
we obtain the following expression of the $2\times2$ Jacobian matrix:

\begin{spacing}{0.8}
\begin{center}
{\small{}
\begin{equation}
\mathcal{J}=\left[\begin{array}{cc}
\mathcal{J}_{EE} & \mathcal{J}_{EI}\\
\mathcal{J}_{IE} & \mathcal{J}_{II}
\end{array}\right]=\left[\begin{array}{cc}
R_{E}J_{EE}g_{E}\left(\mu_{E}\right) & R_{I}J_{EI}g_{I}\left(\mu_{I}\right)\\
R_{E}J_{IE}g_{E}\left(\mu_{E}\right) & R_{I}J_{II}g_{I}\left(\mu_{I}\right)
\end{array}\right],\quad g_{\alpha}\left(\mu_{\alpha}\right)=\frac{1}{\sqrt{2\pi}\sigma_{\alpha}^{\mathcal{B}}}e^{-\frac{\left(\mu_{\alpha}-\theta_{\alpha}\right)^{2}}{2\left(\sigma_{\alpha}^{\mathcal{B}}\right)^{2}}}.\label{eq:Jacobian-matrix}
\end{equation}
}
\par\end{center}{\small \par}
\end{spacing}

\noindent $\mu_{E,I}$ represent the stationary solutions (fixed points)
of the two populations, which are obtained from Eq.~\eqref{eq:mean-field-equations-1}
for constant stimuli and $t\rightarrow\infty$. In particular, $\mu_{E,I}$
satisfy the following system of non-linear equations:

\begin{spacing}{0.80000000000000004}
\begin{center}
{\small{}
\begin{equation}
\begin{cases}
\mu_{E}=\frac{1}{2}R_{E}J_{EE}\left[1+\mathrm{erf}\left(\frac{\mu_{E}-\theta_{E}}{\sqrt{2}\sigma_{E}^{\mathcal{B}}}\right)\right]+\frac{1}{2}R_{I}J_{EI}\left[1+\mathrm{erf}\left(\frac{\mu_{I}-\theta_{I}}{\sqrt{2}\sigma_{I}^{\mathcal{B}}}\right)\right]+I_{E}\\
\\
\mu_{I}=\frac{1}{2}R_{E}J_{IE}\left[1+\mathrm{erf}\left(\frac{\mu_{E}-\theta_{E}}{\sqrt{2}\sigma_{E}^{\mathcal{B}}}\right)\right]+\frac{1}{2}R_{I}J_{II}\left[1+\mathrm{erf}\left(\frac{\mu_{I}-\theta_{I}}{\sqrt{2}\sigma_{I}^{\mathcal{B}}}\right)\right]+I_{I},
\end{cases}\label{eq:mean-field-equations-of-the-stationary-solutions}
\end{equation}
}
\par\end{center}{\small \par}
\end{spacing}

\noindent which describe the black curves in the right panels of Fig.~(10)
(see the main text).

Finally, the eigenvalues of $\mathcal{J}$ are:

\begin{spacing}{0.8}
\begin{center}
{\small{}
\begin{equation}
\lambda_{0,1}=\frac{\mathcal{J}_{EE}+\mathcal{J}_{II}\pm\sqrt{\left(\mathcal{J}_{EE}+\mathcal{J}_{II}\right)^{2}-4\left(\mathcal{J}_{EE}\mathcal{J}_{II}-\mathcal{J}_{EI}\mathcal{J}_{IE}\right)}}{2}.\label{eq:mean-field-eigenvalues}
\end{equation}
}
\par\end{center}{\small \par}
\end{spacing}

\noindent Depending on the conditions satisfied by the eigenvalues,
the network undergoes different kinds of local bifurcations, which
are studied in the next subsections. Similarly to \cite{Fasoli2016},
for simplicity in this paper we do not investigate the non-degeneracy
conditions of the bifurcations.

\subsubsection{Limit-Point and Period-Doubling Bifurcations \label{sub:Limit-Point-and-Period-Doubling-Bifurcations}}

The limit-point and period-doubling bifurcations are described by
the conditions $\lambda_{0,1}=1$ and $\lambda_{0,1}=-1$, respectively
\cite{Haschke2005}. If $s=\pm1$ and $\left(\mathcal{J}_{EE}+\mathcal{J}_{II}\right)^{2}-4\left(\mathcal{J}_{EE}\mathcal{J}_{II}-\mathcal{J}_{EI}\mathcal{J}_{IE}\right)>0$,
from Eq.~\eqref{eq:mean-field-eigenvalues} and the condition $\lambda_{0,1}=s$,
we get:

\begin{spacing}{0.8}
\begin{center}
{\small{}
\begin{equation}
s^{2}-s\left(\mathcal{J}_{EE}+\mathcal{J}_{II}\right)+\mathcal{J}_{EE}\mathcal{J}_{II}-\mathcal{J}_{EI}\mathcal{J}_{IE}=0.\label{eq:limit-point-and-period-doubling-condition}
\end{equation}
}
\par\end{center}{\small \par}
\end{spacing}

\noindent Now, if we define the parameter $\mathfrak{v}\overset{\mathrm{def}}{=}\mu_{E}$,
from Eq.~\eqref{eq:limit-point-and-period-doubling-condition} we
obtain:

\begin{spacing}{0.8}
\begin{center}
{\small{}
\begin{equation}
g_{I}\left(\mu_{I}\right)=\frac{sR_{E}J_{EE}g_{E}\left(\mathfrak{v}\right)-s^{2}}{R_{E}R_{I}\left(J_{EE}J_{II}-J_{EI}J_{IE}\right)g_{E}\left(\mathfrak{v}\right)-sR_{I}J_{II}}.\label{eq:derivative-of-the-activation-function-for-the-limit-point-and-period-doubling-bifurcations}
\end{equation}
}
\par\end{center}{\small \par}
\end{spacing}

\noindent Eq.~\eqref{eq:derivative-of-the-activation-function-for-the-limit-point-and-period-doubling-bifurcations}
can be inverted in order to obtain the expression of the stationary
solution in the inhibitory population:

\begin{spacing}{0.8}
\begin{center}
{\small{}
\begin{equation}
\mu_{I}\left(\mathfrak{v}\right)=\theta_{I}\pm\sqrt{-2\left(\sigma_{I}^{\mathcal{B}}\right)^{2}\ln\left(\sqrt{2\pi}\sigma_{I}^{\mathcal{B}}g_{I}\left(\mu_{I}\right)\right)}.\label{eq:inhibitory-membrane-potential}
\end{equation}
}
\par\end{center}{\small \par}
\end{spacing}

\noindent Finally, from Eq.~\eqref{eq:mean-field-equations-of-the-stationary-solutions}
we get:

\begin{spacing}{0.8}
\begin{center}
{\small{}
\begin{equation}
\begin{cases}
I_{E}\left(\mathfrak{v}\right)=\mathfrak{v}-\frac{1}{2}R_{E}J_{EE}\left[1+\mathrm{erf}\left(\frac{\mathfrak{v}-\theta_{E}}{\sqrt{2}\sigma_{E}^{\mathcal{B}}}\right)\right]-\frac{1}{2}R_{I}J_{EI}\left[1+\mathrm{erf}\left(\frac{\mu_{I}\left(\mathfrak{v}\right)-\theta_{I}}{\sqrt{2}\sigma_{I}^{\mathcal{B}}}\right)\right]\\
\\
I_{I}\left(\mathfrak{v}\right)=\mu_{I}\left(\mathfrak{v}\right)-\frac{1}{2}R_{E}J_{IE}\left[1+\mathrm{erf}\left(\frac{\mathfrak{v}-\theta_{E}}{\sqrt{2}\sigma_{E}^{\mathcal{B}}}\right)\right]-\frac{1}{2}R_{I}J_{II}\left[1+\mathrm{erf}\left(\frac{\mu_{I}\left(\mathfrak{v}\right)-\theta_{I}}{\sqrt{2}\sigma_{I}^{\mathcal{B}}}\right)\right],
\end{cases}\label{eq:parametric-equations}
\end{equation}
}
\par\end{center}{\small \par}
\end{spacing}

\noindent for all $\mathfrak{v}$ such that:

\begin{spacing}{0.8}
\begin{center}
{\small{}
\[
\begin{cases}
0<g_{I}\left(\mu_{I}\right)\leq\frac{1}{\sqrt{2\pi}\sigma_{I}^{\mathcal{B}}}\\
\\
\left(\mathcal{J}_{EE}+\mathcal{J}_{II}\right)^{2}-4\left(\mathcal{J}_{EE}\mathcal{J}_{II}-\mathcal{J}_{EI}\mathcal{J}_{IE}\right)>0.
\end{cases}
\]
}
\par\end{center}{\small \par}
\end{spacing}

\noindent Eqs.~\eqref{eq:parametric-equations} are parametric formulas
in the parameter $\mathfrak{v}$, which describe the limit-point ($s=1$)
and period-doubling ($s=-1$) bifurcations in the codimension two
bifurcation diagram of the network. In the case of the network parameters
reported in Tab.~(7) of the main text and $R_{E}=R_{I}=0.5$, Eqs.~\eqref{eq:parametric-equations}
describe analytically the blue and red curves that we obtained numerically
through the MatCont Matlab toolbox in Fig.~(10).

\subsubsection{Neimark-Sacker Bifurcation \label{sub:Neimark-Sacker-Bifurcation}}

The Neimark-Sacker bifurcation is described by the condition $\lambda_{0,1}=e^{\pm\iota\omega}$
for some $\omega\in\mathbb{R}$, where $\iota=\sqrt{-1}$ \cite{Haschke2005}.
If $\left(\mathcal{J}_{EE}+\mathcal{J}_{II}\right)^{2}-4\left(\mathcal{J}_{EE}\mathcal{J}_{II}-\mathcal{J}_{EI}\mathcal{J}_{IE}\right)<0$
and we set $\frac{1}{2}\left(\mathcal{J}_{EE}+\mathcal{J}_{II}\right)=\cos\left(\omega\right)$,
from Eq.~\eqref{eq:mean-field-eigenvalues} we get $\lambda_{0,1}=e^{\pm\iota\omega}$,
provided the condition:

\begin{spacing}{0.8}
\begin{center}
{\small{}
\begin{equation}
\mathcal{J}_{EE}\mathcal{J}_{II}-\mathcal{J}_{EI}\mathcal{J}_{IE}=1\label{eq:Neimark-Sacker-condition}
\end{equation}
}
\par\end{center}{\small \par}
\end{spacing}

\noindent is satisfied. If we introduce again the parameter $\mathfrak{v}\overset{\mathrm{def}}{=}\mu_{E}$,
from Eq.~\eqref{eq:Neimark-Sacker-condition} we get:

\begin{spacing}{0.8}
\begin{center}
{\small{}
\begin{equation}
g_{I}\left(\mu_{I}\right)=\frac{1}{R_{E}R_{I}\left(J_{EE}J_{II}-J_{EI}J_{IE}\right)g_{E}\left(\mathfrak{v}\right)}.\label{eq:derivative-of-the-activation-function-for-the-Neimark-Sacker-bifurcations}
\end{equation}
}
\par\end{center}{\small \par}
\end{spacing}

\noindent Similarly to SubSec.~\eqref{sub:Limit-Point-and-Period-Doubling-Bifurcations},
we can invert Eq.~\eqref{eq:derivative-of-the-activation-function-for-the-Neimark-Sacker-bifurcations}
to obtain the parametric equations that describe the Neimark-Sacker
bifurcation in the codimension two bifurcation diagram of the network.
These equations are defined for all $\mathfrak{v}$ such that:

\begin{spacing}{0.8}
\begin{center}
{\small{}
\[
\begin{cases}
0<g_{I}\left(\mu_{I}\right)\leq\frac{1}{\sqrt{2\pi}\sigma_{I}^{\mathcal{B}}}\\
\\
-1<\frac{1}{2}\left(\mathcal{J}_{EE}+\mathcal{J}_{II}\right)<1.
\end{cases}
\]
}
\par\end{center}{\small \par}
\end{spacing}

\noindent In the case of the network parameters reported in Tab.~(7)
of the main text (with only the exception of the intra-population
synaptic weights, which now are set to $J_{EE}=-J_{II}=10$ for the
bifurcation to occur) and $R_{E}=R_{I}=0.5$, Eqs.~\eqref{eq:parametric-equations}
and \eqref{eq:derivative-of-the-activation-function-for-the-Neimark-Sacker-bifurcations}
describe analytically the green curves that we obtained numerically
through the MatCont Matlab toolbox in Fig.~(10).

\subsubsection{Fixed Point Cycle Curves \label{sub:Fixed-Point-Cycle-Curves}}

For completeness, we describe the relation between the mean membrane
potentials during the oscillatory states (fixed point cycle curves),
and the parameters of the network. If for example the network undergoes
an oscillation with period $\mathcal{T}=2$ (which are caused by a
period-doubling bifurcation), then the mean membrane potentials satisfy
the periodicity condition:

\begin{spacing}{0.8}
\noindent \begin{center}
{\small{}
\begin{equation}
\overline{V}_{\alpha}\left(t\right)=\overline{V}_{\alpha}\left(t+2\right),\quad\alpha=E,\,I.\label{eq:periodicity-condition}
\end{equation}
}
\par\end{center}{\small \par}
\end{spacing}

\noindent By applying iteratively Eq.~\eqref{eq:mean-field-equations-1},
the condition \eqref{eq:periodicity-condition} can be equivalently
rewritten as follows:

\begin{spacing}{0.8}
\begin{center}
{\small{}
\begin{equation}
\overline{V}_{\alpha}\left(t\right)=\mathfrak{F}_{\alpha}\left(\mathfrak{F}_{E}\left(\overline{\boldsymbol{V}}\left(t\right)\right),\mathfrak{F}_{I}\left(\overline{\boldsymbol{V}}\left(t\right)\right)\right),\quad\alpha=E,\,I.\label{eq:mean-field-equations-of-the-fixed-point-cycle-curves}
\end{equation}
}
\par\end{center}{\small \par}
\end{spacing}

\noindent These equations can be solved only numerically or through
analytical approximations, and their solutions correspond to the brown
curves that we obtained through the MatCont Matlab toolbox (see the
top-right panel of Fig.~(10) in the main text).

The oscillations generated by the Neimark-Sacker bifurcation have
period $\mathcal{T}>2$, therefore the corresponding set of equations
can be derived in a similar way by applying Eq.~\eqref{eq:mean-field-equations-1}
iteratively $\mathcal{T}$ times. The complexity of the resulting
equations and of the corresponding solutions increases rapidly with
$\mathcal{T}$, therefore for simplicity in Fig.~(10) of the main
text we showed only the solutions of the case $\mathcal{T}=2$.

\bibliographystyle{plain}
\bibliography{Bibliography}